\documentclass[final,dissertation]{msudoc}

\usepackage{graphicx}
\usepackage{amsmath,amsthm, amssymb}
\usepackage{array}
\usepackage{wrapfig}
\usepackage{lscape}
\usepackage{verbatim}
\usepackage{multirow}
\usepackage{booktabs}
\usepackage[usenames,dvipsnames]{color}

\usepackage{natbib}
\usepackage{setspace}
\usepackage{gensymb} 
\usepackage{longtable}               
\usepackage{url}

\newcommand{\apj}{ApJ}
\newcommand{\araa}{ARA\&A}
\newcommand{\jgr}{J. Geophys. Res.}
\newcommand{\pra}{Phys. Rev. A}
\newcommand{\aap}{A \&\ A}
\newcommand{\apjl}{ApJ}
\newcommand{\apjs}{ApJS}
\newcommand{\solphys}{Sol. Phys.}
\newcommand{\pre}{Phys. Rev. E}
\newcommand{\jcp}{J. Chem. Phys.}
\newcommand{\grl}{Geophys. Res. Lett.}
\newcommand{\ssr}{Space Sci. Rev.}
\newcommand{\nat}{Nature}
\newcommand{\pasj}{PASJ}

\definecolor{yellowA}{rgb}{0.88,0.66,0}

\title{PATCHY RECONNECTION \\ IN THE \\ SOLAR CORONA}
\author{Silvina Esther Guidoni}
\degreetitle{Doctor of Philosophy}
\department{Physics}
\submissiondate{April 2011}
\copyrightyear{2011}

\chair{Dr.~Dana W. Longcope}
\head{Dr.~Richard J. Smith}
\dean{Dr.~Carl A. Fox}

\begin{document}
\frontmatter
\maketitlepage
\makecopyrightpage
\makeapprovalpage
\makepermissionpage
\begin{dedication}
 \vspace{4cm}
 \begin{verse}
  To Egidijus Zilinskas (Egis)
 \end{verse}
\end{dedication}

\makeacknowledgementpage
\maketableofcontents
\makelistoffigures


\begin{abstract}

Magnetic reconnection in plasmas, a process characterized by a change in connectivity of field lines that are broken and connected to other ones with different topology, owes its usefulness to its ability to unify a wide range of phenomena within a single universal principle. There are newly observed phenomena in the solar corona that cannot be reconciled with two-dimensional or steady-state standard models of magnetic reconnection. Supra-arcade downflows (SADs) and supra-arcade downflowing loops (SADLs) descending from reconnection regions toward solar post-flare arcades seem to be two different observational signatures of retracting, isolated reconnected flux tubes with irreducible three-dimensional geometries. This dissertation describes work in refining and improving a novel model of patchy reconnection, where only a small bundle of field lines is reconnected across a current sheet (magnetic discontinuity) and forms a reconnected thin flux tube. Traditional models have not been able to explain why some of the observed SADs appear to be hot and relatively devoid of plasma. The present work shows that plasma depletion naturally occurs in flux tubes that are reconnected across nonuniform current sheets and slide trough regions of decreasing magnetic field magnitude. Moreover, through a detailed theoretical analysis of generalized thin flux tube equations, we show that the addition to the model of pressure-driven parallel dynamics, as well as temperature-dependent, anisotropic viscosity and thermal conductivity is essential for self-consistently producing gas-dynamic shocks inside reconnected tubes that heat and compress plasma to observed temperatures and densities. The shock thickness can be as long as the entire tube and heat can be conducted along tube's legs, possibly driving chromospheric evaporation. We developed a computer program that solves numerically the thin flux tube equations that govern the retraction of reconnected tubes. Simulations carried out with this program corroborate our theoretical predictions. A comparison of these simulations with fully three-dimensional magnetohydrodynamic simulations is presented to assess the validity of the thin flux tube model. We also present an observational method based on total emission measure and mean temperature to determine where in the current sheet a tube was reconnected.

\end{abstract}

\mainmatter

\chapter{INTRODUCTION}
   \label{chap:intro}

\section{The Nearest Star}
  \label{sec:Nearest_Star}

The Sun sustains life on Earth. We are connected to it in an essential way. It powers our world, warms it to a temperature that allows liquid water to be, and drives our weather, evaporating water from the seas and producing winds. Rain and snow irrigate our lands that together with sunlight allow vegetation to grow. Plants generate oxygen that we breathe and feed the animals that we eat. Solar power, directly or indirectly, fuels machines, energizes the lights of our homes, and powers our cars. It is an amazing cycle. 

We move trough space at the right distance from the Sun for life, the way we know it, to thrive. Had the Earth been closer, we might have been a scorching Venus inferno; farther away, a frozen Mars. If there is life somewhere else, we do not know. We know there is life here, and other planets in our solar system do not seem to support it; we might be inclined to assume that habitable conditions are rare. But we only have one data point, and it is overly optimistic to think that we completely understand the circumstances that make life possible. Is planet Earth unique? I doubt it. There are so many planets in the universe, that the likelihood of finding Earth-like planets seems non trivial. What we definitely know is that the Sun plays a determining factor for life. Other stars may do the same in their planetary systems.

\begin{figure}[ht]
  \centering
   \resizebox{4.0in}{!}{\includegraphics*[0, 0][600, 600]{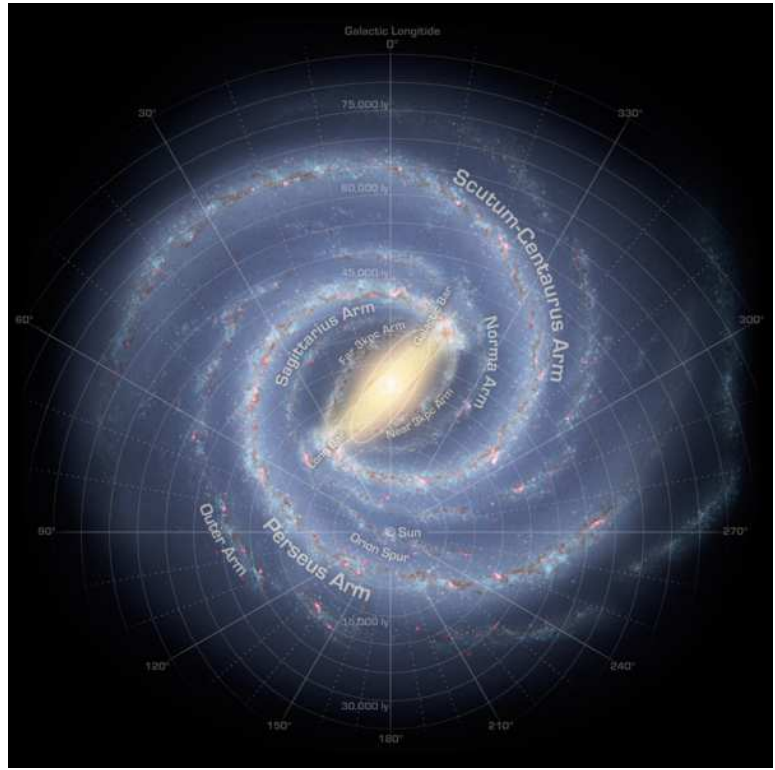}}
   \caption[Milky Way]%
   {Artist's concept illustrating a view of the Milky Way. Our sun lies near a small, partial arm called the Orion Arm, or Orion Spur, located between the Sagittarius and Perseus arms. Image credit: NASA/JPL-Caltech}%
   \label{fig:milky_way} 
\end{figure}

From the universe's viewpoint, our Sun is a fairly ordinary star. Not the brightest. Not the biggest. It is not especially variable or active star, it does not have a peculiar magnetic activity or age. Our solar system is located in a spiral arm of the Milky way galaxy, it is one more among the other stars and planets. Figure \ref{fig:milky_way} shows an artist's concept of our galaxy and the location of our Sun. There are countless galaxies; we are only a speck in the vastness of the cosmos. In addition, the entire human existence is much less than a sigh in the universe's history. Setting this existential lightness aside, it is prudent to understand as much as we can from the star that is the foundation of our existence. We would like to know how it works, why it changes, and how these changes influence us here. 

%
%
%

The Sun is also the key to understand other stars, which are also huge incandescent balls of plasma. Many physical processes that occur elsewhere in the universe can be examined in detail on the Sun. We know its age, radius, mass, and luminosity \citep{Priest_1982}. Through helioseismology we have also been able to learn detailed information about its interior (for a thorough review on helioseismology, see \citep{Christensen_2002}). We now know that we live in the extended solar atmosphere. This outer part of our nearest star is called solar corona and is permeated by magnetic fields generated inside the Sun that extend through the entire solar system. The Sun's visible sharp edge, the photosphere, is something of an illusion. It is the level beyond which the gas in the solar atmosphere is thin enough to be transparent. 

Our star blows itself away sending more than a million tons of electrons and protons \citep{Feldman_1998} into space every second \citep{Withbroe_1977}. This seemingly eternal and stormy solar wind, first predicted by \citet{Parker_1958}, which streams past the Earth at averaged speeds of more than $400$\,km\,s\textsuperscript{-1} \citep{Gosling_1976}, fills the entire solar system. The solar wind was first observed in $1959$ by the Lunik III and Venus I satellites \citep{Priest_1982}, began to be studied in detail by the Mariner II probe in $1962$, and has been studied more recently with Ulysses spacecraft \citep{Aschwanden_SC}. Disturbances in the solar wind shake the Earth's magnetic field and pump energy into the radiation belts \citep{vanAllen_1981}. These belts have harmful effects on satellites and humans working in space \citep{Heliophysics_2010}. 

The Sun's proximity allows us to study it with astonishing resolution. Fortunately, this not only makes it a physical laboratory where plasma theories can be tested, but also makes it a remarkable source of newly observed phenomena. New models are needed to explain observed features that cannot be explained with standard theories. This dissertation is an attempt to unify under the umbrella of a novel model some of such observations. 

\section{The Biggest Explosions in the Solar System}
  \label{sec:Biggest_Explosions}

\begin{figure}[ht]
  \centering
   \begin{center}$
      \begin{array}{c}
      \resizebox{!}{4.0in}{\includegraphics*[0, 0][500, 500]{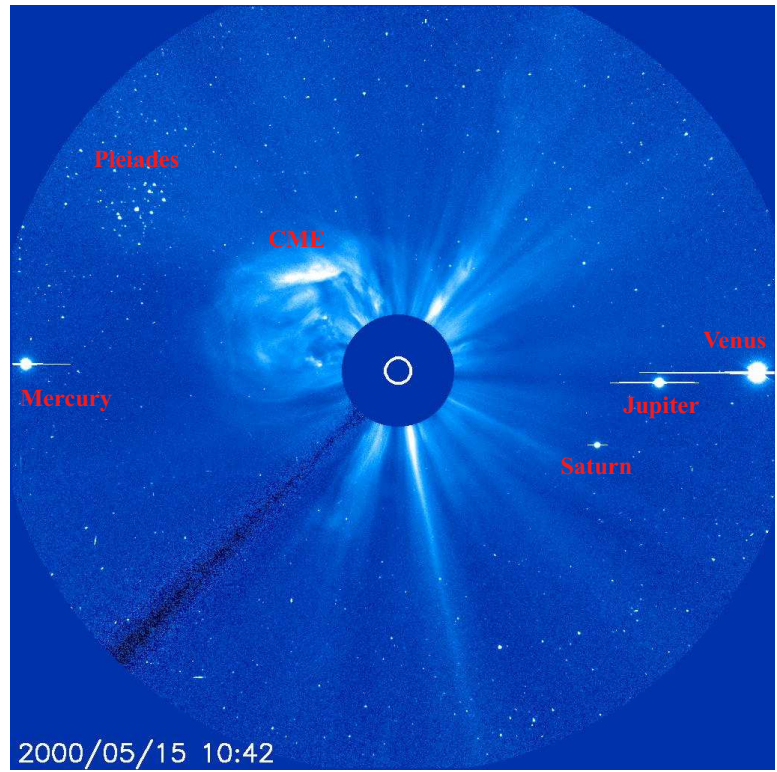}}
      \end{array}$
   \end{center}
   \caption[Coronal mass ejection, planets, and Pleiades]
   {Coronal mass ejection, planets, and Pleiades. A Coronal Mass Ejection erupts from the Sun whose size and position are indicated by the white circle. This image was captured by LASCO C3 coronagraph aboard SOHO, on May 15, 2000. LASCO is able to take images of the solar corona by blocking the light coming directly from the Sun with an occulter disk, creating an artificial eclipse within the instrument itself. The shadow crossing from the lower left corner to the center of the image is the support for the occulter disk. The field of view encompass approximately 31 diameters of the Sun. A rare alignment of four planets is also shown; the bright lines bisecting their disks is an instrumental effect due to the intense sunlight reflected from the planets. The Pleiades star cluster is $408$ light-years away from the Sun. Image credit: Courtesy of SOHO/LASCO C3 consortium. SOHO is a project of international cooperation between ESA and NASA}
   \label{fig:CME_planets} 
\end{figure}

The Sun generates the biggest explosions in the solar system. Enormous eruptions of solar coronal material known as coronal mass ejections (CMEs) send clouds of hot, magnetized plasma out into planetary space (reviews on CMEs can be found, for example, in \citet{Forbes_2000} and \citet{Klimchuk_2001}). They apparently result from a rapid, large-scale restructuring of magnetic fields in the low solar corona. After a few solar radii, they have reached a constant velocity between $400$--$600$\,km\,s$^{-1}$ (gradual CMEs, \citet{Sheeley_1999}), and may exceed $750$\,km\,s$^{-1}$ (impulsive CMEs, \citet{Sheeley_1999}). If directed toward Earth, CMEs typically reach us one to five days after their eruption. They interact with Earth's magnetosphere and may produce powerful geomagnetic storms, intense aurorae, and electric power blackouts. They shake the Earth's magnetic field causing current surges in power lines that destroy equipment and knock out power over large areas. Most of the energy released by the eruption goes to the expelled material, this translates in an average of $10^{30}$ to $10^{31}$\,erg of kinetic energy \citep{Vourlidas_2000}. Energetic protons accelerated by intense solar explosions can inflict serious damage on any astronaut caught in space without adequate shielding \citep{NRC_1996}. Figure \ref{fig:CME_planets} shows a CME erupting from the Sun as observed by the LASCO C3 coronagraph aboard the Solar \&\ Heliospheric Observatory (SOHO) spacecraft \citep{Domingo_1995}. 

%
%

\begin{figure}[ht]
  \centering
   \begin{center}$
      \begin{array}{c}
     \resizebox{5.0in}{!}{\includegraphics*[0, 0][482, 334]{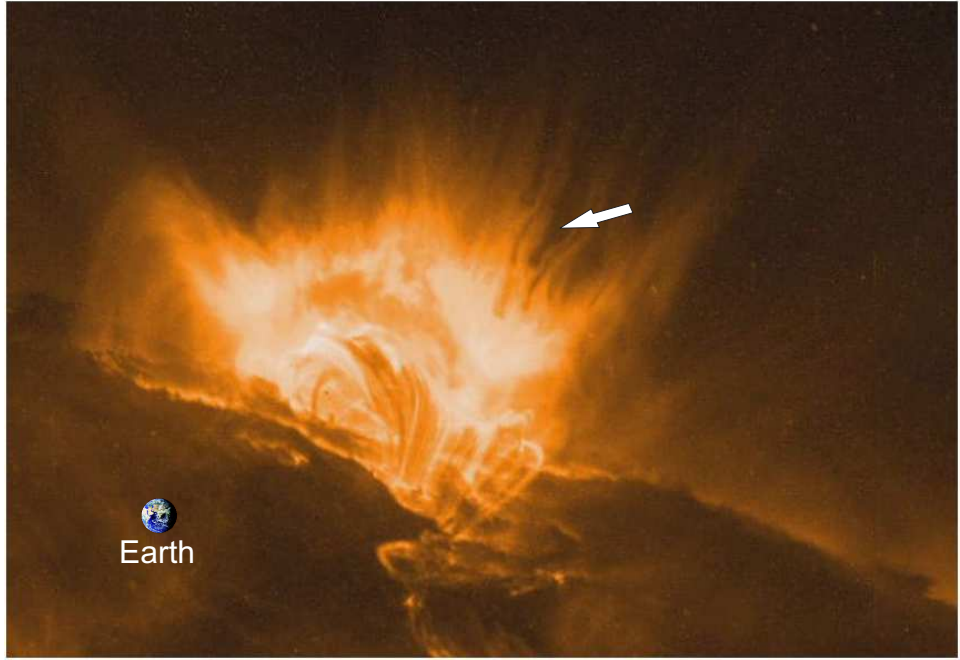}}
      \end{array}$
   \end{center}
   \caption[TRACE flare and dark voids]
   {Flare observed by TRACE. On 21 April 2002, TRACE observed an X1.5 flare in Active Region 9906 at the solar limb (circular horizon). The image is rotated so that north is to the left. The observations, at a high cadence in 195 \AA\ (a mixture of Fe XII at 1.5 million degrees and Fe XXIV at some 10 million degrees) show localized dark blobs that move downward into the bright haze (one of them is shown with a white arrow). An arcade of cooling loops form, with dark, cooled material draining back towards the solar surface. Earth size is shown to scale. Acknowledgement: TRACE is a mission of the Stanford-Lockheed Institute for Space Research, and part of the NASA Small Explorer program}
   \label{fig:trace_flare_21_04_2002} 
\end{figure}

Another dramatic, magnetically energized type of solar explosion are solar flares. They are sudden outbursts that rip through the atmosphere above sunspots, releasing unimaginable amounts of energy, heating solar corona plasma to temperatures between $10$--$20$ MK \citep{Aschwanden_1999}, and sometimes exceeding $40$ MK \citep{Lin_1981, Holman_2003, Caspi_2010}. In just $100$ to $1000$ seconds, the blast can release $10^{32}$ erg in a relatively compact region \citep{Aschwanden_SC, Crosby_1993}. Figure \ref{fig:trace_flare_21_04_2002} shows an image in 195 \AA\ of a flare observed by the Transition Region and Coronal Explorer (TRACE) \citep{Handy_1999}.

In solar flares, most of the magnetic energy is mainly transferred into accelerated particles that subsequently emit intense X-ray, $\gamma$-ray, and radio radiation, but the exact acceleration mechanisms behind this process remain poorly understood \citep{Lin_1971, Lin_1976, Emslie_2004, Emslie_2005, Caspi_2010_PhD}. This radiation from solar flares may heat up the Earth's upper atmosphere, changing its electrical properties, disrupting radio navigation or telecommunications, and making it expand farther into space. Friction between the expanded atmosphere and satellites may shorten mission lifetimes. As we become more dependent upon satellites in space, electronics, and telecommunications, we will increasingly feel the effects of this space weather. The need to understand and predict flares, as well as CMEs is clear. 

%
%


\section{Magnetic Reconnection}
  \label{sec:mag_rec}

The only plausible source of energy for the colossal explosions mentioned in the previous section is the strong magnetic field in the solar corona \citep{Forbes_2000}. Even though the plasma in the lower solar corona is hot (approximately $1$ MK), the magnetic pressure there greatly exceeds the gas pressure. Solar flares occur in active regions associated to sunspots, where magnetic fields are clumped together in intense bundles. Flares and CMEs are synchronized with an $11$-year cycle \citep{Hoyt_1997,Hoyt_1997_I} of solar magnetic activity. The number of active regions, with their energetic and magnetized loops, varies from a minimum to a maximum and back to a minimum with a period of approximately $11$ years. The frequency of solar explosions are greatest at the maximum of the cycle and lowest at the minimum of the cycle \citep{Norquist_2011}. 

The coronal plasma can generally be approximated as a perfectly-conducting fluid \citep{Priest_1982,Low_2003} which thus obeys the principles of ideal magnetohydrodynamics (MHD). This zero-resistivity approximation immediately leads to the``frozen-in'' condition  that guarantees that fluid particles and field lines move together \citep{Priest_1982,Priest_2000,Birn_2007}. Therefore, plasma is confined by the magnetic field and shaped mostly into coronal loops (like the ones shown in Figure \ref{fig:trace_flare_21_04_2002}). 

The magnetic energy necessary to power solar explosions is stored in the form of non-potential magnetic field energy \citep{Priest_1975, Hu_1982, Aly_1989, Birn_2007}. An equilibrium field with no currents except for a current sheet (magnetic discontinuity) is the lowest energy state (for perfectly conducting evolution) subject to a photospheric flux constrain \citep{Longcope_2001}, therefore they are stable under ideal MHD evolution. The current sheet raises the field energy with respect to a potential field with the same photospheric flux. If the perfectly conducting assumption breaks, the field reorganizes itself toward lower energy states. This restructuring of the magnetic field, called magnetic reconnection, is characterized by a change in connectivity of field lines that are broken and connected to other ones with different topology. In this way, large amounts of stored magnetic energy can be released as thermal, kinetic, and radiative energy, as well as accelerated particles \citep{Mandrini_2010}. 

Magnetic reconnection owes its appeal to its ability to unify a wide range of phenomena within a single universal principle \citep{Forbes_1991,Priest_2000}. It is not only present in many theories that undertake to explain the initiation and evolution of flares or CMEs (\citealt{Lin_2000, Forbes_1996, Forbes_2006,Schrijver_2009}; see also \citealt{Forbes_2000} for energy estimations in CMEs), but it is also involved in the interaction of the solar wind with Earth's magnetosphere, where substorms originate \citep{Pudovkin_1985}. (For a recent review on magnetic reconnection in space plasmas and laboratory, see \citet{Yamada_2010}.) 

Some kind of diffusion mechanism initiates the connection of field lines from one side of a current sheet to field lines on the other side. This transfer of flux from one domain to another decreases the energy of the field. The exact nature of this process is not yet well understood, but it has recently become clear that it must produce an electric field restricted to a small portion of the current sheet \citep{Biskamp_2001,Birn_2001}.

\begin{figure}
   \centering
   \begin{center}$
      \begin{array}{c}
         \resizebox{5.0in}{!}{\includegraphics*[0,0][360,410]{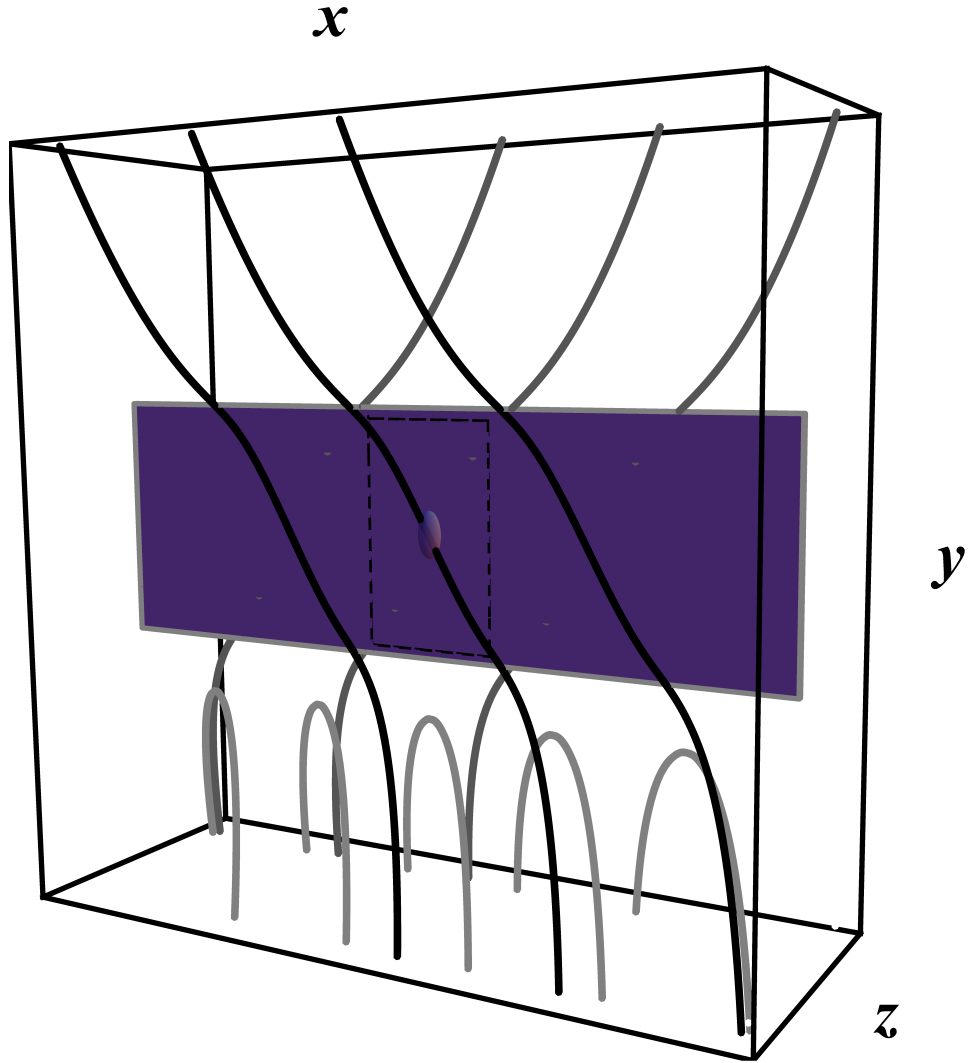}}
      \end{array}$
   \end{center}
   \caption[Flare current sheet cartoon]
   {Flare current sheet cartoon. The solid rectangle represents the current sheet plane located above the post-flare arcade (gray arc-shaped loops at the bottom). The black lines depict some magnetic field lines on the side of the current sheet that is closer to the viewer, and the dark gray ones correspond to field lines on the back side of the current sheet. The lower plane of the box represents the solar surface. The small sphere in the current sheet shows a patchy reconnection region. Only a small bundle of field lines that intersect this region are going to reconnect. The dashed rectangle in the plane of the current sheet corresponds to the region shown in Figure \ref{fig:current_sheet}}
   \label{fig:flare_CS_ch1} 
\end{figure}

A schematic flare-like scenario is depicted in Figure \ref{fig:flare_CS_ch1}, where an arcade similar to the one in Figure \ref{fig:trace_flare_21_04_2002} is represented by gray arc-shaped loops. Reconnection occurs within the current sheet above the arcade of post-flare loops, represented in the figure by a solid rectangle. In this figure, skewed magnetic fields are assumed on each side of the current sheet. Field lines on the side of the current sheet that is closer to the viewer (solid black lines) form a finite angle with field lines on the back side of the current sheet (gray solid lines with endpoints in the top and bottom walls of the box). 

%
%

Magnetic reconnection was first quantitatively modeled in two dimensions and steady-state configurations with uniform resistivity \citep{Sweet_1958,Parker_1957}, but the energy release in this model was too slow to explain the rapid development observed in flares and CMEs \citep{Parker_1963}. This problem was solved by assuming a two dimensional steady electric field {\em localized} to a small diffusion region with four slow-mode shocks (SMSs) attached to it. At these shocks, the magnetic field is deflected and partially annihilated, and the plasma is heated and accelerated \citep{Petschek_1964}. The compressible plasma case was studied by \citet{Soward_1982}.

Localized, {\em non-steady} models began with the work of \citet{Semenov_1983}, followed by \citet{Biernat_1987}, and \citet{Nitta_2002} all of which were two dimensional. For these spatially localized (in only one dimension) and short-lived episodes, four slow-mode shocks (SMSs) extend from the reconnection site as well, but they close back together in a finite region that increases in size as time goes by (as opposed to the infinite length SMSs from steady state). These teardrop-shaped SMSs continue growing and retracting as they sweep up mass, even if reconnection ceases \citep{Semenov_1998}. Therefore, this conversion of magnetic energy into kinetic and thermal energy continues and the dissipated energy can exceed the ohmic dissipation from the reconnection electric field, indicating that a small scale, short lived event can affect the global structure of magnetic fields. 

\begin{figure}
   \centering
   \begin{center}$
      \begin{array}{c}
         \resizebox{5.0in}{!}{\includegraphics*[0, 0][865, 473]{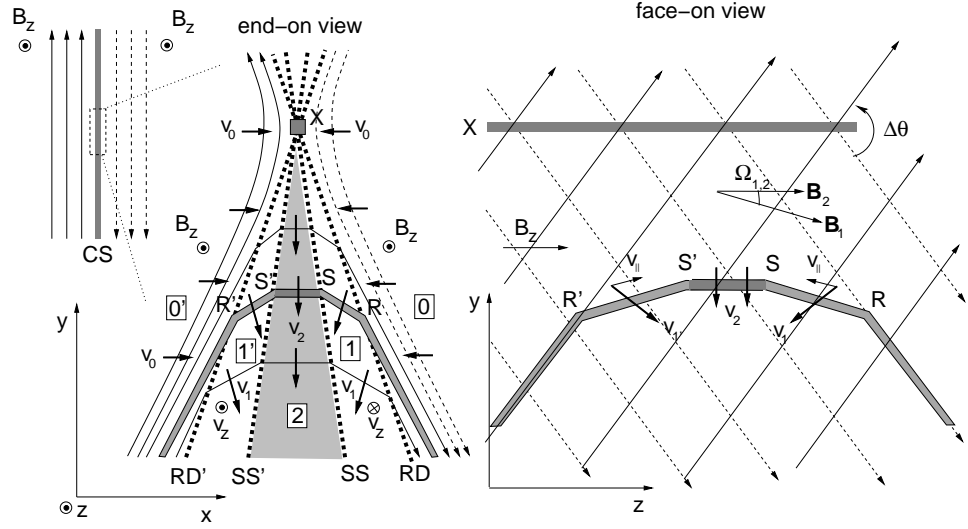}}
      \end{array}$
   \end{center}
   \caption[Petschek reconnection of skewed magnetic fields]
   {Petschek reconnection of skewed magnetic fields. Reconnection occurs within a portion of a large-scale current sheet (far left). Two perspectives of the
reconnection region are shown: the end-on view (center, whose horizontal scale is exaggerated for clarity) and the face-on view (right). Flux transfer occurs along a line X (gray) along the ignorable direction ($\widehat{\mathbf{z}}$). Shocks, RD', SS' (SMS), SS (SMS), and RD (thick dashed lines) originate there, dividing the reconnection region into sub-regions labeled $0'$, $1'$, $2$, $1$, and $0$ (inside boxes). The central region, $2$ (shaded gray) contains plasma which has been heated and compressed by the SSs (SMSs). Reconnected field lines (thin solid lines) bend at each shock. A single flux tube is called out in gray in each panel. It passes though the shocks at points R' , S' , S, and R, left to right. Credit: \citet{Longcope_2010_I}}
   \label{fig:RD_SMS_skewed} 
\end{figure}

All of the models above considered reconnection between magnetic field lines assumed to be perfectly antiparallel. A departure from this special case, called reconnection between {\em skewed} fields, introduces a magnetic field component in the ignorable direction, called a guide field; such models are sometimes called 2.5-dimensional (Figure \ref{fig:flare_CS_ch1} is an example of this kind of configuration). In this case, it is necessary to supplement the SMSs with rotational discontinuities \citep[RDs;][]{Petschek_1967,Soward_1982_II,Heyn_1988,Skender_2003}, as shown in Figure \ref{fig:RD_SMS_skewed}, from \citet{Longcope_2010_I}. In the special case of uniform anti-parallel fields, these two distinct shocks merge into a single SMS of the ``switch-off type'' ($y$-component of the magnetic field is annihilated), often associated with the Petschek model. The effect of the RDs is to produce bulk flow in the ignorable direction in order that the velocity change, magnetic fields, and shock normal may all lie in a single plane; this is a requirement of the SMS often called ``co-planarity'' \citep{Colburn_1966}.  This new flow component is confined to the region between the RD and SMS. Non-steady reconnection between skewed magnetic fields was analyzed by \citet{Semenov_1992} and \citet{Biernat_1998} for dayside magnetospheric events. 

%
%

Common to all the reconnection models is the presence of Alfv\'enic outflows originating at the reconnection site. In steady state models, these form coherent jets. In transient reconnection models, any outflow is more accurately described as a distinct retraction event rather than as a steady jet. It is the RDs rather than the SMSs which accelerate the reconnection flux to the Alfv\'en speed in an outflow jet.

While all of the foregoing reconnection models predict shocks, few investigations of them have explicitly included viscosity or thermal conductivity. This omission is noteworthy since strictly speaking the irreversible nature of any shock demands a non-ideal effect; in the collisional plasma of the solar corona this is usually assumed to be viscosity. For strongly magnetized plasma, viscous diffusion and thermal conduction are highly anisotropic. They are directed along the magnetic field since perpendicular coefficients are many orders of magnitude smaller than their parallel counterparts \citep{Braginskii_1965}. Such strong anisotropy poses challenges for models, especially for numerical solutions. In addition, transport coefficients in a fully ionized plasma have a strong dependence on temperature ($\sim T^{\frac{5}{2}}$) making transport far more significant at high post-shock temperatures than it is assumed to be in general \citep{Spitzer_1962}. It has therefore been recognized as a potentially significant factor in flares \citep{Cargill_1995,Forbes_1986}. In particular, the heat generated at SMSs is predicted to be conducted along reconnected field lines ahead of other reconnection effects, possibly driving chromospheric evaporation \citep{Forbes_1989}. 

Nevertheless, reconnection models generally neglect viscosity altogether to say nothing of accounting for its strong temperature dependence and directional anisotropy. \citet{Craig_2005} and \citet{Litvinenko_2005} studied the effects of temperature-independent viscosity in various reconnection disturbances. Only recently, \citet{Craig_2008} and \citet{Craig_2009} included anisotropic and temperature-dependent viscous dissipation in a three dimensional steady-state model at X-points. Yokoyama \& Shibata (1997, 2001) included anisotropic thermal conduction in two dimensional simulations of reconnection, and \citet{Seaton_2009} presented a detailed theoretical analysis of these simulations. In most of these two-dimensional, anti-parallel models, the heat conduction front extends in front of SMSs, as expected.

Direct observation of the actual reconnection process has been elusive. Its internal structure remains unknown, even for well-diagnosed laboratory experiments \citep{Katz_2010}. In order to explain the short times involved in the mentioned astrophysical explosions, fast reconnection \citep{Petschek_1964} ought to occur in a minuscule diffusion region on a CS where non-ideal field-line transport coefficients are locally enhanced. This requires measurements at prohibitively high-resolution. Several fast reconnection models include localized magnetic resistivity enhancement on the CS \citep{Ugai_1977,Scholer_1987,Erkaev_2000,Ma_2001, Biskamp_2001}. 

Only in the last 15 years or so has technology started to close the gap between theory and observations. Nonetheless, only indirect measurements of solar corona reconnection have been possible. For example, the cusp-shaped, hot and dense flare loops \citep{Tsuneta_1992, Forbes_1996} consistent with the classical CSHKP model (\citealt{Carmichael_1964, Sturrock_1968, Hirayama_1974, Kopp_1976}; see Figures 1 and 2 in \citealt{Forbes_1996} for a modern version including thermal conduction) are considered observational evidence of reconnection. These arcades, slowly growing and whose footpoints separate as their heights rise, are interpreted as closed magnetic loops formed after two-dimensional, steady-state reconnection. They pile up on top of each other to form the arcades. The earlier loops cool down as new loops lie on top of them, and an apparent footpoint motion is observed. The loops are hot, as expected from the conversion of magnetic energy into thermal energy from reconnection. 

There are several other indirect observations of reconnection, like reconnection inflows \citep{Yokoyama_2001_I,Narukage_2006,Lin_2005} with speeds between a few km s\textsuperscript{-1} to approximately $106$ km s\textsuperscript{-1}, as well as Alfv\'{e}nic ($460-3500$ km s\textsuperscript{-1}) reconnection outflows \citep{Lin_2005,Wang_2007} and reconnection in/out pairs \citep{Sheeley_2007, Savage_2010}. Loop shrinkage also seems to indicate reconnection. \citet{Forbes_1996}, using images of flare loops taken with SXT \citep{Tsuneta_1991} aboard Yohkoh \citep{Ogawara_1991}, showed reconnected cusped loops shrinking to form post-flare arcades. This process takes hours and is different from the fast passage ($1$ -- $2$ minutes) of field lines through the reconnection outflows (jets). Hard X-ray (HXR) sources near the apex of soft X-ray (SXR) loops indicate some high-energy process, possibly electron acceleration in that region \citep{Masuda_1994,Krucker_2008,Tomczak_2009}. The acceleration may occur at the reconnection site or might be the result of the interaction of the downward reconnection outflow jet with the top of the SXR arcade, producing superhot plasmas \citep{Shibata_1995}. Plasmoid (magnetic island) formation \citep{Forbes_1983}, as a consequence of reconnection, has been widely observed (e.g., \citet{Shibata_1995,Nishizuka_2010}). For a comparative study of plasmoid in reconnecting CSs (solar and terrestrial contexts), see \citet{Lin_2008}. 

%

\section{Patchy Reconnection}
  \label{sec:patchy_rec}

Even though most of the observations mentioned in Section \ref{sec:mag_rec} can be understood by standard, steady-state, and two-dimensional reconnection theories, there are newly observed phenomena that cannot be reconciled with these representations. Supra-arcade downflows (SADs) and supra-arcade downflowing loops (SADLs) \citep{Savage_2010,Savage_2011} seem to be two different observational signatures of retracting, isolated reconnected flux tubes with irreducibly three-dimensional geometries. 

SADs, dark voids descending from reconnection regions, were first reported by \citep{McKenzie_1999} and since then have been observed by several instruments \citep{McKenzie_2000, McKenzie_2001,Innes_2003, Innes_2003_II, Asai_2004, McKenzie_2009, Savage_2010,Savage_2011}. They are directly related to the path of bright loops that appear lower down in the corona \citep{Sheeley_2004}. These downward-moving features are believed to represent flux tubes, created by patchy reconnection, retracting downward to later form the post-eruption arcade \citep{McKenzie_2009}. They descend through a supra-arcade fan (see \citealt{Svestka_1998} for a description of one of such fans) toward an underlying flare arcade and decelerate close to the arcade apex \citep{Sheeley_2002, Sheeley_2004}. The white arrow in Figure \ref{fig:trace_flare_21_04_2002} points to one of these dark voids. 

\begin{figure}
   \centering
   \begin{center}$
      \begin{array}{cc}
         \resizebox{3.5in}{!}{\includegraphics*[74, 360][394, 735]{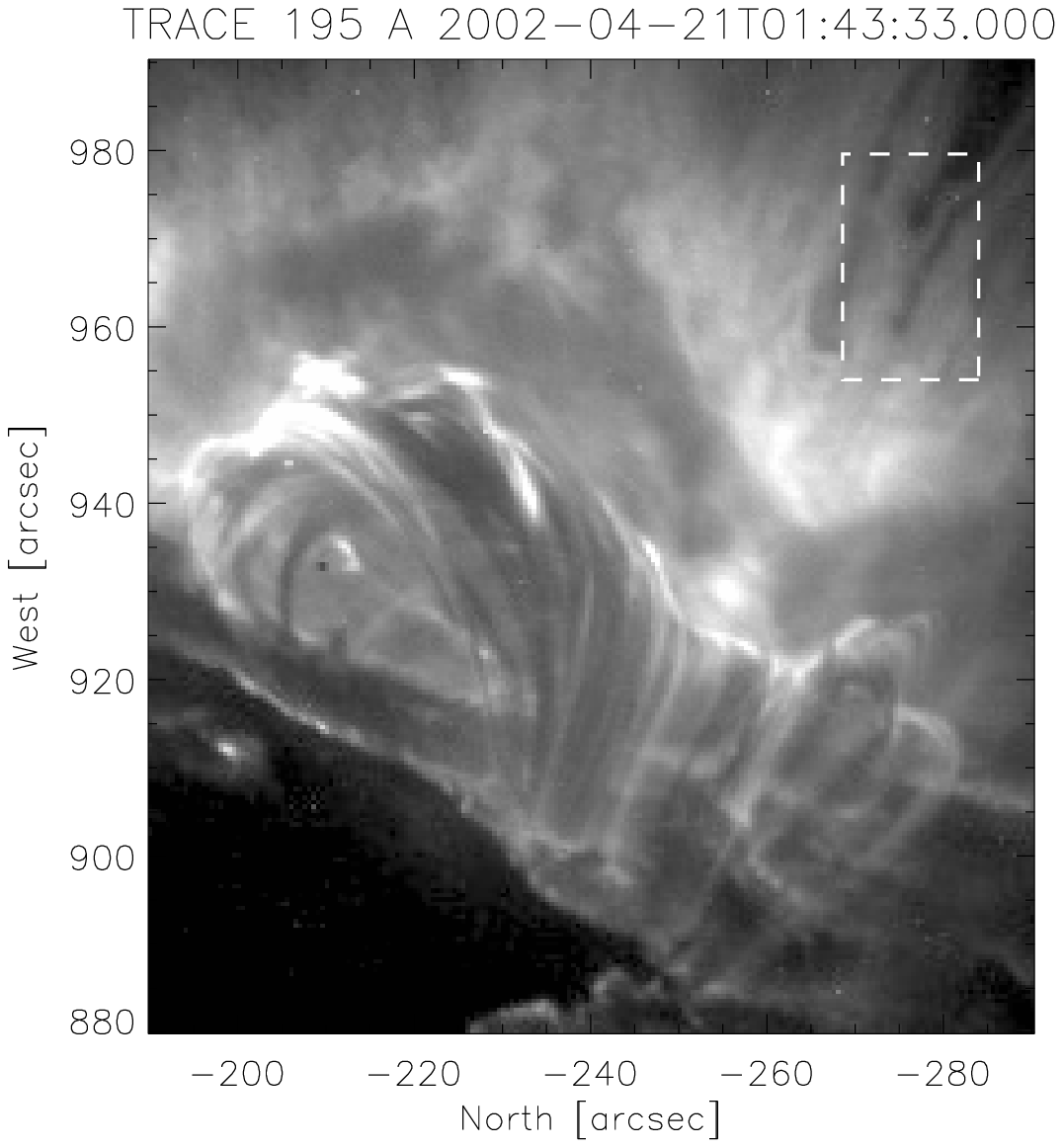}}
         \resizebox{!}{3.9in}{\includegraphics*[114, 350][370, 813]{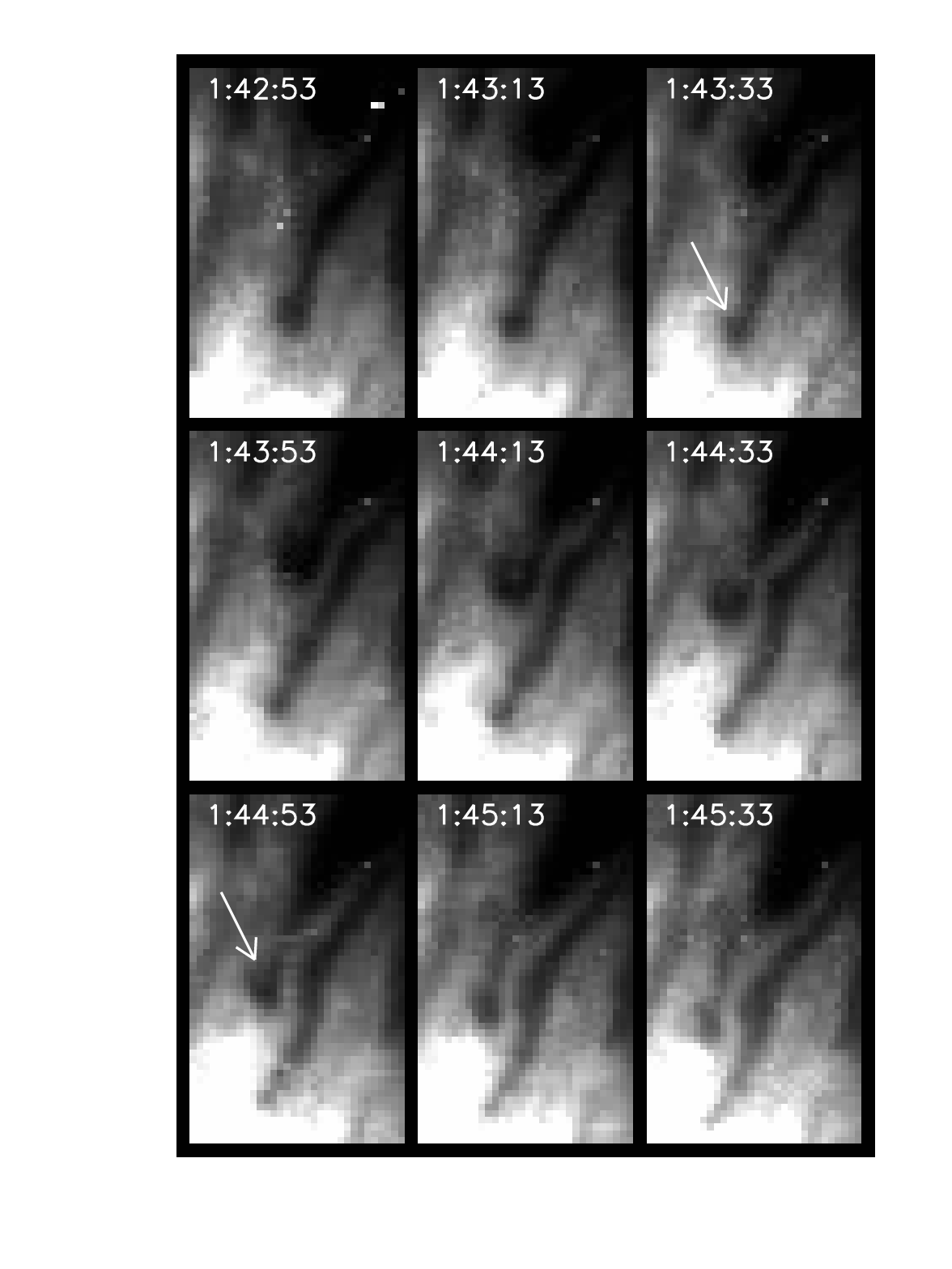}}
      \end{array}$
   \end{center}
   \caption[Dark voids descending toward a flare arcade]
   {2002 April 21 flare. Left panel: rotated TRACE 195 \AA\ image, taken at a time when dark voids can be seen descending toward the flare arcade. The dashed rectangle encloses a handful of them. Right panel: time sequence for the window delineated by the white dashed rectangle in the left panel. The white arrow in the 1:43:33 snapshot shows the position of one of the dark pockets. The other arrow points to a second localized void}
   \label{fig:TRACE_SADs} 
\end{figure}

SADs are spatially localized as dark pockets, followed by a dark wake or lane. These downflows are observed as high as $40-60$ Mm above the top of soft X-ray arcades and interpreted as magnetic reconnection localized outflows \citep{McKenzie_2000}. Therefore, SADs appear to be the result of intermittent or patchy reconnection high in the corona. Three-dimensional and time-dependent models are needed to explain them. Figure \ref{fig:TRACE_SADs} shows examples of SADs, as observed by TRACE \citep{Handy_1999,Schrijver_1999,Golub_1999} in 195 \AA\  passband. The image to the left of the figure shows an arcade for the 2002 April 21 X-class flare \citep{Wang_2002, Gallagher_2002, Innes_2003_II}, where two SADs are enclosed by a white dashed rectangle. Usually, arcades like this one can be described as two dimensional or $2.5$-dimensional, suggesting the presence of a wide spread CS above them (as shown in Figure \ref{fig:flare_CS_ch1}). The image to the right is a time sequence for the location delimited by the dashed rectangle in the left panel. The white arrow on the top right snapshot indicates a dark pocket (SAD's head) that moves toward the arcade. This pocket is followed by a dark wake. The white arrow at time 1:44:53 indicates a second dark pocket (SAD) moving in the same direction. 

%
%

A three-dimensional model of patchy reconnection \citep{Klimchuk_1997}, accompanied by magnetohydrodynamic (MHD) simulations, has been presented by \citet{Linton_2006}. They studied the relaxation of solar-like low plasma-$\beta$ (ratio between thermal and magnetic pressure) reconnected flux tubes. Reconnection is assumed to occur across a uniform CS with skewed magnetic fields. Even though the initial configuration is $2.5$-dimensional, reconnection is initiated within a small patch, producing two distinct bundles of reconnected field lines, flux tubes; the ensuing reconnection dynamics is irreducibly three-dimensional. 

In that study, resistivity is briefly enhanced by an unspecified mechanism within a localized and short-lived region (an illustrative reconnection patch is depicted in Figure \ref{fig:flare_CS_ch1} as a small sphere). This creates a pair of bent, thin reconnected flux tubes that retract in opposite directions. They found that the perpendicular dynamics of a reconnected tube was well described by equations for thin flux tubes \citep[TFT;][]{Spruit_1981}. The equations differed from those used traditionally in convection zone models, by their application to the low-$\beta$ corona. RDs propagate along the legs of the tubes. Only the perpendicular dynamics of the tubes was analyzed by \citet{Linton_2006}, who also argued that the descending coronal voids are reconnected flux tubes descending from a flare site high in the corona. 

\citet{Longcope_2009} extended that study to include the tube's parallel dynamics. The initial sharp angle at the reconnection site creates bends that move along the legs of the tubes as RDs. These bends rotate the magnetic field and generate parallel supersonic inflows toward the center of the tube. There, the collision of these inflows launches outwardly two gas-dynamic shocks (GDSs) that heat and compress the plasma without changing the magnitude of the magnetic field. The GDSs are disconnected from the resistive region and are features of the ideal relaxation following reconnection. The rate of reconnection is not relevant in this case since the diffusion region is short-lived and does not participate in the post-reconnection evolution. 

The GDSs are parallel shocks and close relatives to the nearly parallel SMS in $2.5$-dimensional steady models. In fact, the post-GDS temperatures in the TFT model are almost identical to the post-SMS temperatures in steady-state magnetohydrodynamic MHD models. In both kinds of models, plasma heating occurs through a two-stage process notably different from commonly used switch-off models, applicable to purely anti-parallel reconnection. First, the RD converts magnetic energy to bulk kinetic energy in the form of Alfv\'en-speed flow partly {\em parallel} to the field line. The parallel component of this flow, present in both, is then thermalized in a second shock of large {\em hydrodynamic} Mach number (see Figure 1 in \citealt{Longcope_2010_I} for a single flux tube as it passes through these shocks). 

\begin{figure}
   \centering
   \begin{center}$
      \begin{array}{c}
         \resizebox{5.0in}{!}{\includegraphics*[54, 226][535, 550]{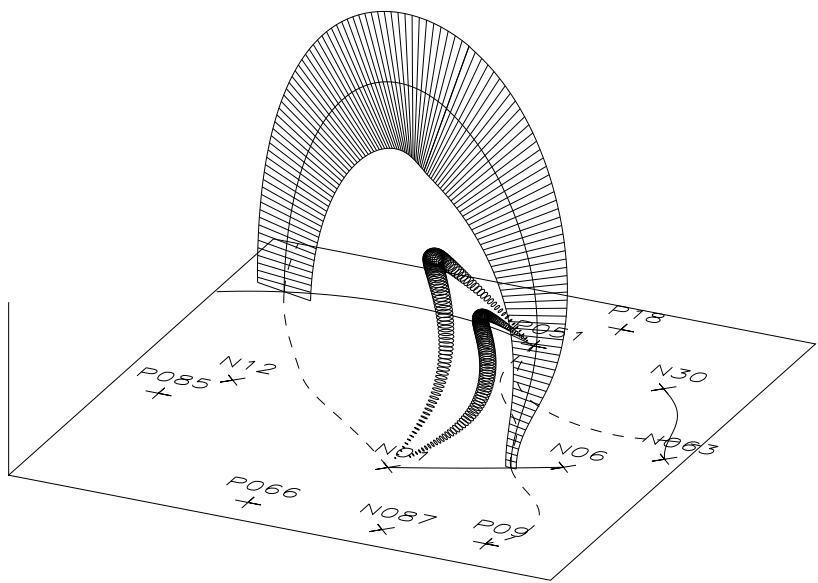}}
      \end{array}$
   \end{center}
   \caption[Illustration in perspective of a model of reconnection]
   {Illustration in perspective of a model of reconnection. The model assumes reconnection between solar active regions 9570 and 9574. The vertical separator current sheet is shown as a skeleton, and a pair of reconnected flux tubes, $4 \times 10^{18}$ Mx each, created by patches of reconnection on the separator current sheet, lie underneath it. Credit: \citet{Longcope_2005}}
   \label{fig:Longcope_2005_fig} 
\end{figure}

Patchy reconnection seems to agree with some observations of reconnection. \citet{Longcope_2005}, through EUV observations and modeling, analyzed an emerging active region in the vicinity of an existing one. They studied non-flaring reconnection along a separator overlying the volume between them. They characterized $43$ loops observed by TRACE that resulted from reconnection between the active regions, and assumed a correspondence between them and modeled flux tubes. The average diameter of the loops is $3.7$ Mm with very small variation, suggesting a sporadic or patchy reconnection. The estimated flux of each one of them is $~4 \times 10^{18}$ Mx for an assumed magnetic field of $37$ G. Figure \ref{fig:Longcope_2005_fig} reproduces Figure $19$ of that paper and illustrates the reconnection model applied to the active regions. The three-dimensional CS is shown, as well as a pair of reconnected flux tubes created by patches of reconnection. The time history of the loops (observed at 171 \AA) with very similar diameters suggests a highly intermittent reconnection process. 

%
%

Although what triggers reconnection is still an open question, several promising mechanisms have been proposed. For example, the MHD CS tearing mode instability \citep{Furth_1963_I,Sturrock_1966,Forbes_1983,Priest_1985, Biskamp_1986} may initiate bursty reconnection. \citet{Shimizu_2009} showed numerically that a two-dimensional fast magnetic reconnection in a simple one-dimensional CS can be unstable to three-dimensional resistive perturbations. Therefore, it is possible to randomly eject three-dimensional magnetic loops along a CS atop a post-flare arcade, resembling SADs. Catastrophe models explain how magnetic configurations are stable for a long time and suddenly release their energy \citep{Cassak_2005}. The Hall reconnection model, extensively studied in the Geospace Environmental Modeling (GEM) magnetic reconnection challenge (\citealt{Birn_2001}, and companion papers), yields energy release rates consistent with the ones observed in the solar corona \citep{Drake_2006_I}.

Even though three-dimensional outflows can be generated from purely two-dimensional configurations, the assumption of perfectly anti-parallel pre-reconnection magnetic fields seems too restrictive for the complex solar corona, where field lines are likely to form finite angles across CSs.  Furthermore, the addition of a guide field is directly related to the temporal evolution of reconnection. \citet{Drake_2006_I} showed that ambient guide magnetic field controls whether magnetic reconnection, once it is established, remains steady (anti-parallel reconnection) or becomes bursty (reconnection with a guide field). Through particle simulations in a system consisting of two Harris CSs \citep{Harris_1962}, they show that magnetic reconnection in a large-scale system with a guide field will break up into many islands, as a result of tearing instability.  Guide fields with a shear angle of $127$\degree\ are sufficient to produce secondary islands that grow to finite amplitude, producing well-developed flux tubes. 

Besides the assumption of being retracting reconnected flux tubes, there are several unanswered questions about SADs. To the best of our knowledge, there has not been a satisfactory explanation of why they seem to be hot and relatively devoid of plasma. \citet{McKenzie_1999}, using an image-ratio method between the SXT Al$12$ and AlMg filters \citep{Tsuneta_1991}, reported temperature ratios between the voids and the surrounding rays of the fan of $9.1:7.9$. The wakes seem to have a slightly lower ratio: $8.6:7.9$. The density ratios between the voids and the rays ranged between $1.0:4.0$ to $1.3:2.3$, depending on the assumptions about the line-of-sight depth. The darkness of the voids in X-ray and extreme-ultraviolet (EUV) images and spectra, together with the lack of absorption signatures in EUV \citep{Innes_2003} is best explained as pockets of very low plasma density. \citet{Innes_2003} ruled out the possibility that the dark tracks are caused by coronal rain (cold plasma falling gravitationally). 

There is another piece of the puzzle that has not been completely understood: SADs' reported speeds ($30$ -- $500$ km s\textsuperscript{-1}) are smaller than the assumed Alfv\'{e}n speed in the corona \citep{McKenzie_1999,McKenzie_2000, Asai_2004, McKenzie_2009, Savage_2010}. This contradicts any standard reconnection theory. While few if any of these events have measured speeds matching the local Alfv\'en speed, they are generally interpreted as evidence of reconnection. Either these outflows/downflows were Alfv\'enic at some earlier time or all reconnection theories must be revised. A possible explanation for reconnected tube's deceleration is given by \citet{Linton_2006}. They showed that a drag effect from the plasma around the tube can reduce its speed considerably. The external fluid deforms and accelerates as the tube moves through it; this requires energy and therefore slows the tube. The ratio between the speed observed in their simulations and the perpendicular Alfv\'{e}n speed (expected speed from the TFT model) increases with the reconnection angle and ranges approximately between $0.1$ and $0.5$ for the low plasma-$\beta$ (ratio between thermal pressure and magnetic pressure) case. This ``added mass'' effect accounts for most of the flux tube drag at high plasma-$\beta$, but is not enough, however, to explain the speed reduction at low plasma-$\beta$ (the expected speeds taking into account added mass effects are double the one observed in the simulations). They conclude that there is an additional source of drag in this case, but they do not speculate about its origin. 

\section{Dissertation Layout}
  \label{sec:layout}

This dissertation describes work done in refining the patchy reconnection model presented by \citet{Linton_2006} and \citet{Longcope_2009}, in order to answer some of the questions stated above. 

Chapter \ref{chap:chap_1} presents generalized TFT equations for the dynamics of reconnected flux tubes that include pressure-driven parallel dynamics as well as temperature-dependent, anisotropic viscosity and thermal conductivity. These additions are essential for self-consistently producing the GDSs predicted analytically by \citet{Longcope_2009} using conservation laws. 

A model for plasma heating produced by time-dependent, spatially localized reconnection within a flare current sheet separating skewed uniform magnetic fields (same reconnection scenario studied by \citet{Longcope_2009}) is presented. The reconnection patch creates flux tubes of new connectivity which subsequently retract at Alfv\'enic speeds from the reconnection site. Heating occurs in gas-dynamic shocks which develop inside these tubes. The evolution of the reconnected tubes is studied through numerical solutions of the TFT equations, for realistic solar coronal conditions. 

The strong gas-dynamic shocks are generated by compressing plasma inside reconnected flux tubes. They develop large velocity and temperature gradients along the tubes, rendering the diffusive processes dominant. These diffusive processes determine the thickness and internal structure of the shocks that evolve up to a steady-state value, although this condition may not be reached in the short times involved in a flare. This internal structure is analized using methods similar to those of \citet{Thomas_1944}, \citet{Grad_1952}, \citet{Gilbarg_1953}, and \citet{Kennel_1988}. For strong shocks at low Prandtl numbers (ratio of viscosity to thermal conductivity), typical of the solar corona, the GDSs consist of a long thermal front where the temperature increases and most of the heating occurs, followed by a narrow isothermal sub-shock where plasma is compressed. The length of each of these sub-regions and the speed of their propagation is theoretically estimated. The thickness of the shocks is comparable to the entire length of the tubes, far greater than the term ``internal structure'' might suggest. The analysis is novel in its focus on solar flare conditions, including very low Prandtl numbers and transport coefficients with strong temperature-dependence. 

\citet{Fowles_1975} showed that for strong shocks, it is impossible to assume inviscid behavior, no matter how large thermal conduction is. \citet{Coroniti_1970} studied dissipation discontinuities in shock waves for temperature-independent transport coefficients. The seemingly harmless assumption of temperature-independent coefficients leads to shock thicknesses on the order of the particle's mean free path, raising concerns about the adequacy of a fluid treatment: when the thickness of the shock is of the order of particle mean free path, the fluid model breaks down \citep{Campbell_1984}. However, if transport coefficients are given an accurate temperature-dependence, we have found that the thickness of the shocks exceeds the particle mean free paths by at least an order of magnitude. Our simulations also show that the ratio between the heat flux and the nominal free-streaming heat flux is smaller than $0.004$. These two results allow us to treat our plasma as a collisional fluid. 

The work described in Chapter \ref{chap:chap_1} has been published in the Astrophysical Journal \citep{Guidoni_2010}. 

In Chapter \ref{chap:chap_2}, through a simple but realistic fluid model of patchy reconnection, it is shown that plasma depletion naturally occurs in flux tubes that are reconnected across Syrovatski{\v i}-type (non uniform) CSs \citep{Green_1965, Syrovatskii_1971} with skewed magnetic fields (similar to the one shown on figure \ref{fig:flare_CS_ch1}). In this kind of CS, the background magnetic pressure has its maximum at the center of the CS plane and decreases toward its edges. The reconnection patch creates two V-shaped reconnected tubes that shorten as they retract in opposite directions, due to magnetic tension. One of them moves upward toward the top edge of the CS, and the other one moves downward toward the top of the underlying arcade. 

Rotational discontinuities (RDs) propagate along the legs of the tubes and generate parallel supersonic flows that collide at the center of the tube. There, gas-dynamic shocks that compress and heat the plasma are launched outwardly. A long lasting, dense central hot region created by the GDSs moves outward from the reconnection region.

While the tube descends toward the top of the arcade, plasma density is decreased by tube's lateral expansions in response to the decreasing external background pressure. This effect may decrease plasma density by $30$\% -- $50$\%\ of background levels. Therefore, retracting reconnected flux tubes may present elongated regions relatively devoid of plasma, resembling SADs. 

A reconnected tube will arrive at the top of the arcade that will slow it to a stop. Here, the perpendicular dynamics is halted, but the parallel dynamics continues along its legs; the RDs are shut down, and the gas is rarefied to even lower densities. After the tubes lie on top of the arcade, rarefaction waves continue decreasing plasma density. The hot post-shock regions continue evolving, determining a long lasting hot region on top of the arcade. We provide an observational method based on total emission measure and mean temperature, that indicates where in the CS the tube has been reconnected. 

Current sheets themselves are notoriously difficult to observe directly, yet they are a necessary component of all magnetic reconnection models. Nevertheless, there are observations providing evidence \citep{Schettino_2010, Savage_2010}. Syrovatski{\v i}-type CSs may be ubiquitous in the solar corona. \citet{Edmondson_2010} show that even a simple multipole topology would form very stable CSs bounded by two Y-type nulls, where an X-type null deforms under stresses due to motions in the photosphere. They did not include a guide field, but suggested that if added, the expected three-dimensional islands formed after reconnection may resemble longer flux-tube-like magnetic islands. 

The work described in Chapter \ref{chap:chap_2} has been published in the Astrophysical Journal \citep{Guidoni_2011}. 

In Chapter \ref{chap:chap_3}, we test the validity of the TFT assumptions by comparing results of the model with those from fully three-dimensional solutions of the general MHD equations. In that chapter, we present three-dimensional MHD simulations carried out in a volume filled with plasma including a Green-Syrovatski{\v i} \citep{Green_1965,Syrovatskii_1971} CS. The reconnection episode is triggered by imposing a short-lived and localized magnetic resistivity patch on the top half of the current sheet. The results of these simulations are compared to the corresponding thin flux tube simulation. 

The work described in Chapter \ref{chap:chap_3} will be submitted for publication in the near future. 

As a final remark, we should mention that the correct answer to the disparity between observations and models regarding the speed of localized reconnection outflows is still not known, but most likely the slowing down is due to the interaction between tubes and their surroundings. In the next chapters, for simplicity, this interaction is not included; the reconnected tubes are assumed to be completely isolated from their background. Therefore, we restrict ourselves to standard reconnection scenarios where the outflows are Alfv\'{e}nic.

\chapter{SHOCKS AND THERMAL CONDUCTION FRONTS IN RETRACTING RECONNECTED FLUX TUBES}
   \label{chap:chap_1}


\newcommand{\sgn}{\hbox{sgn}}

\newcommand{\realTemp}{\mbox{$1$} MK}
\newcommand{\realnumdens}{\mbox{$10^{8}$} cm\textsuperscript{-3}}
\newcommand{\realB}{\mbox{$10$} G}
\newcommand{\realbeta}{\mbox{$0.007$}}
\newcommand{\realL}{\mbox{$100$} Mm}
\newcommand{\realvao}{\mbox{$2.2$} Mm s\textsuperscript{-1}}
\newcommand{\realTao}{\mbox{$290$} MK}
\newcommand{\realReynolds}{\mbox{$2982$}}
\newcommand{\PrandtlSim}{\mbox{$0.01$}}
\newcommand{\realalfvtime}{\mbox{$46$} s}

\newcommand{\timeSim}{\mbox{$2550$} s}
\newcommand{\anglezeta}{\mbox{$45$} degrees}
\newcommand{\doublezeta}{\mbox{$90$} degrees}
\newcommand{\numsegments}{\mbox{$4000$}}
\newcommand{\fintime}{\mbox{$3200$} s}

\newcommand{\MachSim}{\mbox{$3.85$}}
\newcommand{\MachSimShock}{\mbox{$5.3$}}
\newcommand{\LSubShock}{\mbox{$22$} Mm}
\newcommand{\SStime}{\mbox{$532$} s}
\newcommand{\LHeatFront}{\mbox{$93$} Mm}
\newcommand{\HHtime}{\mbox{$188$} s}
\newcommand{\LHeatFrontAlfven}{\mbox{$28.0$} Mm}
\newcommand{\elecmfpunitless}{\mbox{$0.0021$}}
\newcommand{\protmfpunitless}{\mbox{$0.9 \times 10^{-3}$}}

\newcommand{\MachCrit}{\mbox{$1.3$}}
\newcommand{\PrandtlSuperCrit}{ \mbox{$0.01,0.03,0.1,0.3,1.0$}, and \mbox{$3.0$}}
\newcommand{\PrandtlCrit}{ \mbox{$0.05,0.1,0.5$}, and \mbox{$1.0$}}


The layout of the present chapter is as follows. The next section is a brief review of the general transport effects in magnetohydrodynamics. The following section describes the thin flux tube assumptions that lead to the final thin flux tube equations presented in the third section. Section \ref{sec:sims} presents simulations of the retraction of a thin flux tube reconnected in a localized region inside a current sheet with uniform, skewed magnetic fields for a case with low plasma-$\beta$, and parameters relevant to the solar corona. In the fifth section, an analytical analysis of the inner structure of the shocks is presented. Finally, in the last section, the main results of the paper and possible observational implications of the inner structure of the shocks are discussed.

%
%
\section{MHD Equations}
   \label{sec:MHD_equations}

When resistivity, radiation and gravity are neglected, the magneto-hydrodynamic (MHD) equations for a charge-neutral plasma (we will assume sufficient collisionality to justify their use) are \citep[chap. 22]{Shu_VolII}
%
%
\begin{eqnarray}
   \label{eqn:MHD_mass}
      \frac{1}{\rho} \frac{D\rho}{Dt} & = & - \nabla \cdot \mathbf{v}, \\
   \label{eqn:MHD_momentum}
       \rho \frac{D\mathbf{v}}{Dt} & = & 
      -\nabla \left(P + \frac{B^{2}}{8\pi}\right)+ \nabla \cdot 
      \left(\frac{\mathbf{B}\mathbf{B}}{4 \pi}\right)-\nabla \cdot \Pi,  \\
   \label{eqn:MHD_energy}
      \rho T \frac{Ds}{Dt} & = & \rho \frac{D\varepsilon_{I}}{Dt} - \frac{P}{\rho }\frac{D \rho}{Dt} = \Psi -\nabla \cdot \mathbf{q}, \\
   \label{eqn:MHD_induction}
      \frac{\partial \mathbf{B}}{\partial t}
      & = & - \nabla \times \left(\mathbf{B} \times \mathbf{v}\right).
\end{eqnarray}
%
%
\noindent Here, $\rho$, $P$, $\mathbf{B}$, $\mathbf{v}$, $T$, $s$, $\varepsilon_{I}$, $\Pi$, $\Psi$, and $\mathbf{q}$ are the density, plasma pressure, magnetic field, velocity, temperature, specific entropy, specific internal energy, viscous stress tensor, viscous heat, and thermal conduction flux of the fluid, respectively.
The advective derivative is defined as ${D}/{Dt} =
{\partial}/{\partial t}+\mathbf{v}\cdot\nabla$. 
In Equation \eqref{eqn:MHD_momentum}, dyadic notation \citep[chap. 5]{Goldstein_Cl_Mech} is used for the second right-hand term.


When the plasma is strongly magnetized by a field taken in the $\mathbf{\widehat{z}}$ direction, the stress tensor takes a simple gyrotropic form \citep{Braginskii_1965} 
%
%
\begin{eqnarray}
   \label{eqn:MHDstresst}
      \Pi &=& -\eta W_{zz} \left(-\frac{\mathbf{\widehat{x}}\mathbf{\widehat{x}}}{2}-
      \frac{\mathbf{\widehat{y}}\mathbf{\widehat{y}}}{2}
       +\mathbf{\widehat{z}}\mathbf{\widehat{z}}\right), 
\end{eqnarray}
where the $\mathbf{\widehat{z}}\mathbf{\widehat{z}}$ component of the rate-of-stress tensor is
\begin{eqnarray}
   \label{eqn:MHD_rate_of_strebght}
       W_{zz} & = & 2 \frac{\partial v_{
       z}}{\partial z} - 
       \frac{2}{3}\nabla \cdot \mathbf{v}.
\end{eqnarray}
%
%
\noindent Here, $\eta$ is the largest of the dynamic viscosity coefficients, determined essentially by the
ions (the electron-to-ion viscosity ratio is ${\eta_{e}}/{\eta_{i}} \simeq$ \PrandtlSim). 
%
The strongly magnetized limit of the conductive heat flux is
%
%
\begin{eqnarray}
   \label{eqn:MHD_qflux} \mathbf{q} & = & 
      - \kappa \nabla_{\parallel}(k_{B} T),
\end{eqnarray}
%
%
\noindent where $\nabla_{\parallel} = \mathbf{\widehat{z}}\mathbf{\widehat{z}} \cdot \nabla$  is the derivative parallel to the magnetic field, and $\kappa$ is the parallel thermal conduction coefficient, determined mostly by the electrons (the electron-to-ion heat conductivity ratio, 
${\kappa_{e}}/{\kappa_{i}} \simeq 24$). 
%

We close the system using the ideal gas law
%
%
\begin{eqnarray}
   \label{eqn:press_gas}
      P & = & \frac{\rho k_{B} T}{\overline{m}} = 2 n k_{B} T, \\
    \label{eqn:energy_gas}
       \varepsilon_{I} & = & \frac{P}{(\gamma-1) \rho}, \\
   \label{eqn:entropy_gas}
      s & = & c_{V} \ln \left[ \frac{P}{\rho^{\gamma}} \right] + s_{0},
\end{eqnarray}
%
where $\gamma$ is the adiabatic gas constant, $\overline{m}$ is the average particle mass (for a fully ionized Hydrogen plasma, it is approximately equal to half the mass of the proton), $n$ is the electron number density, $c_{V}= {k_{B}}/{(\gamma-1) \overline{m}} $ is the specific heat at constant volume, and $s_{0}$ is a constant.

%
%
%
%
%
%

\section{Thin Flux Tube Equations}

\subsection{The Thin Flux Tube Assumptions}

\begin{figure}[ht]
   \centering
   \resizebox{5.5in}{!}{\includegraphics*[0,0][448,408]{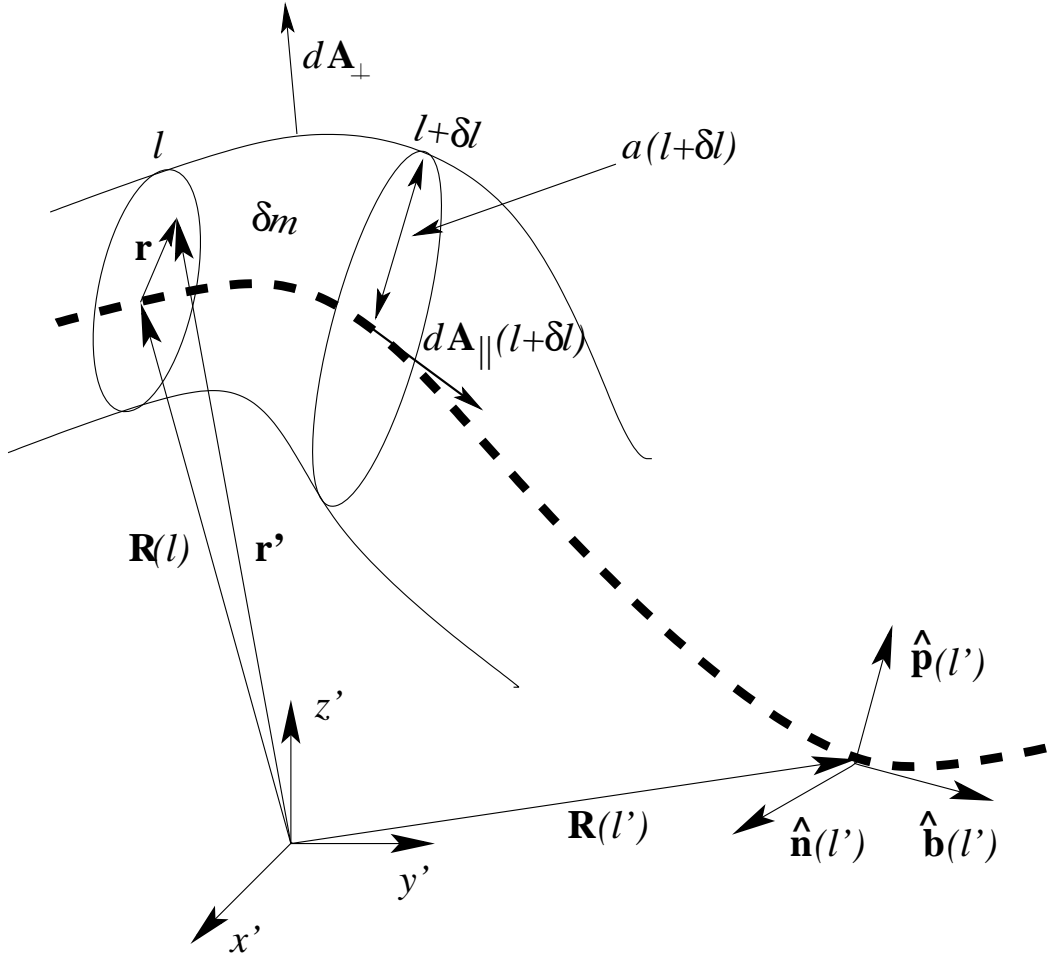}}
   \caption[Thin flux tube element. Frenet coordinate system]
   {Frenet coordinate system. The axis of a thin magnetic flux tube is represented by a central magnetic field line (dashed line) parameterized by vector $\mathbf{R}(l)$. The unit vector $\mathbf{\widehat{b}}$ is parallel to the local magnetic field, unit vector $\mathbf{\widehat{p}}$ points in the direction of the curvature vector $\mathbf{k}$, and unit vector $\mathbf{\widehat{n}}$ completes the system. The two ellipses are the cups of a small tube segment of mass $\delta \hbox{m}$}
   \label{fig:frenetsystem} 
\end{figure}

We consider an untwisted, \textit{thin} tube of magnetic flux $\phi$, like the one depicted in Figure \ref{fig:frenetsystem}, embedded in a strongly magnetized external fluid. The external magnetic field, thermal pressure, and density will be denoted $\mathbf{B}_{e}$, $P_{e}$, $\rho_{e}$, respectively. These quantities may depend on position, but not time, since we will assume the background configuration to be in equilibrium. The tube is considered so thin that the background quantities remain unperturbed as the tube moves through the external plasma. 

%
%

We will assume that the tube can be represented by an internal field 
line, $\mathbf{R}(l)$, the axis (see Figure \ref{fig:frenetsystem}).  The field is untwisted in the sense that all field lines are parallel to the axis.
The time it takes for a fast magneto-sonic wave to establish pressure balance with its surroundings scales as $\tau_{fm}={a(l)}/{\sqrt{c_{s}^{2}(l) + v_{a}^{2}(l)} }$, where $v_{a}$ is the Alfv\'{e}n speed, and $c_{s}$ is the sound speed. Since the fast-magnetosonic time is negligible with respect to the Alfv\'en time over a characteristic scale $L_{ch}$, 
$\tau_{A} = {L_{ch}}/{v_{a}(l)}$, we assume instantaneous, local pressure balance
%
%
\begin{eqnarray}
   \label{eqn:TP_constrain}
       P + \frac{B^{2}}{8 \pi} & = & P_{e} + \frac{B^{2}_{e}}{8 \pi},
\end{eqnarray}
%
\noindent where $P$ and $B$ are pressure and magnetic field inside the tube, respectively. 

It is convenient to define a Frenet coordinate about the axis $\mathbf{R}$,  parameterized by its arc-length $l$, as shown in Figure \ref{fig:frenetsystem}.  The Frenet unit vectors defining the coordinate system are
%
%
\begin{eqnarray}
   \label{eqn:bvect}
      \mathbf{\widehat{b}}(l)&=&\frac{\mathbf{B}[\mathbf{R}(l)]}{B[\mathbf{R}(l)]}
      = \frac{\partial }{\partial l}\mathbf{R}(l), \\
   \label{eqn:kvect} 
      \mathbf{\widehat{p}}(l)&=&\mathbf{k}(l)r_{c}(l)=\frac{\partial \mathbf{\widehat{b}}(l)}{\partial
      l}r_{c}(l), \\
   \label{eqn:nvect}
       \mathbf{\widehat{n}}(l) &=&
       \mathbf{\widehat{p}}(l) \times \mathbf{\widehat{b}}(l).
\end{eqnarray}
%
%
\noindent The unit vector $\mathbf{\widehat{b}} $ is parallel to the
axis, $\mathbf{\widehat{p}}$ is the normal vector and $\mathbf{\widehat{n}}$ is the 
bi-normal completing the system. 
Local derivative operators associated with the unit vectors are
%
%
\begin{eqnarray}
   \label{eqn:dirder}
      \partial_{\parallel} & = & \widehat{\mathbf{b}} \cdot \nabla = \partial_{l}, \nonumber \\
      \nabla_{\bot}  & = & - \widehat{\mathbf{b}} \times (\widehat{\mathbf{b}} \times\nabla).  \nonumber
\end{eqnarray}
%


%
%
We assume that the flux tube is sufficiently isolated from its surrounding that no viscous stress crosses its surface. This is justified by the small cross-field transport coefficients. The stress tensor is therefore
\begin{eqnarray}
   \label{eqn:tft_str_tens}
      \Pi &=& -\eta \left( \widehat{\mathbf{b}} \cdot \frac{\partial \mathbf{v}}{\partial l}  \right)       \widehat{\mathbf{b}}\widehat{\mathbf{b}}. 
\end{eqnarray}
%
For parameters typical at the high corona, $T =$ \realTemp, $n =$ \realnumdens, $B =$ \realB, $L =$ \realL, the Reynold's number is $R_{\eta} \simeq 3 \times 10^{3}$, indicating a relatively small contribution from viscosity. 
For such high Reynolds numbers, the viscous stress tensor may become significant only at steep velocity gradients, like shocks. Shocks in thin flux tubes were predicted by 
\citep{Longcope_2009}. 

The small Prandtl number of an ionized plasma, 
$P_{r} = \eta / \kappa \overline{m} \sim $ \PrandtlSim, indicates that the effect of thermal conduction is larger than viscosity. Temperature gradients will then be even smaller at shocks than velocity gradients.

%
%
%

\subsection{Thin Flux Tube Equations}

The thin flux tube equations are derived by applying the MHD equations to a small piece of the tube with mass $\delta \hbox{m}$, like the one shown in Figure \ref{fig:frenetsystem}. The position of every point in the tube can be expressed as $\mathbf{r'} = \mathbf{R} + \mathbf{r}$. Vector $\mathbf{r}$ belongs to the circular area perpendicular to the tube axis, and can be expressed as $\mathbf{r}=r \sin \theta \hbox{ }\mathbf{\widehat{n}} + r \cos \theta \hbox{ }\widehat{\mathbf{p}} =  r\mathbf{\widehat{r}}$, where $r$ and $\theta$ are the polar coordinates associated to the $\mathbf{\widehat{n}}$-$\mathbf{\widehat{p}}$ plane. 

The local radius of the tube, $a(l)$, is small compared with scales of variation along the tube, permitting Taylor-expansion of any state variable $f$ as follows
%
%
\begin{eqnarray}
\label{eqn:expansion}
   f(\mathbf{r})  =  \mathop{\sum_{k=0}^{\infty}}
   \frac{\left[(\mathbf{r}\cdot\nabla)^{k}f\right](l)}{k!} = \mathop{\sum_{k=0}^{\infty}} \mathop{\sum_{i=0}^{k}} \binom{k}{i}
      \frac{r^{k}}{k!}(\cos \theta)^{k-i}
      (\sin \theta)^{i} \frac{\partial^{k}f(l)}{\partial
      n^{k-i}\partial p^{i}}.
\end{eqnarray}
%
%
The second equality was obtained expanding the binomial term in the local polar coordinates.
The shortness of the tube element allows us to assume that the piece's axis lays on a plane and its radius of curvature is constant. For this case, the differentials of area and volume for the piece can be expressed as
%
%
\begin{eqnarray}
    d\mathbf{A_{\parallel}} & = &
    \widehat{\mathbf{b}}(l) \hbox{ }r\hbox{ } dr\hbox{ } d\theta,    \\
    d\mathbf{A_{\perp}}  & = &  \left.\left[ \frac{\partial
    \mathbf{r'}}{\partial \theta} \times \frac{\partial
    \mathbf{r'}}{\partial l}\right]\right|_{r=a(l)} dl \hbox{ } d\theta, \nonumber \\
    & = &  \left\{\mathbf{\widehat{r}}(l) \left[a(l)-
   \frac{a^{2}(l)\sin\theta}{r_{c}(l)} \right]  -
   \widehat{\mathbf{b}}(l)  a(l)\frac{\partial a(l)
   }{\partial l} \right\} dl d\theta  ,     \\
   d\hbox{\textit{v}}  & = & \left[ \frac{\partial \mathbf{r'}}{\partial
   \theta} \times \frac{\partial \mathbf{r'}}{\partial
   l}\right]\cdot \frac{\partial
   \mathbf{r'}}{\partial r}\hbox{ }  \hbox{ } dl \hbox{ }d\theta \hbox{ } dr \nonumber \\
   \label{eqn:diff_vol}
      & = & \left(r-\frac{r^{2} \hbox{ } \sin\theta}{r_{c}(l)}\right)
      \hbox{ }  \hbox{ } dl \hbox{ }d\theta \hbox{ } dr.   
\end{eqnarray}
%
%
 
An example of the expansion \eqref{eqn:expansion} is the mass of the tube piece. Computing the volume integral using expression \eqref{eqn:diff_vol} for the volume differential, and expanding the density around the center of the tube, gives
%
%
\begin{eqnarray}
   \delta \hbox{m} & = & \int \rho \hbox{ } d\hbox{v} \nonumber \\
   & = & \mathop{\sum_{k=0}^{\infty}}
   \mathop{\sum_{i=0}^{k}} \binom{k}{i} \frac{1}{k!} \int_{l }^{l+
   \delta l} \frac{\partial^{k}\rho(l')}{\partial n^{k-i}\partial
   p^{i}}\hbox{ }dl' \int_{0}^{a(l)} \int_{0}^{2\pi}
   r^{k+1} \times \nonumber \\
   &  &  \times \hbox{ } (\cos \theta)^{k-i} (\sin \theta)^{i}  \hbox{ } \left(1-\frac{r \hbox{ }
   \sin\theta}{r_{c}(l)}\right)  d\theta dr.  \nonumber \\
   & = &  \int_{l }^{l+ \delta l} \rho(l') \pi a^{2}(l') \hbox{ }dl'
   + \int_{l }^{l+ \delta l} \mathcal{O}(a^{4}(l')) dl', \nonumber
\end{eqnarray}
%
%
where $\mathcal{O}(a^{4}(l'))$ represents an error of fourth order
in the local radius. If the arc-length interval, $\delta l$, is small then the above equation can be expressed as
%
%
\begin{eqnarray}
   \label{eqn:delta_mass}
      \delta \hbox{m}(l) & \simeq & \rho(l) \pi a^{2}(l) \delta l.
\end{eqnarray}
%
%
Expanding the dot product $\mathbf{B} \cdot \widehat{\mathbf{b}}$, gives a similar expression for the magnetic flux
%
%
\begin{eqnarray}
   \label{eqn:flux_tft}
      \phi & =& \int_{l} \mathbf{B}\cdot d\mathbf{A_{\parallel}} \nonumber \\
      & =& B(l) \pi a^{2}(l) +  \mathcal{O}(a^{4}(l)) \simeq B(l) \pi a^{2}(l) . 
\end{eqnarray}
%
%
%

For a given tube piece, the ratio between the mass and the magnetic flux, $\delta \mu  = \delta \hbox{m}/\phi = (\rho/B) \delta l$, is constant. This definition, along with Equations \eqref{eqn:delta_mass} and \eqref{eqn:flux_tft}, allow us to define the integrated mass per flux, 
$\mu = \int{(\rho/B) dl } $, to parameterize the tube instead of $l$.
The differential with respect to $\mu$ is
%
%
\begin{eqnarray}
   \label{eqn:deriv_mu}
   \frac{\partial }{\partial \mu}  & =& \frac{B}{\rho} \frac{\partial}{\partial l}.
\end{eqnarray}
%
%

In order to obtain the thin flux tube mass equation we integrate over the piece's volume every term of the corresponding MHD equation, and keep only the first non-vanishing terms. 
To the first order, the velocity at any point in the tube is the sum of the velocity of the center of the tube plus a radial term $v_{r}$. This radial velocity at the surface of the tube is equal to the change in local radius $(D/Dt) a(l)$. Finally, applying the fundamental theorem of calculus to recover integrals along the arc-length, and expression \eqref{eqn:flux_tft}, the final result is
%
\begin{eqnarray}
   \label{eqn:TFT_mass_ch1} 
      \frac{D}{Dt}\left(\frac{B}{\rho}\right) & = & \widehat{\mathbf{b}} \cdot \frac{\partial \mathbf{v}}{\partial \mu}
\end{eqnarray}
%
%

In the integration of the momentum Equation \eqref{eqn:MHD_momentum},
the surface integrals for the perpendicular area vanish since $\mathbf{B} \cdot \mathbf{dA_{\perp}} = 0$ and $\Pi \cdot \mathbf{dA_{\perp}}=0$ (in this last equality we have used definition \eqref{eqn:tft_str_tens}). For the first term of the right-hand side, constraint \eqref{eqn:TP_constrain} is applied and expanded around the center of the tube yielding
%
%
%
\begin{eqnarray}
   \label{eqn:TFT_mom_ch1} 
      \frac{D\mathbf{v}}{Dt} & = & \frac{1}{\rho} \left[
      -\widehat{\mathbf{b}} \frac{\partial P}{\partial l}-\nabla_{\perp}
      \left(P_{e} + \frac{B^{2}_{e}}{8 \pi}\right ) +2
      \mathbf{k}\left(P_{e}-P+ \frac{B^{2}_{e}}{8 \pi}\right) \right] + \nonumber \\
      &  & + \frac{B}{\rho} \frac{\partial}{\partial l} \left[ \frac{\widehat{\mathbf{b}\eta}}{B}             \left(\widehat{\mathbf{b}}\cdot \frac{\partial \mathbf{v}}{\partial l}\right)\right],
\end{eqnarray}
%
%
where every term is evaluated at the center of the tube. In the above equation, the viscous momentum term is a perfect derivative in the integrated mass. There is no viscous momentum interchange between the tube and its surroundings.

To determine the form of the viscous heating, we compute the rate of change in kinetic energy as the dot product between the momentum Equation \eqref{eqn:TFT_mom_ch1} and the velocity. 
Balancing this loss with a heating rate gives 
%
%
\begin{eqnarray}
   \label{eqn:TFT_visc_heat}
      \Psi & = & \eta
      \left( \widehat{\mathbf{b}} \cdot \frac{\partial \mathbf{v}}{\partial l}.\right)^{2}
\end{eqnarray}
%
%
With the above definition, and similar treatment than for the mass and momentum equation, the thin flux tube energy equation becomes
%
%
\begin{eqnarray}
    \label{eqn:TFT_energ}
      \frac{D\varepsilon_{I}}{Dt} & = &  \frac{P}{\rho^{2} }\frac{D \rho}{Dt} + \frac{\eta} {\rho}  \left(\widehat{\mathbf{b}}          \cdot \frac{\partial \mathbf{v}}{\partial l}\right)^{2} +\frac{B}{\rho} \frac{\partial}{\partial l} \left(       \frac{\kappa}{B}  \frac{\partial}{\partial l}(k_{B} T)\right),
\end{eqnarray}
%
%
%
where we see that the thermal conduction term is also a perfect derivative in the integrated mass per flux (there is no thermal conduction across the tube).

For our simulations, it is more convenient to re-write the energy equation by defining a new variable $P_{c}$ in the following way
%
%
%
\begin{eqnarray}
    \label{eqn:TFT_P0}
      P(\mu,t) & = &  P_{c}(\mu,t) \left(\frac{\rho(\mu,t)}{\rho(\mu,0) } \right)^{\gamma} = P_{c}(\mu,t) \left(\frac{\rho(\mu,t)}{\rho_e(\mu) } \right)^{\gamma}. 
\end{eqnarray}
%
%
%
%
In an adiabatic case, $P_{c}$ is constant in time. The change in plasma entropy with respect to the initial state (assumed to be in equilibrium) is directly related to $P_{c}$, where $\Delta s = c_{V} \ln \left[{P_{c}(\mu,t)}/{P_{e}(\mu)}\right]$. With definition \eqref{eqn:TFT_P0}, the energy equation is transformed to an equation for $P_{c}$ 
%
%
\begin{eqnarray}
   \label{eqn:TFT_P0_time}
      \frac{D P_{c}}{Dt} & = & (\gamma -1)  \left( \frac{\rho_{e}}{\rho } \right)^{\gamma} 
      \left[ \eta \left(\widehat{\mathbf{b}} \cdot \frac{\partial \mathbf{v}}{\partial l} 
      \right)^{2} + B \frac{\partial}{\partial l} \left( \frac{\kappa}{B}   \frac{\partial}{\partial l}(k_{B} T) 
      \right) \right].
\end{eqnarray}
%
%
%
When density is known, the above equation allows us to update pressure, and use it in the momentum equation. 

Heating and cooling occur where there is a change in entropy. From Equations \eqref{eqn:MHD_energy}, \eqref{eqn:TFT_energ}, and \eqref{eqn:TFT_P0_time}, the volumetric heating rate can be written as
%
%
\begin{eqnarray}
   \label{eqn:heating}
      \dot{Q} & = & \rho T \frac{Ds}{Dt} =  \frac{1}{\gamma-1} \left( \frac{\rho}{\rho_{e}} \right)^{\gamma}  \frac{D P_{c}}{Dt} \nonumber \\
      & = & \eta \left(\widehat{\mathbf{b}} \cdot \frac{\partial \mathbf{v}}{\partial l} 
      \right)^{2} + B \frac{\partial}{\partial l} \left( \frac{\kappa}{B}   \frac{\partial}{\partial l}(k_{B} T) 
      \right).
\end{eqnarray}
%
%
In this equation, the viscosity term is always positive, and therefore always heats the plasma. On the other hand, the thermal conduction term can have either sign, depending on the temperature's arc-length second derivative. 

The transport coefficients in an ionized plasma depend on temperature as \citep{Spitzer_1962}
%
%
\begin{eqnarray}
   \label{eqn:transp_coeff_eta}
      \eta &=& \eta_{c}T^{\frac{5}{2}}, \\
   \label{eqn:transp_coeff_kappa}
      \kappa &=& \kappa_{c}T^{\frac{5}{2}},
\end{eqnarray}
where $\eta_{c}$ and $\kappa_{c}$ are constants. 

For thin flux tubes, there is no need to solve the induction
equation, since the magnitude of the magnetic field inside the
tube is determined by constraint \eqref{eqn:TP_constrain}, and its direction is given by
the position of the tube piece governed by the momentum equation (resistivity has been neglected, and therefore the magnetic field follows the fluid).

It is possible to describe the TFT equations in dimensionless variables, by defining a characteristic magnetic field  magnitude, $B_{0}$, density, $\rho_{0}$, and length, $L_{0}$. For example, these characteristic variables may represent the magnetic field magnitude and density at the reconnection point, $\mathbf{x}_{R}$, and the length of a typical reconnected tube, respectively. Then, $B_{0} = B_{e}(\mathbf{x}_{R})$ and $\rho_{0}=\rho_{e}(\mathbf{x}_{R})$. With these values, we can construct an associated Alfv\'{e}n speed $v_{A0} = B_{0}/ \sqrt{4 \pi \rho_{0}}$, and an associated temperature $T_{0}^{A} = v_{A0}^{2}\overline{m}/ k_{B}$, to non-dimensionalized velocities and temperatures. 

The new dimensionless variables are
%
%
\begin{eqnarray}
    \label{eqn:unitless_var}
       &   \mu^{*} =\mu \frac{B_{0} }{\rho_{0} L_{0}},\hbox{ } \rho^{*}=\frac{\rho }{\rho_{0}} ,\hbox{ }t^{*} = t \frac{ v_{A0}}{L_{0}}, \hbox{ } l^{*}  =                 \frac{l}{L_{0} }, P^{*} = P\frac{ 4 \pi}{B_{0}^{2} }, \hbox{ } P_{c}^{*} = P_{c}\frac{ 4 \pi}{B_{0}^{2} }, \hbox{ } T^{*} = \frac{T}{T_{0}^{A}}, \nonumber \\
       &   \mathbf{v}^{*} = \frac{\mathbf{v}}{v_{A0}}, \hbox{ } \mathbf{B}^{*} =  \mathbf{B}\frac{ \sqrt{4 \pi}}{B_{0}}, 
       \hbox{ } \mathbf{x}^{*}  =  \frac{\mathbf{x}}{L_{0}}, \hbox{ }\mathbf{k}^{*}=  \mathbf{k} L_{0} , \hbox{ }                    \widehat{\mathbf{b}}^{*}=  \widehat{\mathbf{b}}, \nonumber \\
       &   \eta^{*} =  \frac{\eta}{\rho_{0} L_{0} v_{A0}}=\eta_{c}^{*}\left(T^{*}\right)^{5/2},\hbox{with } \eta_{c}^{*}           =\eta_{c}\frac{\left(T_{0}^{A}\right)^{5/2}}{\rho_{0} L_{0} v_{A0}}, \nonumber \\
       &   \kappa^{*} = \kappa \frac{\overline{m}}{\rho_{0} L_{0} v_{A0}} = \kappa_{c}^{*}\left(T^{*}\right)^{5/2},             \hbox{with } \kappa_{c}^{*} = \kappa_{c}\frac{\overline{m}\left(T_{0}^{A}\right)^{5/2}}{\rho_{0} L_{0} v_{A0}}.      
\end{eqnarray}
%
%
With these new variables, the dimensionless thin flux tube equations have exactly the same form as Equations \eqref{eqn:TFT_mass_ch1}, \eqref{eqn:TFT_mom_ch1}, \eqref{eqn:TFT_energ}, and \eqref{eqn:TFT_P0_time}, except that the Boltzmann constant is no longer present. From now on, we will refer to the thin flux tube equations as the dimensionless ones.


%
%
%
%
%
%

\section{Solutions of Reconnection Dynamics}
\label{sec:sims}\vspace*{-0.5cm}

\subsection{Low Plasma-$\beta$ Uniform Background Case}
\label{sec:Unif_Low_beta}

The results of the last section are quite general. In this section, we will simplify the equations to apply them to the low-$\beta$ solar corona. 
For simplicity, we will also assume uniform background magnetic field. 

\begin{figure}[ht]
  \centering
  \resizebox{5.0in}{!}{\includegraphics*[0,0][457,437]{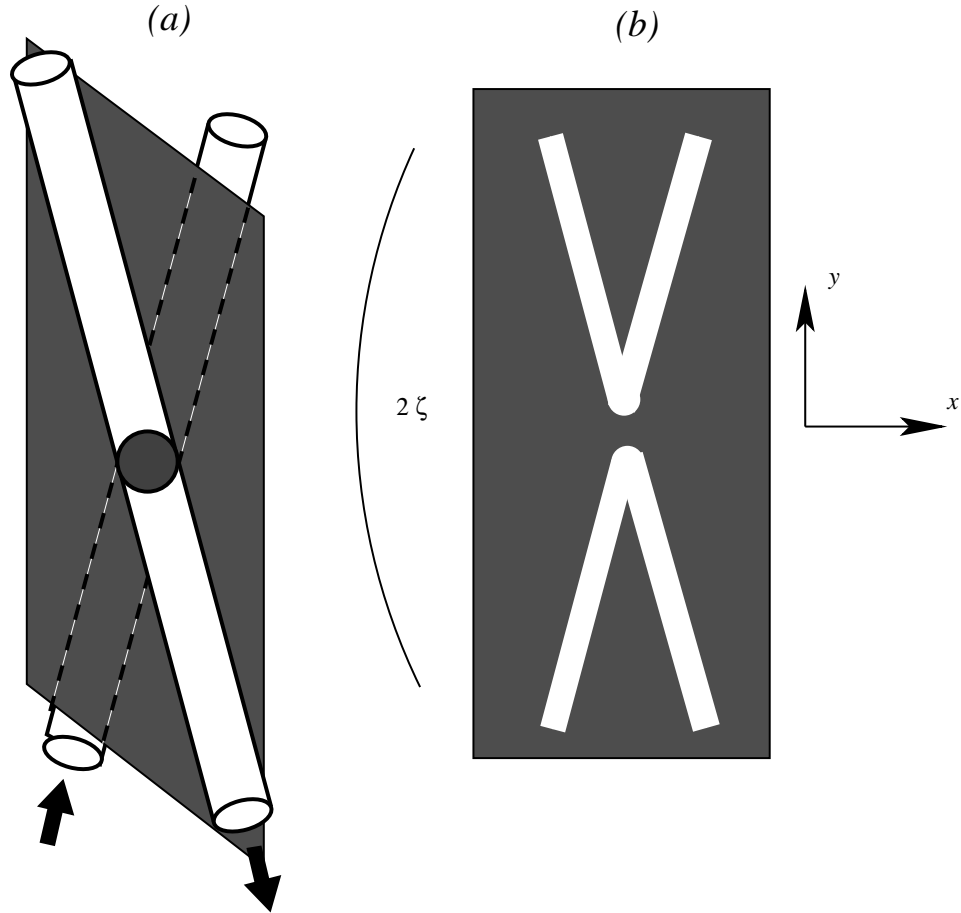}}
  \caption[Current sheet schematic]
   {Current sheet schematic. (a) The two tubes represent the front (solid contour) and back (dashed contour) pre-reconnection tubes. Thick arrows indicate the direction of the magnetic field. The sphere in the middle represents the small reconnection region. The tubes are initially part of the uniform background equilibrium.  (b) Reconnected V-shaped tubes a short time after reconnection. The angle between the initial tubes is $2 \zeta$. Due to magnetic tension they will retract in opposite directions. The end-points of the flux tubes are located near the edges of the current sheet}
   \label{fig:current_sheet}
\end{figure}

We will consider 
skewed, uniform external magnetic field forming an angle $2 \zeta$ on opposite
sides of a static, infinite current sheet, as shown in Figure \ref{fig:current_sheet}. The current sheet is assumed to be infinitely thin, represented by the $x-y$-plane. The initial pressure is also assumed to be uniform everywhere. A spatially and temporally localized reconnection event has occurred (by an unspecified physical mechanism), connecting some field lines from one side of the current sheet to some on the other side. The amount of flux is determined by the size of the reconnection region. This bundle of field lines forms two V-shaped flux tubes (as shown in panel (b) of Figure \ref{fig:current_sheet}) and magnetic tension at their cusps causes them to retract. These tubes will remain as coherent entities, preserving their magnetic flux $\phi$, while they move. No further reconnection is assumed. 

%
%
%

For this uniform case, the characteristic magnetic field magnitude and density at the reconnection point are the same as everywhere else. Then, $B_{0}=B_{e}$, $\rho_{0} = \rho_{e}$, and $L_{0}$ is some characteristic length. The perpendicular gradient of the external total pressure is zero. 

In the solar corona, the ratio between thermal pressure and magnetic pressure, $\beta = 8\pi P/B^{2}$, is small.  For a plasma with magnetic field $B_{e} =$ \realB, number density $n_{e} =$ \realnumdens, and temperature $T_{e} =$ \realTemp, $\beta$ is approximately equal to \realbeta. Consequently, we will neglect the components of the thermal pressure gradient which are perpendicular to the tube's axis with respect to the magnetic pressure gradient in that direction. The initial small internal pressure might increase due to compressions along the tube, but we will consider it small compared to the internal magnetic pressure at all times. In the parallel direction, there is no force related to the magnetic field, therefore the parallel derivative of the thermal pressure cannot be neglected. 

With the above considerations, and making the temperature dependence of the transport coefficients explicit, we can re-write the constraints and thin flux tube equations in the following way
%
%
%
\begin{eqnarray}
    \label{eqn:TFT_low_beta_B}
       B  &= & B_{e}, \\
    \label{eqn:TFT_low_beta_mass}
       \frac{D\rho}{Dt} & = & - \frac{\rho^{2}}{B_{e}} \widehat{\mathbf{b}} \cdot \frac{\partial \mathbf{v}}{\partial \mu}, \\
    \label{eqn:TFT_unif_cons_mom}
       \frac{D\mathbf{v}}{Dt}  &= & \frac{\partial}{\partial \mu} \left\{\left[ \frac{B_{e}}{4 \pi} - \frac{P}{B_{e}}                + \frac{\eta_{c}T^{5/2}}{B_{e}} \left( \widehat{\mathbf{b}} \cdot \frac{\partial \mathbf{v}}{\partial l}                      \right ) \right] \widehat{\mathbf{b}} \right\} + \frac{P}{\rho} \mathbf{k} \\
       &= &   
       \frac{1}{\rho} \left\{ 
                       \frac{B^{2}_{e}}{4 \pi} \mathbf{k} 
                      -\widehat{\mathbf{b}} \frac{\partial P}{\partial l} - P \mathbf{k}
                      + \frac{\partial}{\partial l} \left[ \widehat{\mathbf{b}} \eta_{c}T^{5/2}                                             \left(\widehat{\mathbf{b}}\cdot \frac{\partial\mathbf{v}}{\partial l}\right) \right]
                      \right\} + \frac{P}{\rho}\mathbf{k}, \nonumber \\
    \label{eqn:TFT_unif_cons_ene}
       \frac{D \varepsilon_{T}}{Dt} &= & \frac{\partial}{\partial \mu} \left\{\left[\frac{B_{e}}{4 \pi} -                      \frac{P}{B_{e}} + \frac{\eta_{c}T^{5/2}}{B_{e}}\left( \widehat{\mathbf{b}} \cdot \frac{\partial               \mathbf{v}}{\partial l}\right )\right ] \left( \widehat{\mathbf{b}} \cdot \mathbf{v}\right ) + 
       \frac{\kappa_{c}T^{5/2}}{B_{e}}\frac{\partial T}{\partial l}\right\}  \\
&  &    + \frac{P}{\rho} \mathbf{k} \cdot \mathbf{v}, \nonumber
\end{eqnarray}
%
%
where
%
%
\begin{eqnarray}
    \label{eqn:TFT_spec_ene_I}
       \varepsilon_{T} = \varepsilon_{I} + \varepsilon_{K} + \varepsilon_{M},
\end{eqnarray}
with specific internal, kinetic, and magnetic energies
\begin{eqnarray}
    \label{eqn:TFT_spec_ene_II}
       \varepsilon_{I} = \frac{P}{(\gamma -1) \rho},\hbox{ } 
       \varepsilon_{K} = \frac{|\mathbf{v}|^{2}}{2},\hbox{ } 
       \varepsilon_{M} = \frac{B_{e}^{2}}{4 \pi \rho}.
\end{eqnarray}
%
%
%
The specific magnetic energy of the tube doubles because there is work done by the background magnetic field expanding into volume vacated by motion of the tube. We have assumed that the external field returns to its original value after the tube has passed. In  more realistic scenarios this value may be slightly different. 

Conservation of energy improves stability of any numerical solution. Toward this end, we will neglect the small last term in Equation \eqref{eqn:TFT_unif_cons_ene} (proportional to $\beta$), that is not part of the perfect derivative in the integrated mass. With this simplification, Equation \eqref{eqn:TFT_P0_time} remains unchanged, and the change of the tube's total energy per flux is equal to
%
%
%
\begin{eqnarray}
   \label{eqn:TFT_cons_Energy} 
   \frac{D }{Dt} \left( \frac{E_{T} }{\Phi}\right ) = \left. \left\{ \left[\frac{B_{e}}{4 \pi} - \frac{P}{B_{e}} + \frac{\eta_{c}T^{5/2}}{B_{e}}\left      
   (\widehat{\mathbf{b}} \cdot \frac{\partial \mathbf{v}}{\partial l}\right )\right ] \left( \widehat{\mathbf{b}} \cdot \mathbf{v}\right ) + 
   \frac{\kappa_{c}T^{5/2}}{B_{e}}\frac{\partial T}{\partial l} \right \} \right |^{\mu_{0}}_{\mu_{L}}, 
\end{eqnarray}
%
%
%
where ${\mu_{0}}$ and ${\mu_{L}}$ represent the end-points of the tube, respectively. If the-end points are fixed, and there is no temperature gradient there, the total energy of the tube is conserved.  The neglected term came from the change in kinetic energy, calculated from Equation \eqref{eqn:TFT_unif_cons_mom}, and to be consistent, the correspondent perpendicular term will be neglected in the momentum equation, as well. Then, the change in total momentum per flux of the tube is equal to 
%
%
\begin{eqnarray}
  \label{eqn:TFT_cons_momentum}
   \frac{D }{Dt} \left( \frac{\mathbf{P}_{T} }{\Phi}\right ) &=& \left. \left\{\left[\frac{B_{e}}{4 \pi} - \frac{P}{B_{e}}   
   + \frac{\eta_{c}T^{5/2}}{B_{e}}\left( \widehat{\mathbf{b}} \cdot \frac{\partial \mathbf{v}}{\partial l}\right )
   \right ]  \widehat{\mathbf{b}} \right \} \right |^{\mu_{0}}_{\mu_{L}}. 
\end{eqnarray}
%
%
With the neglect of the proportional to $\beta$ small term, and the fact that $P_{e}$ is uniform, the ideal part of Equation \eqref{eqn:TFT_unif_cons_mom} reduces to the thin flux tube equation presented by \citet{Longcope_2009}. Therefore, gas-dynamic shocks and rotational discontinuities are expected along the reconnected flux tubes.

The volumetric heating rate when the magnetic field magnitude is uniform becomes
%
%
\begin{eqnarray}
   \label{eqn:heating_uniform}
      \dot{Q} & = &  \eta_{c} T^{\frac{5}{2}} \left(\widehat{\mathbf{b}} \cdot \frac{\partial \mathbf{v}}{\partial l} 
      \right)^{2} +  \frac{2}{7} \kappa_{c} \frac{\partial^{2}T^{\frac{7}{2}}}{\partial l^{2}}.
\end{eqnarray}
%
In the above equation, viscosity always contributes to heat the plasma. Depending on the curvature of temperature to the $7/2$ power, and the value of the Prandtl number, the thermal conduction term may contribute to plasma cooling or heating.  

%
%
%
\subsection{Simulations}
   \label{sec:Simulations}

We developed a computer program, called Dynamical Evolution of Flux Tubes (DEFT), which solves the thin flux tube equations for the retraction of the two reconnected thin flux tubes described in the last section. The program takes into account the effect of the transport coefficients, including their strong dependence on temperature. It also can perform simulations using real coronal dimensionless parameters, as well as implement a strong anisotropy in the transport coefficients (heat conduction and momentum transport are allowed only in the direction parallel to the magnetic field). This is far more difficult to achieve in fully three dimensional MHD programs. The DEFT program can be implemented for different background conditions as well, but we will restrict ourselves in this paper to the low-$\beta$ uniform background case from the last section. In this case, the two tubes moving in opposite directions are symmetric.  In cases with non-uniform background it might not be so. 
We use the conservative form of the thin flux tube equations introduced in the last section. 

The DEFT program uses a staggered mesh where each tube piece is represented by grid points at its ends and by its mass per flux. The code implements a Lagrangian approach where each tube piece is followed, guaranteeing mass conservation. 
Due to mirror symmetry, only half of the tube is simulated, the other half is a mirror image in the $x$-direction.

The initial state corresponds to two opposite V-shaped flux tubes, connected at the reconnection region, as shown in Figure \ref{fig:current_sheet}. The initial angle between the magnetic field lines at each side of the current sheet is equal to $2 \zeta$. Each tube is divided in $N$ segments separated by $N+1$ tube-point positions. The initial tubes are clearly out of equilibrium due to the sharp angle between the field lines. To avoid introducing length scales at the limit of resolution, the central parts of the initial tubes are smoothed. The end-points of the tube are assumed to be fixed, and to have no temperature gradient with the external boundaries (this last condition ensures no heat transfer from the end-points).

Since the reconnected tubes originated as background plasma they are initialized with uniform density, $\rho = \rho_{e}$, and thermal pressure $P = P_{e}$. The mass per flux of each piece is equal to $\delta m / \Phi = \rho_{e} \delta l_{e} /B_{e}$, where $\delta l_{e}$ is the segment's initial length. The magnetic field magnitude, $B_{e}$, is constant everywhere throughout the entire simulation.
The initial velocities are zero (equilibrium state), even for the central bend (we are not interested in the details of the initial transient evolution). Viscosity and thermal conductivity are calculated using Equations \eqref{eqn:transp_coeff_eta} and \eqref{eqn:transp_coeff_kappa}, where $T = P/\rho$ (unitless quantities). The unit parallel vectors are calculated at segment's centers from definition \eqref{eqn:bvect}, and curvature vectors are calculated at grid points from definition \eqref{eqn:kvect}. Arc-length derivatives are computed using centered differences.

At every time, the $N+1$ positions and velocities, as well as the $N$ pressure pre-factors $P_{c}$, are advanced using the current state variables in Equations \eqref{eqn:TFT_unif_cons_mom} and \eqref{eqn:TFT_P0_time}, following a simple Eulerian approach, for example $ \mathbf{v}(t + \delta t) = \mathbf{v}(t) + (D/Dt)\mathbf{v}(t) \delta t$. The time interval is chosen to satisfy the Courant-Friedrichs-Lewy condition \citep{CFL_1967}.
 
All the terms in Equation \eqref{eqn:TFT_unif_cons_mom} are evaluated at grid points, while the terms in 
Equation \eqref{eqn:TFT_P0_time} are evaluated within segments. The state variables and the unit tangent vector $\widehat{\mathbf{b}}$ are defined inside the segments. They are averaged between neighbors when needed at grid points. A similar approach is used when vectors defined at tube points are needed at segment centers. 
From the advanced positions, velocities, and pressure pre-factors $P_{c}$, it is possible to define all the advanced state variables. The density of each segment at a given time is calculated as $\rho = (\delta m/ \Phi) (B_{e} / \delta l) $, where $\delta l$ is the updated length of the segment. Pressure follows from Equations \eqref{eqn:TFT_P0} and \eqref{eqn:TFT_P0_time}, and the process is repeated until the end of the simulation.

As the two tubes retract they move in opposite directions, but due to the symmetry of the background configuration their dynamics are identical. Therefore, we present results only for the downward moving one. The simulated equations are dimensionless, but for concreteness the results are re-dimensionalized using characteristic magnetic field magnitude, density, and length.  For this illustrative purpose, we choose $B_{0} =$ \realB, density $\rho_{0} = n_{0} \overline{m}$ with electron number density $n_{0} =$ \realnumdens, and $\overline{m}$ equal to half the mass of the proton. The initial angle between the pre-reconnection field lines is $2 \zeta=$ \doublezeta. The associated Alfv\'{e}n speed and temperature are $v_{A0} \simeq$ \realvao, and $T_{0}^{A} \simeq$ \realTao. We will consider a plasma at $T_{e}$ = \realTemp, and the plasma-$\beta$ in this case is equal to $\beta = {2 T_{e}}/{T_{0}^{A}}\simeq$ \realbeta. For a typical length of $L_{0} =$ \realL, the Reynolds number is $R_{\eta}\simeq$ \realReynolds. An Alfv\'{e}n time (time to move one characteristic length at the Alfv\'{e}n speed) would be approximately \realalfvtime. As mentioned before, the Prandtl number of the solar corona is typically equal to \PrandtlSim, and we will use this value in our simulations. The total number of segments in the tube is \numsegments. 

The simulation is carried out for approximately \fintime, to allow for the full development of shocks (discussed in detail in the next section). The initial length of the tube was chosen to be unrealistically long in order to achieve the steady state for the shocks. As we will see in the next section, the thickness of the shocks, when fully developed, is of the order of the length of an entire coronal loop. The fluid inside the tube evolves toward the steady state, which it probably would not attain within a realistic flare time. 

\begin{figure}
   \centering
   \resizebox{5.5in}{!}{\includegraphics*[0,0][504,360]{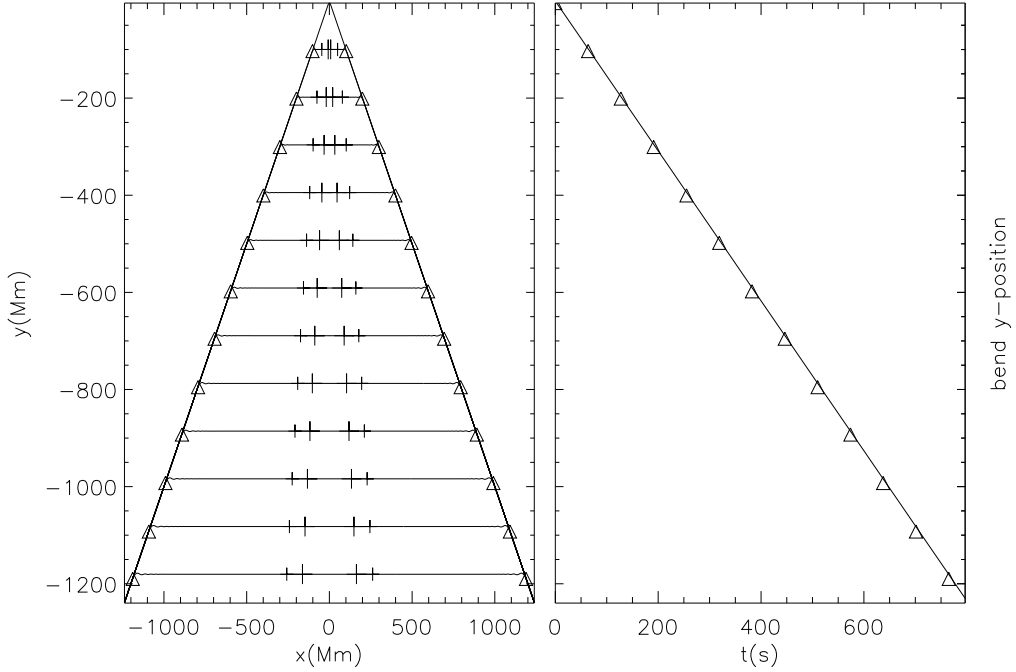}}
   \caption[Simulation results for a downward moving reconnected tube]
   {Left panel: simulation results for the temporal evolution of a downward moving reconnected tube. Each horizontal segment corresponds to a different time. Vertical ticks account for the thickness of the two outward moving gas-dynamic shocks. The width of the shock evolves up to a steady value. Triangles indicate bend positions. Right panel: right bend $y$-position as function of time (same times chosen for the left panel). The solid line corresponds to a line with a slope equal to $- v_{Aey}$. Both panels share the same $y$-axis scale}
   \label{fig:tube_evolution}
\end{figure}

The left panel of Figure \ref{fig:tube_evolution} shows the downward moving tube at different times. The tube is seen to be composed of three main segments, one horizontal segment moving downward at the Alfv\'{e}n speed related to the $y$-component of the magnetic field at the reconnection point ($v_{Aey} = v_{Ae} \sin (\zeta)$, with $v_{Ae}=v_{A0}$), and two legs from the unperturbed parts of the initial tube. The two corners where these three pieces come together move at the Alfv\'{e}n speed along the legs of the initial tube and will be called {\em bends} (triangle shapes in the figure). In the right panel, the $y$-position of the right bend is plotted as a function of time for the same times as in the left panel. The solid line correspond to a line with a slope equal to $- v_{Aey}$.

%
%
%

\begin{figure}[t]
  \centering
  \resizebox{5.5in}{!}{\includegraphics*[0,0][504,360]{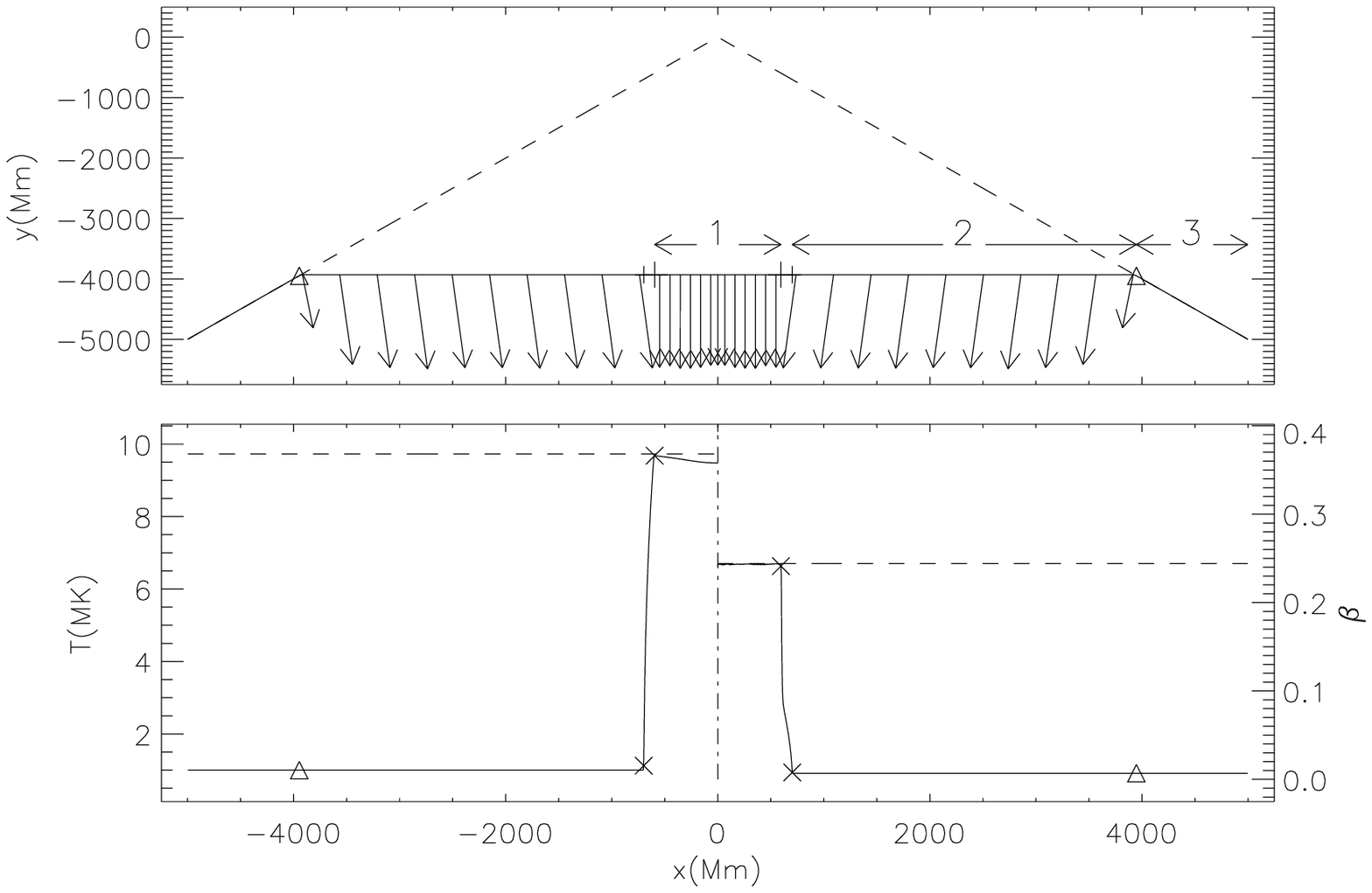}}
  \caption[Jump conditions along the tube]
   {Top panel: dashed line represents initial tube. Solid line represents the tube at time equal to \timeSim \ (vertical scale is compressed). Horizontal arrows delimit regions 1, 2 and 3 (post-shock plasma, inflow region, and unperturbed leg of the tube, respectively). Arrows show velocity directions in the tube (not to scale). Plasma is deflected in the bisector direction at the bends. Large and small vertical ticks indicate the beginning and end of the width of the shock, respectively. Bottom panel: the left side depicts temperature profile in half of the tube, and the right side, the plasma-$\beta$ profile for the same time as in the top panel. Dashed lines show the theoretical steady-state Rankine-Hugoniot post-shock values. Crosses are the same as vertical ticks in the top panel. In both panels triangles indicate the position of the bends. There is no change in the state variables at the bends (Alfv\'{e}n waves). The tube has mirror symmetry in the $x$-direction. Both panels share the same $x$-axis scale}
   \label{fig:profile_T_Beta}
\end{figure}

The magnitude of the magnetic field remains constant along the tube due to constraint \eqref{eqn:TFT_low_beta_B}, therefore the area of the tube also remains constant (Equation [\ref{eqn:flux_tft}]). At the bends, the initially stationary plasma is deflected along the bisector of the angle between the two adjacent straight segments (average direction of the curvature force), as shown in Figure \ref{fig:profile_T_Beta}. In this figure, a time approximately equal to \timeSim\ was chosen to show the shape and internal structure of the tube long after reconnection. 

%
%

The velocity of the deflected plasma (region 2 in Figure \ref{fig:profile_T_Beta}) is \citep[see][]{Longcope_2009}
%
%
\begin{eqnarray}
    \label{eqn:inflow_speed}
       \mathbf{v}_{2}  = - 2 v_{Ae} \sin^{2}\left( \frac{\zeta }{2} \right) \widehat{\mathbf{x}} - v_{Ae} \sin (\zeta)  \widehat{\mathbf{y}}.
\end{eqnarray}
%
%
This velocity is depicted in the top panel of Figure \ref{fig:profile_T_Beta} (the left side is a mirror image). 

The parallel component ($\widehat{\mathbf{x}}$ direction) of this velocity is proportional to the Alfv\'{e}n speed, and is therefore generally supersonic ($v_{Ae} = c_{se} \sqrt{2/\gamma \beta}$). The tube gets shorter as it moves, creating parallel supersonic flows at the bends. These supersonic flows are directed toward the center of the tube where they collide. The small, but non-zero thermal pressure there prevents plasma from piling up and stops it in the parallel direction. The two colliding flows generate two gas-dynamic shocks (the magnetic field remains constant across these shocks) that move outward from the center of the tube. In the top panel of Figure \ref{fig:profile_T_Beta}, velocity discontinuities in the parallel direction between region 1 (post-shock region) and region 2 (inflow region), are evident. The vertical ticks indicate the beginning and end positions of these shocks. In the bottom panel of the figure, the left side depicts the temperature profile along the tube, and the right side the plasma-$\beta$ profile, both for $t =$ \timeSim. We see that temperature rises almost one order of magnitude from the pre-shock value, and the plasma-$\beta$ more than $30$ times its initial value, and neither changes at the bends since these corners are Alfv\'{e}n waves.

\begin{figure}
  \centering
  \resizebox{5.5in}{!}{\includegraphics*[0,0][504,360]{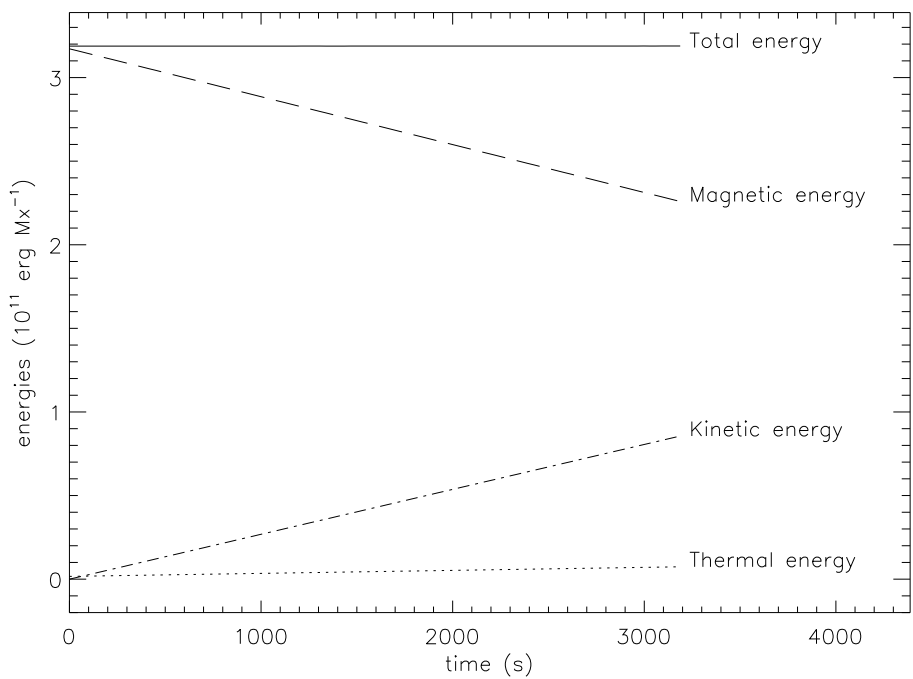}}
  \caption[Energy evolution]
  {Energies' temporal evolution for the simulation presented in Section \ref{sec:Simulations}. Each line represents the temporal evolution of the corresponding specific energy integrated over the entire tube. The percentage of magnetic energy converted to thermal is small (less than 10 \%). The rest is converted to kinetic energy}
   \label{fig:energy_convertion}
\end{figure}

From Equation \eqref{eqn:TFT_cons_Energy} and the boundary conditions at end-points of the tube, we see that the total energy of the tube is conserved (as shown in Figure \ref{fig:energy_convertion}). Even though, temperature changes significantly at the shock, only a small fraction (less than 10 \%) of the available magnetic energy is converted to thermal energy (Figure \ref{fig:energy_convertion}). The magnetic energy of the tube decreases with time as the tube shortens (the magnitude of the field remains constant in the tube, but its length decreases), and is mostly converted to kinetic energy at the bends.

%
%
%
%

%
%
\section{Gas-Dynamic Shock Internal Structure}
   \label{sec:Inner_Structure}

Transport coefficients (viscosity and thermal conduction) and their temperature dependence determine the inner structure of the gas-dynamic shocks. As the flux tubes retract, they follow a time-dependent evolution until they reach a steady-state characterized by a transition between upstream and downstream states dictated by the Rankine-Hugoniot conditions \citep{Rankine_1870,Hugoniot_1887}. The diffusive terms smooth out the transitions but do not alter the upstream or downstream values provided these regions are asymptotically uniform (${\partial \mathbf{v}}/{\partial l}\to0$ and ${\partial T}/{\partial l}\to0$). These asymptotic values are governed by conservation laws observed by the diffusion.

Gas-dynamic shocks in reconnected thin flux tubes were introduced by \citet{Longcope_2009}, in the case of ideal thin flux tube equations. It was shown there that only two kind of discontinuities are supported by reconnected thin tubes embedded in a low plasma-$\beta$ uniform background: one that corresponds to a bend moving at the Alfv\'{e}n speed, where temperature, density and pressure are continuous, and a second one that corresponds to a gas-dynamic shock within a straight section of the tube. Both of these discontinuities were present in our simulation. In this section, we will focus on the later one, where heating and cooling occur (the heating at the bends is negligible). 

To determine the effect of diffusive terms in the inner structure of the shocks, we follow an approach similar to the one used by several authors \citep{Thomas_1944,Grad_1952,Gilbarg_1953,Kennel_1988}. We consider a straight section 
($\partial \widehat{\mathbf{b}}/ \partial l = 0$) and further assume the presence of a 
transition (shock) moving to the right at constant speed $v_{s}$. The inflow region will be denoted by subscript ``2'', and the post-shock region by subscript ``1''. In our reconnection scenario, the inflow region corresponds to the plasma moving supersonically after being accelerated  by the bend (region 2 in Figure \ref{fig:profile_T_Beta}), and the post-shock region corresponds to the center of the tube (region 1 in the same figure).  

\subsection{Steady-state Solution}

The Mach number (the ratio between the fluid velocity and the local sound speed) in the inflow region is determined by the initial plasma-$\beta$ and the initial angle between the field lines \citep{Longcope_2009}, namely
%
%
\begin{eqnarray}
   \label{eqn:Mach_num} 
      M_{2} = \sqrt{ \frac{8 }{\gamma \beta_{e}} } \sin^{2} \left( \frac{\zeta }{2} \right),
\end{eqnarray}
%
%
and the Mach number in the reference frame of the shock is 
%
\begin{eqnarray}
   \label{eqn:Mach_num_shock} 
      M_{2,s} = M_{2} + \frac{v_{s}}{c_{s,2}}=\sqrt{\frac{u_{2}^{2}}{\gamma T_{2}}}, 
\end{eqnarray}
%
%
where $c_{s,2}$ is the sound speed in region 2, and $u_{2}$ and $T_{2}$ are the pre-shock velocity in the shock reference frame and temperature, respectively. For our simulation in the last section, the Mach numbers were $M_{2} \simeq$ \MachSim\ and $M_{2,s} \simeq$ \MachSimShock.

In the reference frame of the shock, the inflow velocity corresponds to plasma moving at $u_{2} = v_{x,2} - v_{s}$ ($< 0$), and the post-shock flow is $u_{1} = -v_{s}$ (plasma is stopped in the parallel direction). The ratio between the shock speed and the inflow speed magnitude, $|v_{x,2}|$, is given by
%
\begin{eqnarray}
   \label{eqn:shock_vs} 
       \frac{v_{s}}{|v_{x,2}|}  = \sqrt{ \frac{1}{M_{2}^{2}} + \frac{(\gamma + 1)^{2}}{16} } - \frac{3-\gamma}{4} .
\end{eqnarray}
%
%
If constant states are assumed, $\partial / \partial t = 0$ for all the quantities calculated in the moving frame of this discontinuity. In this reference frame, the steady-state, conservative, unitless TFT equations for uniform background case and low beta can be written as 
%
%
\begin{eqnarray}
    \label{eqn:Int_Str_mass}
       \rho_{1} u_{1} & = & \rho_{2} u_{2} = \rho u , \\
    \label{eqn:Int_Str_mom}
         \rho_{1} u^{2}_{1} + P_{1} & = &  \rho_{2} u^{2}_{2} + P_{2}    = \rho u^{2} + P - \eta \frac{\partial u}{\partial l}, \\
    \label{eqn:Int_Str_ene}
         \rho_{1} u^{3}_{1} + u_{1} P_{1} \frac{2 \gamma}{\gamma-1} & = & \rho_{2} u^{3}_{2} + u_{2} P_{2} \frac{2 \gamma}{\gamma-1} =   \\ 
        & = &  \rho u^{3} + u P \frac{2 \gamma}{\gamma-1} - 2 u \eta \frac{\partial u}{\partial l} - 2 \kappa \frac{\partial T}{\partial l}, \nonumber
\end{eqnarray}
%
%
%
The second equalities in each of the above equations refer to the state variables inside the shock. The usual Rankine-Hugoniot conditions can be found from the first equalities in each of the above equations. 

It is convenient for our analysis of the internal structure of the shock, to define a characteristic velocity $u^{*} = \frac{1}{2} \left(u_{2} + {T_{2}}/{u_{2}} \right)= ({u_{2}}/{2})\left( 1 + 1/{\gamma  M_{2,s}^{2} } \right )$. With this definition, the  Rankine-Hugoniot conditions can be written as 
%
\begin{eqnarray}
   \label{eqn:jump_dens} 
      \frac{\rho_{2}}{\rho_{1}} & = &  \frac{u_{1}}{u_{2}} = \frac{(4 \gamma u^{*}/\gamma+1) - u_{2}}{u_{2}}, \\
   \label{eqn:jump_T}
      \frac{T_{1}}{T_{2}} & = & \frac{u_{1} (2u^{*}- u_{1})}{T_{2}}.
\end{eqnarray}
%
%
The plasma-$\beta$ ratio can be calculated as ${\beta_{1}}/{\beta_{2}} = 
({\rho_{1}}/{\rho_{2}})({T_{1}}/{T_{2}})$. In Figures \ref{fig:profile_T_Beta}, \ref{fig:shoulder_zoom}, and \ref{fig:shock_heating_cooling}, horizontal dashed lines show the theoretical steady-state jump values for our simulation. Even though the post-shock plasma-$\beta$ is more than $30$ times the initial value, it is still below unity. 

The second equalities in Equations \eqref{eqn:Int_Str_mom} and \eqref{eqn:Int_Str_ene} are differential equations for the shock's internal state variables. Their boundary conditions are given by the steady-state Rankine-Hugoniot conditions. Combining these two equations we obtain
%
%
\begin{eqnarray}
   \label{eqn:inner_prandtl} 
      \frac{2(\gamma-1)}{P_{r} \hbox{ } u} \frac{\partial T}{\partial u} & = & \frac{2 T - u^{2} (\gamma-1) + 4 (\gamma-1) u^{*}u + u_{2}^{2} (\gamma+1) - 4 \gamma u_{2} u^{*}}{u^{2} - 2 u u^{*} + T} \\
   & = & \frac{h(T,u)}{g(T,u)}, \nonumber
\end{eqnarray}
%
%
where we have defined functions $h(T,u)$ and $g(T,u)$ as the numerator and denominator of the above differential equation. $P_{r}$ indicates the Prandtl number. This equation is valid for any temperature dependence of the transport coefficients, as long as both are the same; and can be numerically integrated for different Prandtl numbers and shock strengths. Numerical solutions of this equation in conjunction with Equations \eqref{eqn:Int_Str_mom} or \eqref{eqn:Int_Str_ene} provide a method to determine the thickness of the shock.

\begin{figure}[t]
  \centering
   \begin{center}$
      \begin{array}{c}
         \resizebox{4.5in}{!}{\includegraphics*[0,0][504,360]{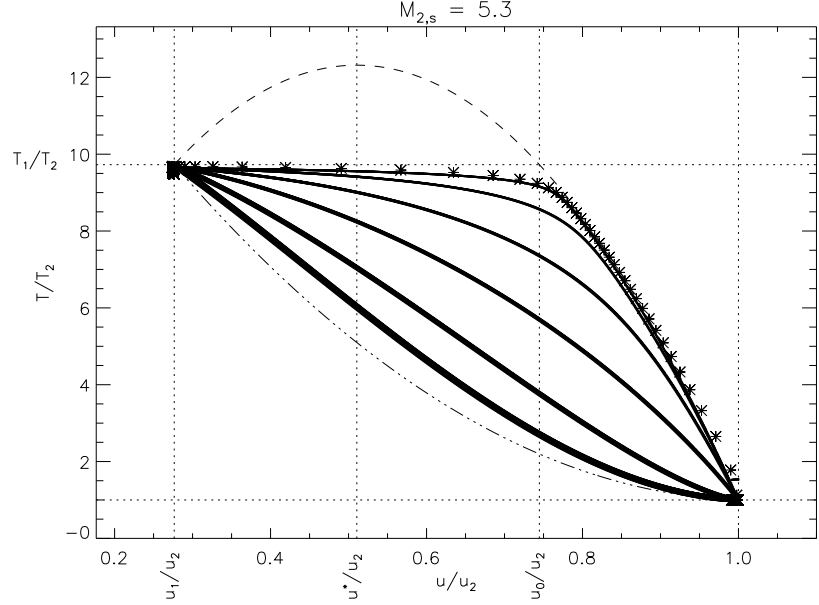}}
      \end{array}$
   \end{center}
  \caption[Fixed Point Diagram for supercritical Mach number]
  {Fixed Point Diagram for super-critical Mach number equal to \MachSimShock\ (the same one than for simulation in Section \ref{sec:Simulations}). Temperature (ordinate) and velocity (abscissa) are scaled to the pre-shock values. For this case, $|u^{*}|>|u_{1}|$. Top and bottom horizontal dotted lines represent post- and pre-shock temperatures, respectively. The leftmost vertical dotted line corresponds to the post-shock speed, and the rightmost dotted line the pre-shock speed. The dashed line represents the curve $g(T,u)=0$, and the mixed dotted and dashed line depicts the $h(T,u) =0$ curve (the latter is the limit solution for an infinite Prandtl number). Each solid line represents a numerical solution for Equation \eqref{eqn:inner_prandtl} for different Prandtl numbers (increasing thickness of the line corresponds to increasing value of  the Prandtl number). The shown Prandtl numbers are \PrandtlSuperCrit. The second vertical dotted line from the left shows the characteristic speed $u^{*}$ (maximum of the curve $g(T,u)=0$), and the second dotted line from the right shows velocity $u_{0}$ (separation between sub-shock and heat front for $P_{r} = 0$). Asterisks corresponds to grid points inside the shock for the simulation presented in Section \ref{sec:Simulations} ($P_{r}=$ \PrandtlSim) for time = \timeSim\ (the same time chosen for Figure \ref{fig:profile_T_Beta}). The internal structure of the shock is well resolved by the DEFT program (there are several grid points in each shock section)}
   \label{fig:fixed_point}
\end{figure}

\begin{figure}[t]
  \centering
  \resizebox{5.0in}{!}{\includegraphics*[0,0][504,360]{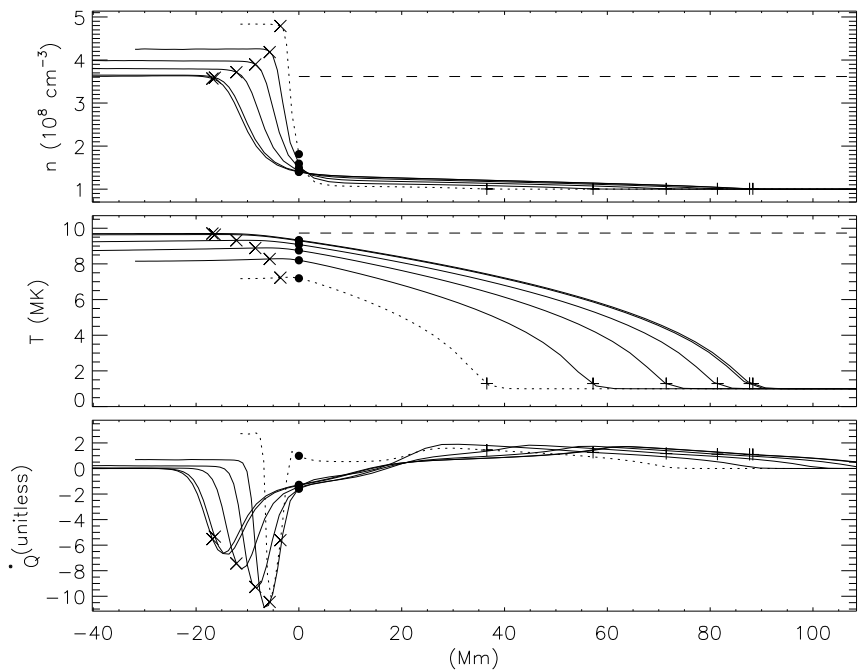}}
  \caption[Electron number density, temperature, and volumetric heating rate]
   {Top panel: electron number density evolution in the shock frame for times $t = 63$ s, $159$ s, $319$ s, $638$ s, $1913$ s, and $3156$ s. Middle panel: temperature evolution in the shock frame for the same times as in the top panel. Bottom panel: volumetric heating rate in the shock frame for the same times as in the top panel. In these three panels, plus signs indicate the position of the heat front, crosses indicate the position of the isothermal sub-shock, and dark dots indicate the position of the transition between the isothermal sub-shock and the thermal front. The dotted lines indicate the earliest time. Horizontal dashed lines are the post-shock Rankine-Hugoniot correspondent values}
   \label{fig:shock_heating_cooling}
\end{figure}

In Figure \ref{fig:fixed_point}, numerical solutions to the above equation are plotted for the same Mach number as in our simulation from last section (temperature and velocity are normalized to the pre-shock values). The pre- and post-shock values are fixed by the strength of the shock, independently of the transport coefficients, therefore we will call this kind of diagrams {\em fixed point diagrams} (they correspond to a velocity contrast-temperature plane described by  \citet{Kennel_1988}, or Navier-Stokes direction fields by \citet{Grad_1952}). Each solid line corresponds to a numerical solution to the above equation for a given Prandtl number (increasing thickness of the line represents increasing value in the Prandtl number). In this figure, the mixed dotted and dashed line shows the curve $h(T,u) = 0$ that corresponds to the limit of infinite Prandtl number $P_{r}$. As the Prandtl number increases from low values, the solutions (solid lines) approach this limit case. The dashed line represents the $g(T,u)=0$ curve, and asterisks indicate grid points (inside the shock region) from our simulation ($P_{r} = $ \PrandtlSim) for the same time chosen for Figure \ref{fig:profile_T_Beta}. Simulations coincide with the steady-state theoretical predictions. There are several grid points in each sub-region of the shock, which shows that the DEFT program can resolve the internal structure of the shock.

%
%
%
Velocity $u^{*}$ corresponds to the maximum value of the curve $g(T,u)=0$ in the $T$-$u$ space, as shown in Figure \ref{fig:fixed_point}. When the magnitude of this speed is smaller than the magnitude of the post-shock speed $u_{1}$, the temperature increases monotonically from $T_{2}$ to $T_{1}$, without presenting an isothermal region, for all Prandtl numbers \citep[\S Section 88]{Landau_1959}. This occurs in weak shocks, called sub-critical, which have Mach numbers $M_{2,s}$ smaller than a critical Mach number $M_{2,cr}=\sqrt{({3\gamma-1})/{(3-\gamma)\gamma}}$. 

On the other hand, if the shock is strong, the fixed point diagram looks like the one depicted in Figure \ref{fig:fixed_point}. In this case, if the Prandtl number is small (as in the solar corona and our simulation), the solution is composed of two regions \citep[\S Section 88]{Landau_1959}. In the first region, called the heat front or thermal conduction front, temperature is increased approximately along the $g(T,u) = 0$ curve, from $T_{2}$ up to approximately the steady-state value $T_{1}$. The second region, called isothermal sub-shock or isothermal discontinuity (or jump), consists of an almost isothermal (${\partial T}/{\partial u}$ = 0) change of velocity from $u_{0} = {T_{1}}/{u_{1}}$ to the steady-state value $u_{1}$. This last region corresponds to an isothermal compression of the plasma; most of the compression across the shock occurs in this region (see top panel of Figure \ref{fig:shock_heating_cooling}). The heat front increases the temperature of the plasma without changing the density significantly. The compression ratio between the initial density $\rho_{2}$ and the density at the transition between the sub-shock and the heat front $\rho_{0}$ is bounded between the values $({\gamma+1})/{2}$ (limit for a infinitely strong shock) and $({3 \gamma-1})/({\gamma +1})$ (at the critical Mach number), for a Prandtl number equal to zero. For a gas constant equal to $\gamma = \frac{5}{3}$, this corresponds to a density ratio between $1.33$ and $1.5$.

%
%
%

\subsection{Physical Size}
 
The thickness of the two internal shock regions, sub-shock and heat front, is determined by Equations (\ref{eqn:Int_Str_mom}) and (\ref{eqn:Int_Str_ene}), respectively. For the case of strong shocks at low Prandtl number, the thickness of the isothermal sub-shock is determined mostly by viscosity, and the heat front length by the thermal conduction, as shown below. To calculate these thicknesses analytically, we approximate our solution by the case where $P_{r}$ is zero. In Figures \ref{fig:tube_evolution} and \ref{fig:profile_T_Beta}, large and small vertical ticks indicate the beginning and end of the thickness of the shock that correspond to the width of the sub-shock in addition to the width of the thermal front. 

\begin{figure}[ht]
  \centering
  \resizebox{5.0in}{!}{\includegraphics*[0,0][504,360]{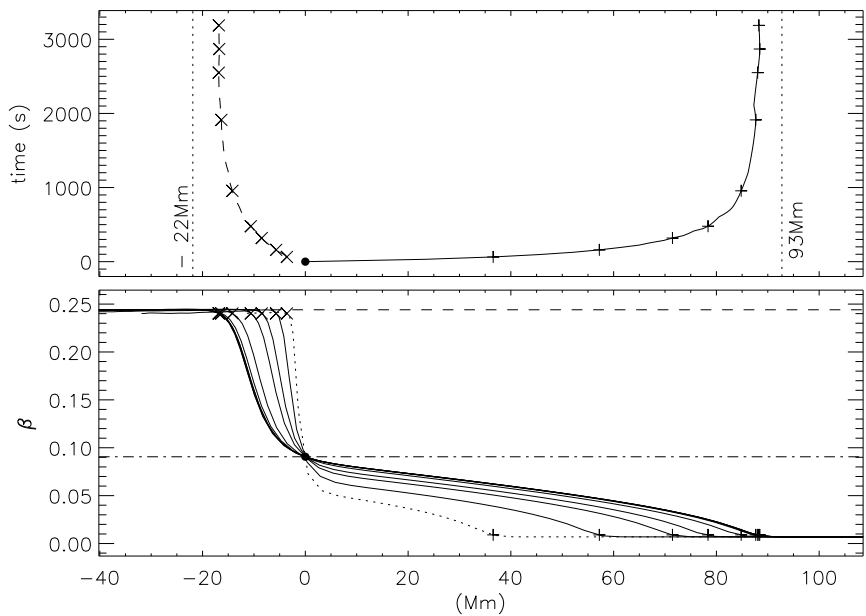}}
  \caption[Internal shock structure in the reference frame of the shock]
  {Internal shock structure in the reference frame of the shock, for the simulation presented in Section \ref{sec:Simulations}. The origin is the transition between the sub-shock and the heat front, indicated with a solid circle. Both panels share the same $x$-axis. Bottom panel: plasma-$\beta$ as function of position in the tube for different times ($64$ s, $159$ s, $319$ s, $478$ s, $956$ s, $1913$ s, $2550$ s, $2869$ s, and $3188$ s). The dotted line represents the earliest time. Horizontal dotted line represents the value of $\beta_{0}$ that corresponds to temperature $T_{1}$, velocity $u_{0}$, and density $\rho_{0}$. Horizontal mixed dotted and dashed line represent the theoretical post-shock values. Plus signs indicate the position of the heat front (they correspond to small vertical tick marks in top panel of Figure \ref{fig:profile_T_Beta}). Crosses indicate the position of the isothermal sub-shock (they correspond to large tick marks top panel of Figure \ref{fig:profile_T_Beta}). Top panel: temporal evolution of the heat front (solid line) and sub-shock (dashed line) lengths. Plus signs and crosses indicate the same positions as in the bottom panel. Leftmost vertical dotted line is the theoretical value for the length of the sub-shock calculated from Equation \eqref{eqn:sub_shock_length}, and rightmost vertical dotted line is the theoretical value for the length of the heat front} 
   \label{fig:shoulder_zoom}
\end{figure}

In the bottom panel of Figure \ref{fig:shoulder_zoom}, the evolution of the plasma-$\beta$ of the tube is shown (in the reference frame of the shock). The thermal front extends to the right, and the isothermal sub-shock extends to the left. The top panel shows the evolution of the length of each region. 

The length (unitless) of the isothermal sub-shock, $\Delta l_{\mathrm{SS} (1 \rightarrow 0)}$, is calculated replacing $T = T_{1}$ in Equation \eqref{eqn:Int_Str_mom}, then
%
%
\begin{eqnarray}
   \label{eqn:sub_shock} 
      \Delta l_{\mathrm{SS} (1 \to 0)} & = & \frac{\eta(T_{1})}{\rho_{2}u_{2}} \int_{u_{1}}^{u_{0}} \frac{u \hbox{ }du}{(u-u_{1})(u-u_{0})}, 
\end{eqnarray}
%
where integration is calculated from velocity $u_{1}$ to velocity $u_{0}={T_{1}}/{u_{1}}=2u^{*}-u_{1}$ (this latter speed corresponds to the velocity at the transition between the sub-shock and the heat front as shown in Figure \ref{fig:fixed_point}). The plasma-$\beta$ at this point can be calculated as $\beta_{0}=\beta_{2} u_{1} {u_{2}}/{T_{2}}$. In the bottom panel of Figure \ref{fig:shoulder_zoom}, the horizontal dotted line represents $\beta_{0}$. This value is used to approximate the position of the transition between the sub-shock and the heat front (zero horizontal coordinate in this figure). 

The integrand in Equation \eqref{eqn:sub_shock} diverges at both limits, therefore we will slightly modify them. We will define a smaller centered interval corresponding to a $95$\% of the original integration interval, and the integration will be performed with new limits. Defining a small interval $u_{\mathrm{int}}= \left({u_{0}}/{u_{1}}-1 \right)/40$, the analytical solution for the length of the sub-shock is prescribed as
%
\begin{eqnarray}
   \label{eqn:sub_shock_length} 
      \Delta l_{\mathrm{SS}} & = &  \frac{L^{\mathrm{ion}}(T_{2})}{\sqrt{2 \gamma} M_{2,s}} \left. \left[ \frac{\log(u-1) -  \frac{u_{0}}{u_{1}} \log(|u-\frac{u_{0}}{u_{1}}|) }{\frac{u_{0}}{u_{1}}-1} \right] \right |^{\frac{u_{0}}{u_{1}} - u_{\mathrm{int}}}_{1+u_{\mathrm{int}}}  \\
 &  &  \times 
\left\{ 
  \begin{array}{c l}
  1 & (\eta \sim T^{0}), \\
  \left( \frac{T_{1}}{T_{2}} \right)^\frac{5}{2}  & (\eta \sim T^{\frac{5}{2}}). 
\end{array}
\right. \nonumber
\end{eqnarray}
%
%
where $L^{\mathrm{ion}}(T_{2}) = \eta(T_{2})/\rho_{2} v_{\mathrm{th}}(T_{2}) L_{0}$ is the unitless mean free path of the ions at temperature $T_{2}$ (all the lengths are non-dimensionalized to characteristic length $L_{0}$); and $v_{\mathrm{th}}(T_{2})$ is the thermal speed of the ions at $T_{2}$. For the simulation presented in Section \ref{sec:Simulations}, $L^{\mathrm{ion}}(T_{2}) \simeq$ \protmfpunitless. 

%
%
%

Expression \eqref{eqn:sub_shock_length} presents values for two cases, as shown in the last part of the equation. The first one corresponds to temperature independent $\eta$, and the second one corresponds to $\eta \sim T^{\frac{5}{2}}$. The length of the sub-shock is considerably different depending on which model is used for viscosity. For the simulation we presented in the last section (temperature dependence of viscosity given by Equation \eqref {eqn:transp_coeff_eta}), the ratio between post- and pre-shock temperatures was almost one order of magnitude, therefore the sub-shock length would be approximately $10^{5/2}$ times the one predicted by a constant viscosity model. 

For the parameters of our simulation, the bottom part of Equation \eqref{eqn:sub_shock_length} gives a length of the sub-shock approximately equal to \LSubShock, which is in good agreement with our simulation (see top panel of Figure \ref{fig:shoulder_zoom}). In this figure, the length of the sub-shock (negative positions) increases up to a steady-state value well approximated by the analytical solution for Prandtl number equal to zero.

\begin{figure}[t]
  \centering
  \resizebox{5.0in}{!}{\includegraphics*[0,0][504,360]{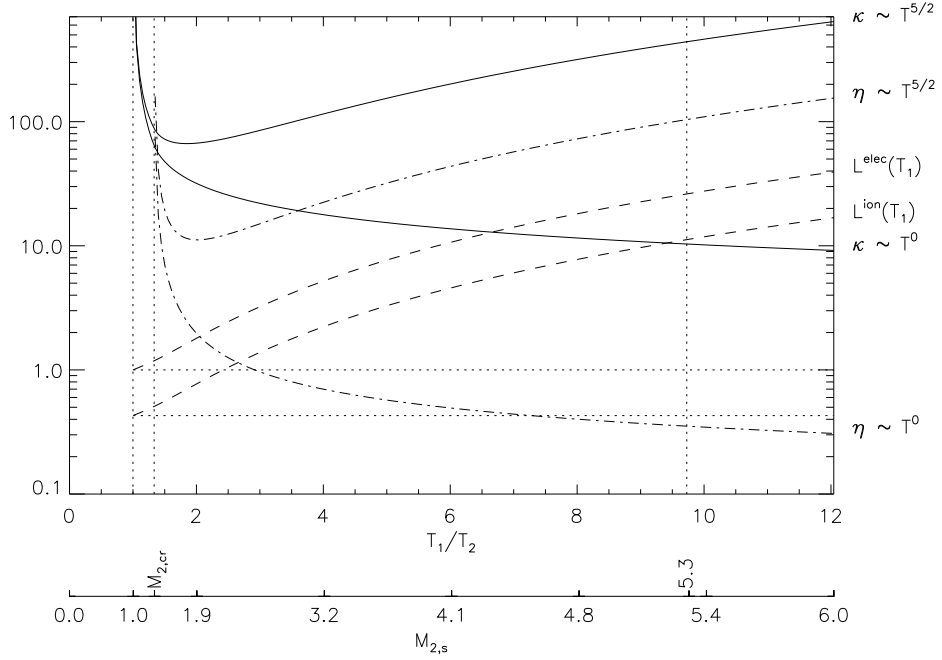}}
  \caption[Different lengths scaled with the pre-shock electron mean free path]
    {Different lengths scaled with the pre-shock electron mean free path $L^\mathrm{elec}(T_{2})$, as function of the ratio of the post-shock to the pre-shock temperature. The second axis at the bottom shows the corresponding Mach number $ M_{2,s}$. Vertical dotted lines represent cases $M_{2,s} =1$ (no shock), $M_{2,s} = M_{2,cr}$ (critical case), and $M_{2,s} =$ \MachSimShock \ (shock strength for the simulation presented in Section \ref{sec:Simulations}). The lengths in  this figure correspond to the same Reynolds and Prandtl number used for the simulation presented in Section \ref{sec:Simulations}. Bottom horizontal dotted line represents the ratio of the ion to electron mean free path at $T_{2}$, and top horizontal dotted line is the unity (electron mean free path at $T_{2}$). Solid lines depict the length of the heat front $\Delta l_{\mathrm{HF}}$ (calculated from Equations \eqref{eqn:H_F_length_const} and \eqref{eqn:heat_front}), for the cases labeled to the right. Mixed dotted and dashed lines represent the length of the isothermal sub-shock $\Delta l_{\mathrm{SS}}$ (calculated from Equation \eqref{eqn:sub_shock_length}), for the same cases, and dashed lines show the mean free path of electrons and ions at post-temperature $T_{1}$}
   \label{fig:internal_lengths}
\end{figure}

When a constant viscosity model is used, the length of the sub-shock is smaller than the mean free path of the ions for temperature ratios larger than approximately $7$ (see Figure \ref{fig:internal_lengths}). This is not true for the case of more realistic temperature dependencies like the one used in our simulation ($\eta \sim T^{\frac{5}{2}}$). For this case, the length of the sub-shock is always more than one order of magnitude larger than that of the ion mean free path, supporting the use of fluid equations. The length of the sub-shock diverges when $u_{0} = u_{1}$ (critical case), as shown in Figure \ref{fig:internal_lengths} (mixed dotted and dashed lines), and there is no sub-shock for Mach numbers below the critical one. 

%
%
%

To determine the length of the heat front, we will consider the portion of $g(T,u) = 0$ between velocities $u_{0}$ and $u_{2}$, and replace the corresponding expression for the velocity $ u = u^{*}\left(1 + \sqrt{1-\frac{T}{(u^{*})^{2}}} \right) $ in Equation \eqref{eqn:Int_Str_ene}. If we define a unitless variable $\tau = \sqrt{1-T/(u^{*})^{2}}$, and integral limits $\tau_{2}(M_{2,s}) = \sqrt{1-\frac{T_{2}}{(u^{*})^{2}}}$, and $\tau_{1}(M_{2,s}) = \sqrt{1-\frac{T_{1}}{(u^{*})^{2}}}$, the integral equation for the (unitless) thickness of the heat front becomes
%
\begin{eqnarray}
   \label{eqn:heat_front} 
      \Delta l_{\mathrm{HF} (0 \to 2)} & = & \frac{4}{\rho_{2}u_{2}} \left(\frac{\gamma-1}{\gamma+1} \right) \int_{\tau_{1}}^{\tau_{2}} \frac{\kappa(\tau) \tau \mathrm{ }d \tau}{(\tau \pm \tau_{1})(\tau-\tau_{2})}, 
\end{eqnarray}
%
%
where the ``$+$'' sign in the denominator corresponds to super-critical cases, and the ``$-$'' sign corresponds to the sub-critical cases. In the sub-critical cases the above integral diverges at both integral limits, and for super-critical cases, only at $\tau_{2}$. The critical case corresponds to $\tau_{1}=0$, and the integral is continuous at that point. To define a unique length for all the cases, we will integrate expression \eqref{eqn:heat_front} in a centered interval corresponding to a $95$\% of the original integration interval by defining a small interval $\tau_{\mathrm{int}} = (\tau_{2}-\tau_{1})/40$, analogous to the one use for the isothermal sub-shock. 

The case of temperature independent thermal conductivity can be easily integrated since $\kappa(\tau)$ is a constant and can be taken outside the integral. The analytical solution for this case is 
%
%
\begin{eqnarray}
   \label{eqn:H_F_length_const} 
      \Delta l_{\mathrm{HF}}  -4 \left(\frac{\gamma-1}{\gamma+1} \right) \sqrt{\frac{m_{p}}{m_{e}}} \frac{L^{\mathrm{elec}}(T_{2})}{\sqrt{2 \gamma} M_{2,s}} \left. \left[ \frac{ \tau_{1} \log(\tau \pm \tau_{1}) \pm \tau_{2} \log(\tau_{2}-\tau)}{\tau_{1} \pm \tau_{2}} \right] \right |^{\tau_{2}-\tau_{\mathrm{int}}}_{\tau_{1}+\tau_{\mathrm{int}}}, 
\end{eqnarray}
%
%
Here, $L^{\mathrm{elec}}(T_{2}) = \kappa(T_{2})/\rho_{2} v_{\mathrm{th}}(T_{2}) L_{0}$ is the unit-less mean free path of the electrons at pre-shock temperature $T_{2}$, and $m_{p}$ and $m_{e}$ represent the proton and electron masses, respectively. For the simulation presented in Section \ref{sec:Simulations}, $L^{\mathrm{elec}}(T_{2}) \simeq$ \elecmfpunitless. 

The temperature-dependent thermal conductivity case has a slightly more complicated integral where $\kappa(\tau) = \kappa_{c} |u^{*}|^5 (1 - \tau^{2})^{\frac{5}{2}}$. For infinitely strong shocks ($M_{2,s} = \infty$), expression \eqref{eqn:heat_front} has a simpler form since $\tau_{1}(\infty) = \frac{3-\gamma}{\gamma+1}$ and $\tau_{2}(\infty) = 1$ (the divergence at $\tau_{2}$ is removed). In this case the integral converges to a finite number, and we see that the length of the heat front is $\infty$ since the factor $|u^{*}|^5/\rho_{2}u_{2}$ (that can be taken outside the integral) increases with the Mach number of the shock. The opposite occurs for the constant thermal conduction case.

For finite shocks, Equation \eqref{eqn:heat_front} can be integrated analytically or numerically. For the realistic coronal parameters used in our simulation, the length of the heat front is approximately equal to \LHeatFront, in good agreement with our simulation (see top panel of Figure \ref{fig:shoulder_zoom}). This length is of the order of the length of a typical coronal loop, and one order of magnitude more than the length of the mean free path of the electrons at the post-shock temperature. Therefore, for a real coronal situation, the heat front will probably never reach its steady-state value since it will arrive at the end-points before completing its development. At one Alfv\'{e}n time (\realalfvtime), the length of the heat front is already approximately \LHeatFrontAlfven\ long (the evolution of the heat front is shown in the top panel of Figure \ref{fig:shoulder_zoom}). 

\subsection{Justifying a Fluid Treatment}

In Figure \ref{fig:internal_lengths}, the lengths of the isothermal sub-shock and heat front for temperature independent and temperature dependent models are plotted as function of the post-shock to pre-shock temperature ratio. All the lengths are scaled to the electron mean free path at temperature $T_{2}$. Solid lines depict the length of the heat front, mixed dotted and dashed lines show the length of the isothermal sub-shock, and dashed lines display the particle's mean free path at $T_{1}$. 

Despite its complicated analytical form, the length of the heat front for the temperature dependent case follows a simple curve that for relatively strong shocks has a shape similar to the electron mean free path at $T_{1}$ (dashed line). This gives some insight into the physics involved. The electron mean free path increases with post-shock temperature, as $T^{2}$ (Coulomb potential), while the thermal conduction also increases with post-shock temperature as $T^\frac{5}{2}$. The interplay between particle's mean free path and transport coefficient's temperature dependence determines the thickness of the shock. Whenever the diffusive coefficient's growth with temperature is faster than the particle's mean free path, the thickness of the shock is going to be larger than the particle's mean free path. Heat is transported larger distances ahead of the shock as the temperature ratio increases, maintaining a heat front length always larger than the mean free path. A similar analysis can be done for the sub-shock and the ion mean free path. When temperature independent transport coefficients are used, the results are several orders of magnitude smaller than the corresponding mean free paths, which in the past created some concern about the use of fluid equations for shock studies. 

The validity of assuming fluid equations, as opposed to kinetic theory, can also be a concern if the ratio between the diffusion heat flux and the value for streaming particles is larger than a certain value \citep{Campbell_1984}. We tested the collisionality of the plasma in our simulation, and obtained ratios smaller than $0.004$. This, and the above paragraph  conclusion, support the use of continuum fluid equations. 

For the short times involved in a flare, the shocks may not achieve their steady state. The evolution of the isothermal sub-shock and the thermal front toward this state can be approximated by an exponential (see top panel of Figure \ref{fig:shoulder_zoom}). With this approximation, the characteristic time for the development of the sub-shock is approximately equal to \SStime, and its initial (maximum) speed is less than $10$\% that of the ion thermal speed. The characteristic time for the heat front is of the order of \HHtime\ (four times the Alfv\'{e}n time), and its initial development speed is $3$\% of the electron thermal speed. We have tested the speed of the thermal front for a range of reconnection angles between $\zeta = 20$\degree\ to $\zeta = 45$\degree. For all these cases, this speed is between $3$\% and $5$\% that of the electron thermal speed. 

The electron number density and temperature evolution toward the steady state of our simulation are shown in Figure \ref{fig:shock_heating_cooling}. The electron number density approaches its steady state from above the Rankine-Hugoniot value (top panel), and temperature approaches from below (middle panel). A short time after reconnection, the two colliding fluids accelerated at the bends get overcompressed at the center of the tube (generating densities well above the expected value  for infinitely strong shocks), and they relax toward the equilibrium value. From the top panel of the figure, it is evident that most of the compression occurs at the sub-shock. The middle panel shows that most of the increase in temperature occurs at the thermal front. From the bottom panel of Figure \ref{fig:shoulder_zoom}, we see that pressure achieves its steady state in a characteristic time much smaller than the ones for density and temperature. 

The bottom panel of Figure \ref{fig:shock_heating_cooling} shows the volumetric heating rate. Most of the cooling occurs at the sub-shock where the plasma is isothermally compressed. Some cooling and heating occur at the thermal front. In this region, the plasma increases its temperature without considerably changing its density. In the sub-shock region, the velocity changes are large (see Figure \ref{fig:fixed_point}), and it would be expected to have a strong positive contribution to the heating, however the cooling and heating are dominated by thermal conduction since the Prandtl number is small. The cooling comes from the negative curvature of the temperature to the power of $7/2$ (not shown in the figure) as described in Equation \eqref{eqn:heating_uniform}. Even for a small temperature gradient in the sub-shock, the heating is negative. Without viscosity, however, it is not possible to develop the needed sub-shock to connect the pre-shock state to the post-shock state. On the other hand, it is possible to only have viscosity and no thermal conduction (limit of Prandtl number equal to infinity).

%
%
\section{Conclusions}
   \label{sec:conclusions_ch1}

We have presented new general fluid equations that describe the dynamics of thin flux tubes. These tubes are assumed to be isolated from their surroundings, except by total pressure balance with their exterior (their small local radius permits almost instantaneous equilibrium with the surroundings via fast-magnetosonic waves). This assumption neglects any drag force due to the external fluid. The ideal part of these equations is applicable to a wide range of conditions for external magnetic fields and pressure. We also proposed realistic field aligned diffusive terms for viscosity and thermal conduction, valid for strongly magnetized plasmas, that include their strong dependence on temperature ($\sim T^\frac{5}{2}$). 

We solve these equations numerically for the post-reconnection dynamics of a thin flux tube embedded in an uniform background with small plasma-$\beta$ ($\beta =$ \realbeta). This tube was assumed to have been reconnected at an impulsive, small patchy region in a current sheet with skewed magnetic fields. Due to magnetic tension, the tube retracts at Alfv\'enic speeds reducing its length. Realistic coronal parameters (large Reynolds number and small Prandtl number) were used in the simulations. 





As the tube evolves, the initially stationary plasma is deflected at the tube bends (rotational discontinuities) and accelerated to super-sonic speeds toward the center of the tube where they collide. The small, but non-zero thermal pressure at the center of the tube stops the inflows, and the two colliding flows generate gas-dynamic shocks that move outward from the center of the tube. This shows the importance of including the parallel thermal pressure gradient in the thin flux tube equations since there is no magnetic force in that direction. The magnitude of the magnetic field remains constant across the shocks (these shocks differ from the Petschek case since they are not switch-off shocks). The presence of diffusive terms in the thin flux tube equations provides a mechanism for the shock development.

In the simulation, temperature rises almost one order of magnitude from the pre-shock value (\realTemp) to the post-shock value ($\sim$ 10MK) across the gas-dynamics shocks. Nevertheless, only a small fraction (less than 10 \%) of the available magnetic energy is converted to thermal energy. The magnetic energy of the tube decreases with time as the tube shortens, and is mostly converted to kinetic energy at the bends. This suggest that the rate of the reconnection mechanism has nothing to do with the final temperature achieved. This final value only depends on the strength of the shock determined by the initial angle between the field lines. 


The inner structure of the shock is related to the transport coefficients and strength of the shock (Mach number). For solar coronal conditions (strong shocks at low Prandtl numbers), the inner structure of the shocks present two sub-regions, a heat front where most of the temperature increase occurs (its width determined by thermal conduction), and an isothermal sub-shock where the fluid is compressed (its width depends on the viscosity). For these coronal conditions, the existence of a shock transition with thermal conduction alone is not thermodynamically permissible. We estimate the thickness of these regions assuming zero Prandtl number and solving the governing differential equations analytically. The thicknesses of the sub-shock and heat front are many times the mean free path of the ions and electrons, respectively. This result, and our calculations regarding the free-streaming heat flux, lead us to conclude that the plasma is sufficiently collisional to use fluid equations. 


The higher density post-shock region should be readily observable. Post-shock density is almost four times greater than the initial background density, leading to sixteen times more emission measure (emission measure scales as the square of the density). The size of this region increases at a rate equal to the speed of the shock (Equation \eqref{eqn:shock_vs}), and at one Alfv\'{e}n time would be of the order of $10$ Mm. There are several types of observations showing bright features of this scale \citep{Masuda_1994,Tsuneta_1997_II,Warren_1999,Jiang_2006}. The radius of our tube is assumed small compared with this length, suggesting that any observable features will consist of multiple reconnection events in a small region, and several thin flux tubes being retracted outward from that region.

The jump condition for thermal pressure is achieved quite rapidly. On the other hand, density and temperature achieve their steady-state values on characteristic times longer than that for thermal pressure. The latter approaches from lower values than the corresponding Rankine-Hugoniot post-shock temperature, and the former from higher values. For large Mach numbers (large initial reconnection angles), this temporary density overshoot may be larger than the superior limit for the Rankine-Hugoniot post-shock density ratio ($\frac{\rho_{1}}{\rho_{2}} = \frac{\gamma + 1}{\gamma-1}$). The value of this ratio may exceed one order of magnitude which would manifest as even stronger X-ray emissions near the reconnection site, that would decrease in intensity as the tube evolves. Most of the cooling occurs in the isothermal sub-shock region, and most of the heating in the thermal front. 

For real coronal situations, the steady-state thickness of the shock might never be achieved since the shock will arrive at the end-points before completing its development (the steady-state thickness value is of the order of the length of the loop). For a wide range of initial pre-reconnection angles (we tested angles ranging from $\zeta = 20$\degree\ to $\zeta = 45$\degree) the speed at which the heat front develops is between $3$\% and $5$\% that of the post-shock thermal electron speed. The heat front barely compresses the plasma, since most of the compression occurs in the isothermal region. The speed at which these fronts move along the legs of the tube depends on the reconnection angle. For a magnetic field of $B_{0} =$ \realB, and electron number density $n_{0} =$ \realnumdens, the range in speed for the thermal fronts is between 
$450$ km s$^{-1}$ and $650$ km s$^{-1}$ for angles between $\zeta = 20$ degrees and $\zeta = 45$ degrees. For larger angles, it is possible for the heat front to overtake the bends.

Heat fronts are possibly relevant for chromospheric evaporation in flares. This is not directly treatable in our model since it is confined to the current sheet whose edge lays high above the chromosphere (see Figure \ref{fig:flare_CS_ch1}). We expect the heat fronts found in our model to continue along the field lines toward their footpoints in the chromosphere. The temperature at the heat front increases approximately up to the post-shock Rankine-Hugoniot value, which could be one order of magnitude larger than the initial temperature of the loop. At these high temperatures chromospheric material can be evaporated, and X-ray emissions from the foot points might be observed. The upward flow of evaporated chromospheric material would eventually encounter the sub-shock that is descending along the legs of the tube. What would happen after is a subject for further investigation.

We neglected radiation in our model. The radiative cooling time at the pre-reconnection initial conditions is several orders of magnitude larger than the Alfv\'{e}n time. The density increase by as much as a factor of four would by itself decrease the radiative cooling time. Temperature, however, also increases one order of magnitude which dramatically increases the cooling time since the radiative loss function decreases with increasing temperature at temperatures higher than $1$ MK, \citep{Cook_1989,Martens_2000,Rosner_1978}. Therefore, radiation is a negligible effect in the post-reconnection process.

Our model assumed a very simple initial geometry: uniform background conditions with pre-reconnection field lines that form an angle at both sides of the reconnection. In future work, we will investigate a more realistic model with, for example, a Green-Syrovatski{\v i} \citep{Green_1965,Syrovatskii_1971} current sheet with a guide field. This model presents new phenomena as the magnetic field magnitude changes along the tube.


\newcommand{\CSHalfLength}{\mbox{$100$} Mm}

\newcommand{\realReynoldsII}{\mbox{$2982$}}
\newcommand{\PrandtlSimII}{\mbox{$0.01$}}


\chapter{DENSITY ENHANCEMENTS AND VOIDS FOLLOWING PATCHY RECONNECTION}
   \label{chap:chap_2}

The layout of this chapter is as follows. The first section describes the background configuration where reconnection is assumed to happen. Three different locations of reconnection across the CS are described. The next section presents the TFT equations that govern the evolution of the reconnected flux tubes, as well as analytical equilibrium solutions to the perpendicular part of the equations. Section \ref{sec:simul} describes simulations of the evolution of reconnected tubes for each reconnection location. The parallel dynamics of the tube is described in detail before the tube arrives at the top of the underlying arcade. In Section \ref{sec:Top_arcade}, the arrival at the arcade apex is simulated by an overdamped spring force exerted by the compressed arcade that slows down and stops the tube. After this arrival, the parallel dynamics changes and strong rarefactions develop. In Section \ref{sec:rarefaction}, the temporal evolution of the tube's total emission measure and mean temperature are shown, and a method is described to determine where in the CS a tube was reconnected. The last section presents a discussion of our results and their possible observational consequences.

%
\section{Flare Current Sheet in the Solar Corona}
  \label{sec:Green_Syrovatskii}

\begin{figure}[ht]
   \centering
   \begin{center}$
      \begin{array}{cc}
         \includegraphics{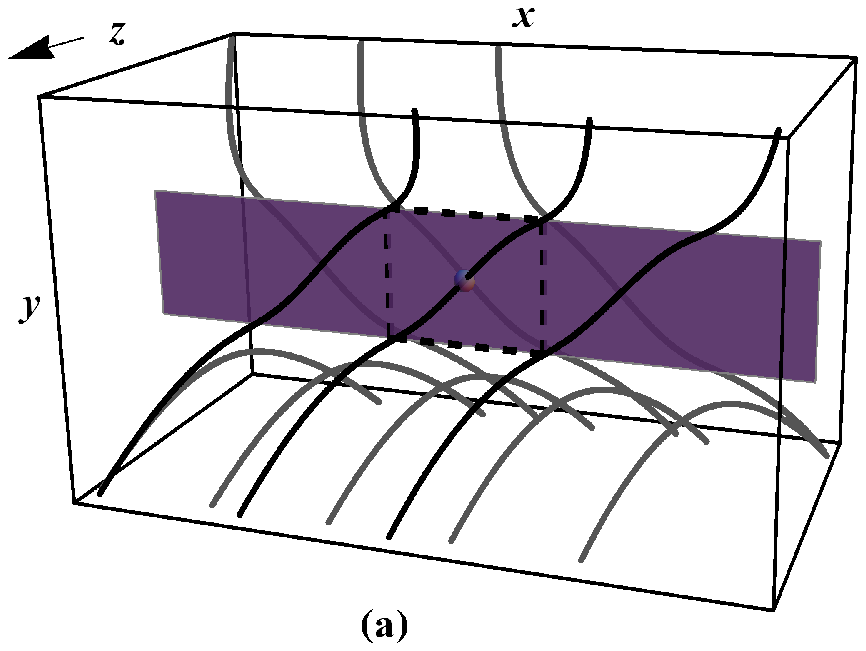}
         \includegraphics{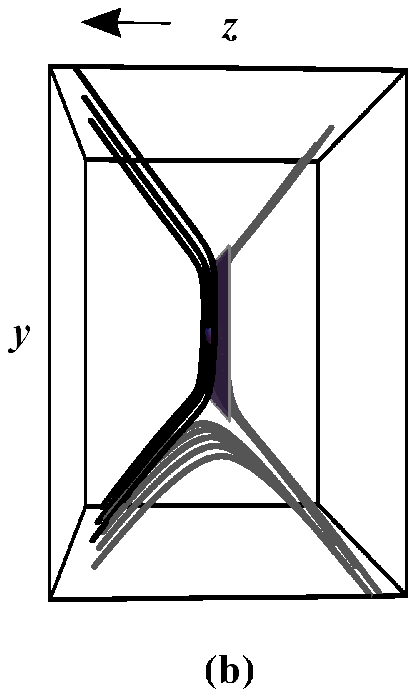}
      \end{array}$
   \end{center}
   \caption[Flare current sheet geometry]
   {Flare current sheet geometry. (a) Field line configuration for a Green-Syrovatski{\v i} current sheet with a guide field in the positive $x$-direction. The transparent gray rectangle represents the current sheet in the $x-y$-plane, located above the post-flare arcade (gray arc-shape loops at the bottom). The current sheet is finite in the $y$-direction, but extends infinitely in the $x$-direction (2.5-dimensional symmetry). The bottom plane of the box represents the solar surface. Black lines depict some magnetic field lines on the side of the current sheet that is closer to the viewer ($z > 0$), and the gray ones whose endpoints touch the top side of the box correspond to field lines on the back side of the current sheet ($z < 0$). The small sphere in the current sheet shows a generic patchy reconnection region. Field lines that intersect this region form a small bundle that reconnects. The dashed rectangle in the plane of the current sheet corresponds to the region shown in Figure \ref{fig:CS_lines}. (b) Different view of the same field lines in panel (a), showing the double Y-type configuration}
   \label{fig:flare_CS} 
\end{figure}

For the location of the reconnection episode in the solar corona, we assume a realistic magnetic field configuration like the one shown in Figure \ref{fig:flare_CS}. This initial equilibrium background presents a CS in the $x-y$-plane, located high in the corona, and skewed magnetic fields on each of its sides. The mathematical description of this configuration is usually called Green-Syrovatski{\v i} \citep{Green_1965,Syrovatskii_1971} CS, or ``double Y-type'' due to the shape of its separatrices when projecting field lines in the $y-z$-plane (see panel (b) of Figure \ref{fig:flare_CS} in the present chapter, and Figure 3(b) in Syrovatski{\v i}'s paper). We included a uniform guide field in the ignorable $x$ direction to provide an arbitrary angle between the pre-reconnection field lines. We also assume uniform initial values of density $\rho_{e}$, and pressure $P_{e}$, on each side of the CS. The CS extends infinitely in the $x$-direction, but is finite in the $y$-direction. The bottom edge of the CS marks the top of the flare arcade (shown in Figure \ref{fig:flare_CS} as arc-shaped loops) with footpoints in the solar surface (bottom panel of the box). We will denote the half length of the CS in the $y$-direction as $L_{e}$. 

%
%

The entire region is assumed to be free of electrical resistivity, with exception of a short-lived small patch, somewhere in the CS where reconnection occurs (small sphere in Figure \ref{fig:flare_CS}). The horizontal and vertical positions of the reconnection site will be denoted by $x_{R}$ and $y_{R}$, respectively. Any horizontal position of the reconnection site is equivalent due to the initial 2.5-dimensional symmetry of the magnetic configuration described above. For simplicity, we will assume it to be at the origin of the horizontal axis, then $x_{R} = 0$. We choose the $y$-axis origin to be located at the center of the current sheet.

\begin{figure}[ht]
  \centering
   \includegraphics{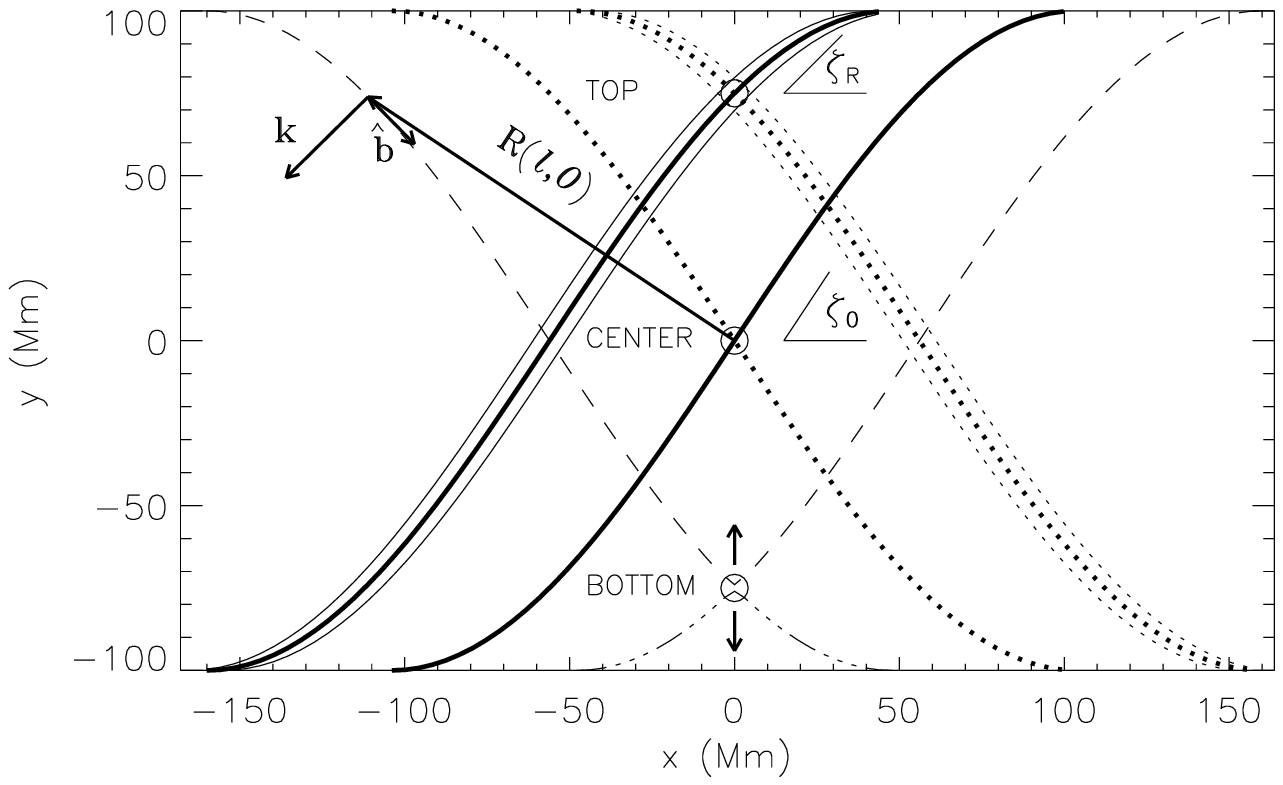}
   \caption[Plane of the current sheet, with three possible reconnection sites]
   {Excerpt of the plane of the current sheet, with three possible reconnection sites. This graph corresponds to the dashed box in Figure \ref{fig:flare_CS}. Pre-reconnection field lines in front of the current sheet ($z > 0$) are represented by solid lines, and pre-reconnection field lines on the back side of the current sheet ($z < 0$) are shown as dotted lines. The edges of the current sheet are located at $y = \pm L_{e} = \pm 100$ Mm. The top circle shows a TOP reconnection site at $y_{R}=\frac{3}{4} L_{e}$. The set of field lines that intersect this region (pre-reconnection tube) is represented by a central line (thicker). At this reconnection site, representative field lines on each side of the current sheet are skewed by $90$\degree. The half reconnection angle ($\zeta_{R} = 45$\degree) is shown to the side of the TOP reconnection circle. The center circle represents a CENTER reconnection site with $\zeta_{R} \simeq 57$\degree. For this site, only the pre-reconnection representative field lines are drawn. The circle in the bottom half of the CS plane, located at $y_{R}= - \frac{3}{4} L_{e}$, represents a BOTTOM reconnection site with $\zeta_{R} = 45$\degree. Here, the initial shape of the already reconnected tubes is drawn, and arrows indicate their direction of motion. The dotted and dashed line shows a reconnected tube that will move downward (DOWNWARD moving tube) due to magnetic tension at its center, and the dashed reconnected tube will move upward (UPWARD moving tube). For this last tube, a parameterization vector $\mathbf{R}(l,0)$ is shown for a given arc-length $l$. Also at this arc length, the unit vector $\widehat{\mathbf{b}}$ is shown, as well as the curvature vector $\mathbf{k}$. For all three reconnection cases, $\zeta_{0} \simeq 57$\degree, therefore field lines on each side of the current sheet are parallel to each other for all reconnection positions}
   \label{fig:CS_lines} 
\end{figure}

We will distinguish three reconnection cases, based on their location on the CS. If reconnection happens in the top half of the CS ($0 < y_{R} < L_{e}$), it will be called TOP reconnection, if it happens in the middle of the CS ($y_{R}=0$), it will be denoted as CENTER reconnection, and finally if reconnection occurs somewhere in the bottom half of the CS, it will be referred to as BOTTOM reconnection. Figure \ref{fig:CS_lines} represents the region of the CS enclosed by a dashed rectangle in Figure \ref{fig:flare_CS} and shows examples of the three different reconnection positions. For the sake of concreteness, we have assumed that the half length of the CS is $L_{e} = $ \realL, a typical length in the solar corona.  

%
%
%

The magnetic field near the CS can be described as the limit as $z \rightarrow 0$ of the skewed Green-Syrovatski{\v i} magnetic field, as follows 
%
\begin{eqnarray}
   \label{eqn:B_syrovatskii}
     \mathbf{B_{e}} &=& \widehat{\mathbf{x}} B_{ex} \pm \widehat{\mathbf{y}} B_{ey} \sqrt{1-\left(\frac{y}{L_{e}}\right)^{2}}  \nonumber \\ 
       &=& \widehat{\mathbf{x}} B_{ex}  \pm \widehat{\mathbf{y}} B_{ey} \sqrt{1-y^{\prime 2}}  ,\hbox{    for } z \gtrless 0.
\end{eqnarray}
%
Here, $B_{ex}$ and $B_{ey}$ are constants. $B_{ex}$ is the guide field, and $B_{ey}$ is the $y$-component of the usual Green-Syrovatski{\v i} field. The positive (negative) sign corresponds to the front (back) side of the CS. For simplicity, we defined unitless prime-variables, $y^{\prime} = y/L_{e}$ and $x^{\prime} = x/L_{e}$.  The lines $y^{\prime} = \pm 1$ represent the edges of the CS. The magnitude of the external field is maximum at the center of the CS and decreases toward the edges, as does the local Alfv\'{e}n speed.

Only field lines in the vicinity of the plane of the CS that intersect the resistive patch will be reconnected, forming two opposite thin flux tubes. In Figure \ref{fig:CS_lines}, spherical resistive patches are represented by their circular projections in the plane of the CS. Reconnected flux tubes have a finite thickness, as also does the CS. The tube's cross-sectional area at the reconnection region, $A$, is determined by the radius of the reconnection patch. This area and the magnitude of the magnetic field at the reconnection site determine the flux of the tube, $\phi = B A$. 

The actual size of the reconnection patches in the solar corona has not been determined, but if SADs represent the cross-sectional area of reconnected flux tubes, this value can be estimated to be only a few megameters in radius, as shown in Figure \ref{fig:TRACE_SADs}. \citet{McKenzie_2009} reported dark voids sizes of approximately $10^{7}$  km\textsuperscript{2}, with significant range for variation.  The areas of the dark pockets can only give an estimation because they may change as tubes move, due to pressure balance with the background. Typical flare coronal loops extend for a hundred megameters, therefore it is reasonable to assume that the reconnected tubes are thin compared with their length.

Field lines in front of the CS that intersect a given reconnection region will connect at this site with field lines on the other side of the CS. Figure \ref{fig:CS_lines} shows a sample of three pre-reconnection field lines on each side of the CS that intersect a TOP reconnection site located at $y_{R}^{\prime}=3/4$. We will assume that the pre-reconnection flux tubes are parameterized by one central field line (their axes, thicker lines in the graph), that crosses the center of the reconnection region. In this figure, only pre-reconnection representative field lines are drawn for the CENTER case. 

The half angle $\zeta_{R}$ between the pre-reconnection representative field lines at a given reconnection location is a parameter in our model and can be calculated as
%
\begin{eqnarray}
   \label{eqn:B_rec_angle}
     \tan{\left(\zeta_{R}\right)} = \frac{B_{ey}}{B_{ex}} \sqrt{1-y_{R}^{\prime 2}} = \tan{\left(\zeta_{0}\right)} \sqrt{1-y_{R}^{\prime 2}},
\end{eqnarray}
%
where $\zeta_{0}$ represents the half angle that field lines make at the center of the CS. Both angles are shown in Figure \ref{fig:CS_lines}. 

Any pre-reconnection field line near the CS can be described as 
%
\begin{eqnarray}
   \label{eqn:initial_tube}
      y^{\prime}(x^{\prime}) &=& \pm \sin \left[ \tan{\left(\zeta_{0}\right)} x^{\prime} + c \right], \hbox{  } z \gtrless 0,
\end{eqnarray}
%
with $c$ being a constant that can be determined by any point that intersects the curve. For field lines that cross the reconnection point $(0,y_{R}^{\prime})$, $c = \pm \arcsin(y_{R}^{\prime})$, where the top sign corresponds to field lines in front of the CS ($z > 0$, thick solid lines in Figure \ref{fig:CS_lines}), and the bottom sign corresponds to field lines at the back of the CS ($z < 0$, thick dotted lines in Figure \ref{fig:CS_lines}). 

The $x$-position at which a tube (its representative field line) is the closest to an edge of the CS can be found replacing $y^{\prime} = \pm 1$ in Equation \eqref{eqn:initial_tube} (the positive sign corresponds to the top edge of the CS, and the negative sign corresponds to bottom edge of the CS), to obtain
%
\begin{eqnarray}
   \label{eqn:end_points}
      x_{\mathrm{top}}^{\prime} &=& \pm \left[\frac{\frac{\pi}{2} - \arcsin \left({y_{R}^{\prime}} \right)}{\tan{\left(\zeta_{0}\right)}} \right] , \hbox{  } z \gtrless 0, \nonumber \\
      x_{\mathrm{bottom}}^{\prime} &=& \mp \left[\frac{\frac{\pi}{2} + \arcsin \left({y_{R}^{\prime}} \right)}{\tan{\left(\zeta_{0}\right)}} \right] , \hbox{  } z \gtrless 0.
\end{eqnarray}
%
At these locations, the tubes are tangent to the edge of the CS since the only remaining component of the magnetic field is the guide field (Equation \eqref{eqn:B_syrovatskii}). The tubes have legs that continue downward to the solar surface and these sections of the tubes do not generally lie in the same plane as the CS (see Figure \ref{fig:flare_CS}). 

After reconnection, the reconnected flux tubes are symmetric in the $x$-direction, but the initial $2.5$-dimensional symmetry of the system is lost. The localization of the reconnection episode makes the problem three dimensional. Fortunately, the reconnected tubes are mostly two dimensional as they lie in the plane of the CS, and their retraction is along this plane (magnetic tension force is parallel to the plane of the CS). In Figure \ref{fig:CS_lines}, representative lines for reconnected flux tubes in a BOTTOM reconnection site are shown. The dashed line represents the initial configuration for a reconnected flux tube whose left side was part of a tube on the back side of the CS and whose right side was part of a tube in the front side of the CS. This new tube is sharply bent near the reconnection site and will move upward as a transient feature that slides along the CS, between the flux layers. We call these tubes UPWARD moving tubes. The mixed dotted and dashed line in the figure is the initial configuration of the DOWNWARD moving reconnected tube that will retract in the negative $y$-direction. The directions of motion for each tube are indicated in the figure by arrows pointing outward from the BOTTOM reconnection site.  The UPWARD and DOWNWARD moving tube definitions are valid for all reconnection cases. Any reconnected UPWARD moving tube is a mirror image of a DOWNWARD moving tube if both tubes have the same absolute value $y_{R}^{\prime}$ reconnection location, but with opposite sign. They will move in opposite directions, but their dynamics and shapes are equal. One difference between them is that the downward-moving ones are connected to a large mass reservoir (the sun's surface), whereas the upward-moving ones are not. Since the tubes are assumed to be initially in equilibrium with their surroundings, this is only relevant if the tube footpoints are somehow perturbed and then interact with their boundaries. Our present analysis and simulations stop before this occurs. Therefore, in this paper we will only focus on DOWNWARD moving tubes.

The shape of a reconnected tube changes with time mainly as a result of magnetic tension. This motion can be characterized by its representative field line arc length parameterization, $\mathbf{R}(l,t)$. At each point of the tube, the local tangent unit vector and curvature vector can be calculated from the parameterization as $\widehat{\mathbf{b}} = \partial \mathbf{R} /\partial l$ and $\mathbf{k}= \partial \widehat{\mathbf{b}} /\partial l$, respectively. 
Figure \ref{fig:CS_lines} shows an example of the parameterization vector for an initial UPWARD moving tube, at a generic arc length $l$. Also, at this point of the curve, $\widehat{\mathbf{b}}$ and $\mathbf{k}$ are shown. Pieces of the tube like this one, that are far enough from the reconnection site, are initially in equilibrium since they were part of the original background configuration. There, the force due to the curvature of the tube is balanced by the force due to the external gradient of the magnetic field. 

\clearpage

%
%
%
\section{Low Plasma-$\beta$ Thin Flux Tube Equations}
  \label{sec:TFT_Syr}

Reconnection is assumed to happen in a small and short-lived region, therefore the reconnected tubes are thin and their evolution can be described by the TFT equations from the previous chapter \citep{Guidoni_2010}. These equations assume total pressure balance between the tubes and their surrounding plasma, and the tubes are assumed to be isolated from the background (no viscous momentum and heat exchange). The plasma in these tubes satisfies the frozen-in field condition since the resistivity that may have originated the reconnection episode was only non-zero at the reconnection region. 

In the solar corona, the plasma is strongly magnetized, ergo its plasma-$\beta$ is small. For this reason, thermal pressure can be neglected with respect to magnetic pressure. This approximation and the assumption of total pressure balance constrain the magnitude of the magnetic field inside the tube; at all times, it is equal to the magnitude of the external magnetic field $B_{e}$ \citep{Guidoni_2010} and can be calculated from Equation \eqref{eqn:B_syrovatskii}. 

The cross-sectional area of each element of the tube changes as the element moves between the flux layers of the CS, satisfying $A = \frac{\phi}{B_{e}}$. The density of each tube element of mass $\delta m$ varies according to $\rho = \delta m B_{e} / \phi \delta l$, where $\delta l$ is the length of the element. 

The magnetic field direction is completely described by the positions of the tube elements since the frozen-in condition guarantees that fluid particles and field lines move together. These positions can be obtained integrating the velocity satisfying the low plasma-$\beta$ TFT momentum equation \citep{Guidoni_2010}
%
\begin{eqnarray}
   \label{eqn:TFT_mom} 
      \rho \frac{D\mathbf{v}}{Dt} & = &  -\widehat{\mathbf{b}} \frac{\partial P}{\partial l}-\nabla_{\perp}
      \left(\frac{B_{e}^{2}}{8 \pi}\right ) + \mathbf{k}\left(\frac{B_{e}^{2}}{4 \pi}\right) +  B_{e} \frac{\partial}{\partial l} \left[ \frac{\widehat{\mathbf{b}}\eta}{B_{e}}\left(\widehat{\mathbf{b}}\cdot \frac{\partial \mathbf{v}}{\partial l}\right)\right]. 
\end{eqnarray}
%
%
\noindent Here, $\rho$, $\mathbf{v}$, $P$, and $\eta$ are the mass density, velocity, plasma pressure, and viscosity of the fluid inside the tube, respectively. $D/Dt$ is the advective derivative in the tube element's reference frame and $\nabla_{\perp}$ represents the gradient in the direction perpendicular to the flux tube. 

The inviscid part of the above equation has a parallel-to-the-magnetic-field term with a derivative of the thermal pressure that cannot be neglected with respect to any magnetic pressure term since magnetic forces act only in the perpendicular direction (as shown in the second and third terms of the right-hand side). The viscosity term has components in both directions but is usually small since viscous Reynolds numbers in the corona are very large. The exception to this rule are shocks, where velocity gradients become significant. For strongly magnetized plasmas, the viscosity depends on temperature as $\eta = \eta_{c}T^{\frac{5}{2}}$ with $\eta_{c}$ being a constant. 

The contribution of thermal conductivity to the change in entropy of the plasma is approximately two orders of magnitude larger than the viscosity. The change in entropy of the plasma (assumed to be an ideal gas) of any given tube element can be expressed as $\Delta s = \ln \left( {P_{c}/P_{e}} \right)$, where $P_{c}$ is related to the tube element's pressure and density as 
%
\begin{eqnarray}
    \label{eqn:TFT_Pc}
      P(t) & = & P_{c}(t) \left(\frac{\rho(t)}{\rho_{e}} \right)^{\gamma}. 
\end{eqnarray}
Here, $\gamma$ is the adiabatic gas constant.
%
%

The TFT entropy equation can be transformed into an equation for $P_{c}$, as follows \citep{Guidoni_2010}
%
%
\begin{eqnarray}
   \label{eqn:TFT_Pc_time}
      \frac{D P_{c}}{Dt} & = & (\gamma -1)  \left( \frac{\rho_{e}}{\rho } \right)^{\gamma} 
      \left[ \eta \left(\widehat{\mathbf{b}} \cdot \frac{\partial \mathbf{v}}{\partial l} 
      \right)^{2} + B_{e} \frac{\partial}{\partial l} \left( \frac{\kappa}{B_{e}}   \frac{\partial}{\partial l}  (k_{B} T) 
      \right) \right].
\end{eqnarray}
%
The temperature dependence of the thermal conductivity is the same as for the viscosity, $\kappa = \kappa_{c}T^{\frac{5}{2}}$, with $\kappa_{c}$ being a constant. The ideal gas constitutive relation is $P = \rho k_{B} T/ \overline{m} = 2 n k_{B} T$, with $k_{B}$ being the Boltzmann constant, $\overline{m}$ the average particle mass, and $n$ the electron number density.

The TFT mass conservation has the following expression \citep{Guidoni_2010} 
%
%
\begin{eqnarray}
   \label{eqn:TFT_mass} 
      \frac{D}{Dt}\left(\frac{B_{e}}{\rho}\right) & = & \frac{B_{e}}{\rho} \widehat{\mathbf{b}} \cdot \frac{\partial \mathbf{v}}{\partial l}.
\end{eqnarray}
%

One of our goals is to explain how reconnected flux tubes may become regions of density depletion, like in the dark voids mentioned in the introduction. To that end, the above equation can be re-arranged to show that the density of each tube element may change due to three different effects, 
%
%
\begin{eqnarray}
   \label{eqn:density change} 
      \frac{D}{Dt}\ln \left(\rho \right) & = & -\frac{D}{Dt}\ln \left( A \right) - \frac{\partial v_{\parallel}}{\partial l} + \mathbf{v}_{\perp} \cdot \mathbf{k},
\end{eqnarray}
%
where $v_{\parallel}$ is the parallel component of the velocity and $ \mathbf{v}_{\perp} $ is the perpendicular one. 

The first term in the right-hand side corresponds to a change in area of the tube as the tube element moves to regions of different confining magnetic field magnitude. In addition, if neighboring elements are slowing down ahead of the direction of motion, a pile-up occurs and the density of the element increases (its length gets reduced, and its area is fixed by the external field). The opposite effect occurs if the neighboring elements are slowing down upstream. The second term in the right-hand side reflects this effect, which is similar to what happens in traffic jams. The third term in the right-hand side is related to the expansion or contraction of a curved tube in its perpendicular direction. One example of this effect is a circular tube with fixed area expanding or contracting radially. In this case, the curvature vector always points toward the center of the tube, and the perpendicular velocity is parallel (contraction) or antiparallel (expansion) to this vector. Each tube element expands or contracts resulting in a change in density. We will show the relevance of some of these terms in Section \ref{sec:parall_dyn}.


\subsection{Joined Equilibrium Solution}
  \label{sec:JES}

A simpler magnetic background configuration than the one assumed in the present chapter was studied by \citet{Guidoni_2010}. There, the background field is skewed and uniform, and the shape of the reconnected TFTs, as they evolve, can be described by a joined equilibrium solution (JES). We will call this uniform background model \textit{Guidoni10}. In this case, the reconnected tubes, at any instant, are composed by three straight segments joined at two corners (the bends) that move at the Alfv\'{e}n speed along the initial tube. There, the magnetic field is rotated without changing its magnitude (RD). Each straight segment of the tube is an equilibrium solution of the inviscid perpendicular part of the momentum TFT equation, but the joined solution is not in equilibrium due to the sharp angle at the bends. The two lateral segments are at rest and the central segment moves at the projection of the Alfv\'{e}n speed in the $y$-direction. Since the magnetic field magnitude and the initial density are uniform, the Alfv\'{e}n speed is the same everywhere. 

The actual parallel dynamics in Guidoni10 does not satisfy an equilibrium solution (constant pressure and temperature) since strong GDSs develop inside the tubes. Nevertheless, outside the shocks, the density and temperature are approximately constant. To completely satisfy an equilibrium solution of the general TFT equations, the change in velocity should satisfy $\widehat{\mathbf{b}} \cdot \frac{\partial \mathbf{v}}{\partial l}  =  0$; a condition that is not satisfied at the shocks. 

For our current non-uniform magnetic configuration (Equation \eqref{eqn:B_syrovatskii}), it is also possible to find an analytical solution for the inviscid perpendicular part of the momentum TFT Equation \eqref{eqn:TFT_mom}, where each mass element is in equilibrium ($ \frac{D }{Dt} = 0 $). In this case, the analytical solution differs quantitatively with the actual tube dynamics, as we will see in the next section, but the general aspects of the solution are maintained.  

It is straightforward to see that a tube satisfying the following equation constitutes such an equilibrium solution
%
%
\begin{eqnarray}
   \label{eqn:Equil_B} 
      \mathbf{k} & = & \frac{\partial^{2} \mathbf{R}} {\partial l^{2}} = \frac{1}{2} \nabla_{\perp} \ln{\left( B_{e}^{2} \right)} . 
\end{eqnarray}
%
A possible general equilibrium solution of the TFT equations is given by a tube having uniform density and temperature that satisfies the above equation, as well as $\widehat{\mathbf{b}} \cdot \frac{\partial \mathbf{v}}{\partial l}  =  0$. In this section, we will only focus on the shape of the tube given by Equation \eqref{eqn:Equil_B}, as the parallel dynamics is not expected to be in equilibrium. 

Since the reconnected tubes lie in the plane of the CS, the derivative with respect to the arc length of the tube can be expressed as $\partial/ \partial l = (1 / \sqrt{1+ \dot x^{\prime 2}}) \partial/ \partial y^{\prime}$, where $\dot x^{\prime} = \partial x^{\prime}  / \partial y^{\prime}$. With this expression, after some algebra, Equation \eqref{eqn:Equil_B} becomes 
%
\begin{eqnarray}
   \label{eqn:Equil_der} 
      \frac{\partial}{\partial l} \ln{B_{e}^2} & = & \frac{- 2 \ddot x^{\prime}}{\dot x^{\prime} (1 + \dot x^{\prime 2})} = \frac{\partial}{\partial y^{\prime}} \ln{\left(\frac{1 + \dot x^{\prime 2}}{\dot x^{\prime 2}}\right)}.
\end{eqnarray}
%
The initial background configuration satisfies the general MHD equation's equilibrium. Its field lines (Equation \eqref{eqn:initial_tube}) also satisfy the above equation. Therefore, pre-reconnection tubes are in equilibrium with respect to the general TFT equations. Immediately after reconnection, every tube element is also in equilibrium, except the central element that is sharply bent at the reconnection site.

The reconnected flux tubes are symmetric in the $x$-direction and smooth at their center. A solution to Equation \eqref{eqn:Equil_der} that satisfies these conditions can be expressed as
%
\begin{eqnarray}
   \label{eqn:Equil_sol} 
      y^{\prime}(x^{\prime}) & = & y_{c}^{\prime} \cos{\left[\frac{\tan{\left(\zeta_{0}\right)} x^{\prime}}  {  \sqrt{   1 + \left(1 - y_{c}^{\prime 2}\right)\tan^{2}{\left (\zeta_{0} \right)} } } \right]}.
\end{eqnarray}
%
This solution is a one-parameter family of curves whose parameter is the $y$-position of the center of the tube, $y^{\prime}_{c}$.

\begin{figure}
  \centering
   \resizebox{5.0in}{!}{\includegraphics{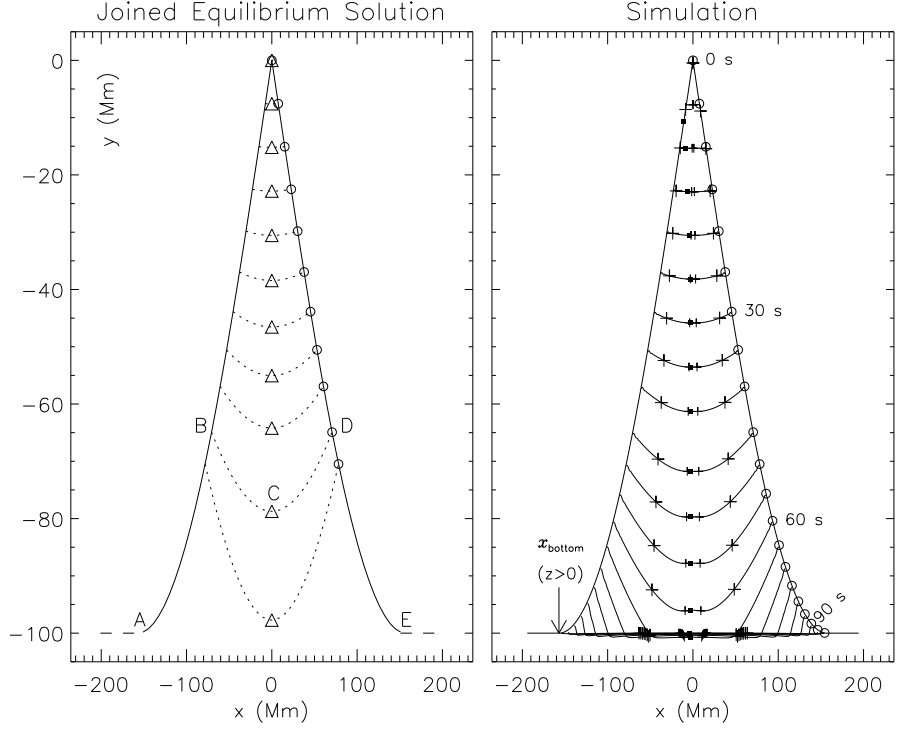}}
   \caption[DOWNWARD moving tube for the CENTER reconnection case]
   {DOWNWARD moving tube for the CENTER reconnection case and reconnection angle $\zeta_{R}=\zeta_{0}=45$\degree\ (note that $y$-axis and $x$-axis are not drawn to scale). Both panels share the same $y$-axis. Left panel: solid line represents the shape of the reconnected tube at $t = 0$. Each dotted line corresponds to a different JES with center position parameter $y_{c}^{\prime} = 0.0,-0.08,-0.15,-0.23,-0.30,-0.38,-0.47,-0.55,-0.64,-0.79,$ and $-0.98$ (shown as triangles), respectively. Circles indicate the bends for the right side ($z < 0$) of the tube. Letters ``ABCDE'' join a complete JES curve for $y_{c}^{\prime} = -0.79$. The horizontal dashed lines show extensions of the tube that simulate the legs that connect it to its footpoints at the photosphere. Right panel: DEFT simulation of the same initial reconnected tube as in left panel. Each central curved line represents the shape of the tube at a different time. The time interval between each curve is approximately 5 s. Circles represent the theoretical positions of the bends (assumed to move at the local Alfv\'{e}n speed). Time in seconds is shown for some selected tube configurations to guide the eye. Large plus signs at the center of the tube indicate the beginning of the gas-dynamic shocks (heat front) and the smaller plus signs indicate the end of the shock (sub-shock). The solid black squares indicate the positions of a tube element as it moves along the tube. The arrow points to the location where the tube becomes tangent to the edge of the current sheet ($x_{\mathrm{bottom}}$) for the front side of the tube ($z > 0$). To the left of this point, the tube is extended as a straight line to simulate the leg that connects it to the photosphere (it corresponds to the horizontal dashed line to the left in the left panel). A similar straight line is added to the other end of the tube}  
   \label{fig:CENTER_JES_sim} 
\end{figure}

It is possible to construct a JES where the shape of the reconnected tubes is composed of three different equilibrium regions: two unperturbed equilibrium side sections given by Equation \eqref{eqn:initial_tube} and a central curve described by the above equation. The left panel of Figure \ref{fig:CENTER_JES_sim} shows such solutions for a CENTER reconnection case with half reconnection angle $\zeta_{R}=\zeta_{0}=45$\degree and DOWNWARD moving tube. Each dotted line corresponds to a different parameter (position of the center of the tube, represented in the figure by triangles). The unperturbed side sections are part of the initial reconnected tube (solid lines in the figure). For example, a complete JES for the parameter $y^{\prime}_{c}$ labeled ``C'' in the figure corresponds to two side sections of the initial tube (sections ``AB'' and ``DE'') plus a central symmetric curve (``BCD''). For a given center position, the intersection between the dotted curve and the solid line (the bends, shown as circles in the figure) can be obtained setting Equations \eqref{eqn:Equil_sol} and \eqref{eqn:initial_tube} equal to each other. 

%
%
%

\begin{figure}
  \centering
   \resizebox{5.5in}{!}{\includegraphics{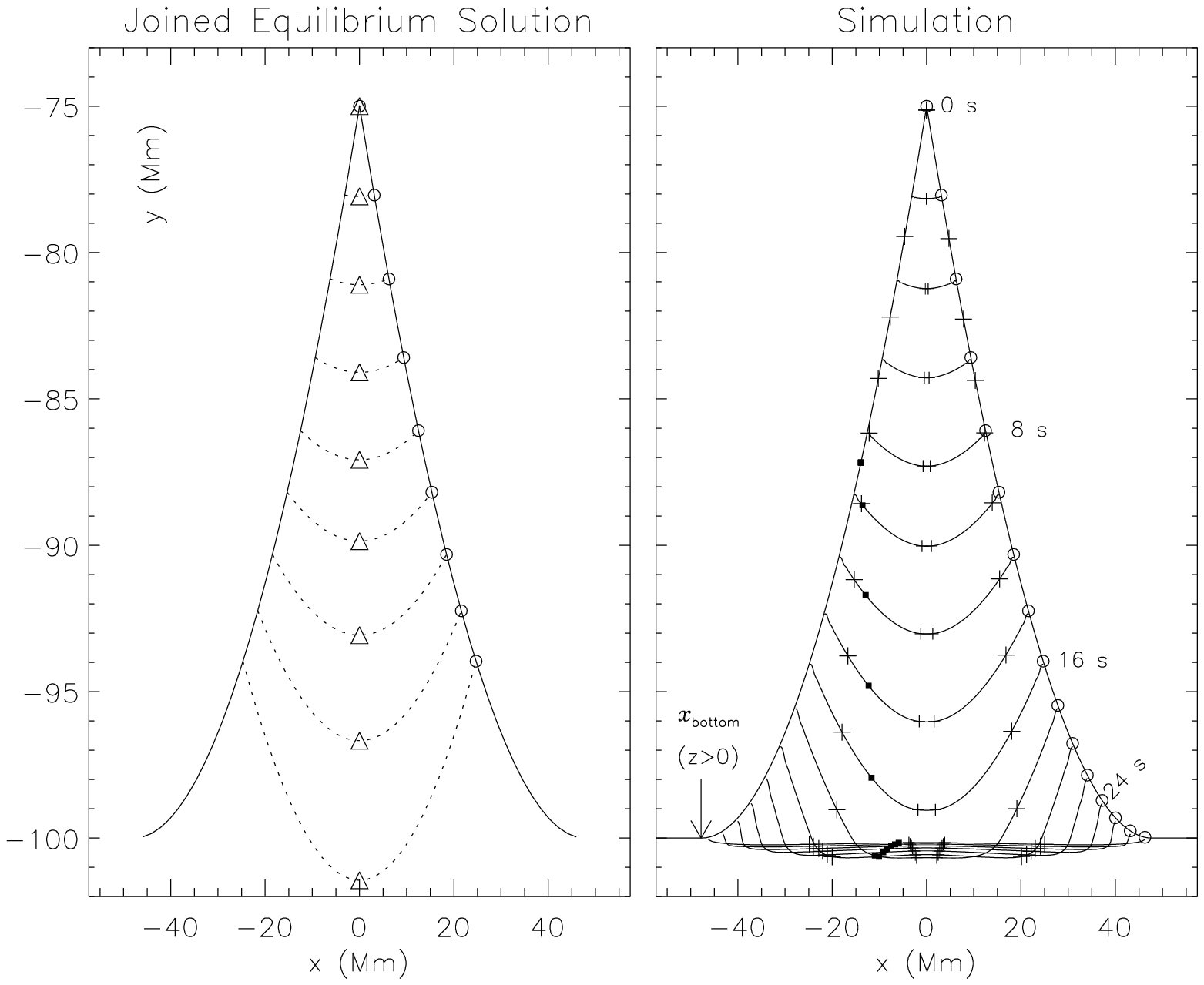}}
   \caption[DOWNWARD moving tube for the BOTTOM reconnection case]
   {DOWNWARD moving tube for the BOTTOM reconnection case and reconnection angle $\zeta_{R}=45$\degree\ and $\zeta_{0} \simeq 57$\degree, with the same format as Figure \ref{fig:CENTER_JES_sim}. Left panel: the center position parameters are $y_{c}\prime = -0.75,-0.78,-0.81,-0.84,-0.87,-0.90,-0.93,-0.97,$ and $-1.01$, respectively. Right panel: the time interval between each curve is approximately 2 s}
   \label{fig:BOTTOM_JES_sim} 
\end{figure}

\begin{figure}
  \centering
   \resizebox{5.5in}{!}{\includegraphics{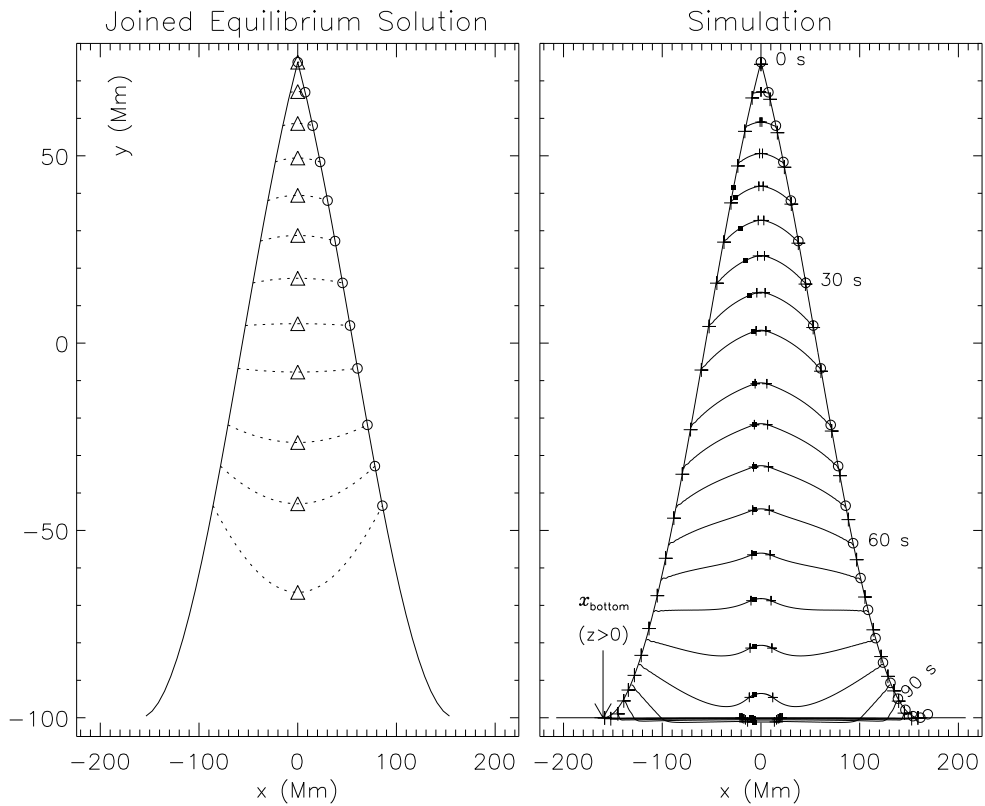}}
   \caption[DOWNWARD moving tube for the TOP reconnection case]
   {DOWNWARD moving tube for the TOP reconnection case and reconnection angle $\zeta_{R}=45$\degree\ and $\zeta_{0} \simeq 57$\degree, with the same format as Figure \ref{fig:CENTER_JES_sim}. Left panel: the center position parameters are $y_{c}^{\prime} = 0.75,0.67,0.59,0.50,0.40,0.29,0.17,0.05,-0.08,-0.27,-0.43,$, and $-0.67$, respectively. Right panel: the time interval between each curve is approximately 5 s}
   \label{fig:TOP_JES_sim} 
\end{figure}

The whole set of these JESs describes a reconnected tube that descends from the reconnection region. At each time, the tube is shorter than the original tube. Even though each section of the joined solution is an equilibrium solution, the joined solution is not. The bends are clearly out of equilibrium, therefore a complete curve like ``ABCDE'' does not satisfy equilibrium Equation \eqref{eqn:Equil_der}. Left panels of Figures \ref{fig:BOTTOM_JES_sim} and \ref{fig:TOP_JES_sim} show similar joined equilibrium solutions for BOTTOM and TOP reconnection cases, respectively. 
%
%
%

%
%
%

For comparison purposes, all cases analyzed in the present section have a half reconnection angle $\zeta_{R}=45$\degree\ (note that this is different for the CENTER reconnection case shown in Figure \ref{fig:CS_lines} that has $\zeta_{R} \simeq 57$\degree; this choice visually simplified the graph making pre-reconnection field lines parallel to each other on each side of the CS, for all the reconnection cases). From now on, CENTER reconnection case implies $\zeta_{R}=45$\degree. 

In the CENTER and BOTTOM reconnection cases, the concavity for the JESs is upward for all the central regions. On the other hand, for the TOP reconnection case, the concavity of the central regions changes from downward to upward when the center position changes from positive to negative (at $y^{\prime}_{c} = 0$, the equilibrium central region corresponds to a straight horizontal line). 

Equilibrium solutions \eqref{eqn:Equil_sol} are valid as long as the position of the center of the tube is above the edge of the CS ($y^{\prime}_{c} > -1$, or $y_{c} > $ \CSHalfLength). This corresponds to the arrival of the center of the tube to the top of the arcade. Since our analysis was restricted to field lines near the plane of the CS, equilibrium solutions beyond this point have no real significance.

%
%
%
\section{Simulations}
  \label{sec:simul}

The low plasma-$\beta$ TFT Equations \eqref{eqn:TFT_mom}, \eqref{eqn:TFT_Pc_time}, and \eqref{eqn:TFT_mass} are non-linear, hence a time-dependent analytical solution of these equations is hardly a possibility. The JES, although far from being a general solution, presents the correct general aspects of the shape of the tube, as we will see in this section. The parallel dynamics are a different matter. The strong compressions of the plasma due to the shortening of the tube generate GDSs along the tubes that present discontinuities in the state variables \citep{Guidoni_2010}. 

To study the full dynamical solution of the TFT equations, we developed a computer program called Dynamical Evolution of Flux Tubes (DEFT) which solves the TFT Equations \eqref{eqn:TFT_mom} and \eqref{eqn:TFT_Pc_time} in dimensionless form for the two reconnected tubes (DOWNWARD and UPWARD moving). Mass conservation is automatically satisfied since the DEFT program implements a Lagrangian approach where each mass element of the tube is followed. The program uses a one-dimensional staggered mesh where each tube piece is represented by grid points at its ends. A rather detailed description of this computer program was presented in the previous chapter.

In this chapter, we will present three different simulations that correspond to the three cases described in Section \ref{sec:JES}. For all of them, the half reconnection angle is $\zeta_{R}=45$\degree. Results depend quantitatively on this angle, but we chose this intermediate value to show the general aspects of the solutions that are common for a wide range of angles. For these chosen cases, the half angle that field lines make at the center of the CS is $\zeta_{0} \simeq 57$\degree\ for the BOTTOM and TOP cases, and $\zeta_{0} \simeq 45$\degree\ for the CENTER case. 

The central initial part of the tubes are clearly out of equilibrium due to the sharp angle at the reconnection site. To avoid introducing length scales at the limit of resolution, the central part of the initial tube is smoothed. The end points of the tube are assumed to be fixed, and to have no temperature gradient (this last condition ensures no heat transfer from the end points). 

For concreteness, we will present the results from all simulations in units that assume a magnitude of the background magnetic field at the reconnection site of \realB, an initial uniform electron number density of $n_{e} =$ \realnumdens, and initial uniform temperature $T_{e}$ = \realTemp. The Alfv\'{e}n  speed $v_{aeR}$ at the reconnection site is then approximately $2200$ km s\textsuperscript{-1}. As in the above section, the half length of the CS is $L_{e} =$ \realL. We assume $\overline{m}$ equal to half the mass of the proton. Each tube has $1100$ points. 

The unitless numbers for this simulation correspond to plasma-$\beta \simeq $ \realbeta, Prandtl number (ratio between the viscosity and thermal conductivity)  $P_{r} = $ \PrandtlSimII, and viscous Reynolds number $R_{\eta} = L_{e} v_{aeR} \rho_{e} / \eta \simeq $ \realReynoldsII. These values are typical values for the high solar corona. 

The simulated evolution of a DOWNWARD moving tube reconnected at the CENTER site is shown in the right panel of Figure \ref{fig:CENTER_JES_sim}. This simulation has similarities with the corresponding JES (left panel). For each given time, the tube consists of three sections: two unperturbed ones on its sides and a central one that moves downward. Times shown were chosen to coincide with the same bend positions as in the corresponding JES (circles are located in the same positions). The concavity of the central portion has the same sign as the JES, although the JESs are more curved. 

The flattening of the center part of the tube is due to the perpendicular component of the viscous force density in Equation \eqref{eqn:TFT_mom}. In the figure, the small crosses near the center of the tube correspond to the location of sub-shocks, where parallel velocities have large gradients and viscosity becomes relevant. At those locations, the perpendicular viscous force acts in the direction opposite to the local curvature vector; therefore, the total force in the curvature direction is reduced. This increases the curvature of the tube, resulting in a change of curvature sign at the center of the tube. A thorough analysis of the shocks is presented in the next section.

In the simulation, the center of the tube moves with approximately constant speed equal to the Alfv\'{e}n speed related to the $y$-component of the magnetic field at the reconnection site, which is of the order of $1500$ km s\textsuperscript{-1}. This can be understood, at least for earlier times, by comparison with the Guidoni10 model where this is also the case. The initial tube can be approximated as the initial conditions for skewed uniform magnetic fields (same as in Guidoni10). The magnitude of the local Alfv\'{e}n speed decreases toward the bottom edge of the CS, therefore there are sections of the tube, like the center one, whose perpendicular speed is super-Alfv\'{e}nic. The ratio between some tube elements' perpendicular speed and the magnitude of the local Alfv\'{e}n speed can reach values above $1.1$, although this value is relative since the direction of the background magnetic field is different than the one of the tube. 

In over a minute, the central part of the tube arrives at the lower edge of the CS which marks the beginning of the top of the arcade, located at $y=-100$ Mm. The evolution of the tube after its arrival at the top of the arcade is discussed in detail in Section \ref{sec:Top_arcade}.
 
In the Guidoni10 model, the bends are RDs that move at the Alfv\'{e}n speed. We expect similar behavior in our present case. The positions of the circles in the right panel of Figure \ref{fig:CENTER_JES_sim} were theoretically calculated as moving at the local Alfv\'{e}n speed along the tube. They coincide with the actual bend positions in the simulation. They continuously slow down in the $y$-direction as they approach the edge of the CS. Their speed in the $x$-direction, determined by the guide field, is constant. Being faster than the bends in the $y$-direction, the tube's center is the first to arrive at the top of the arcade. This difference in speed is what determines the general upward concavity of the moving part the tube. 

At the bends, the initially stationary plasma is deflected along the bisector of the angle between the two adjacent tube portions (average direction of the curvature force). The resulting speed is Alfv\'{e}nic and has a parallel component toward the center of the tube. These bends, as in Guidoni10, are RDs where only the magnetic field direction and the plasma velocity are changed. Notably, the angle at the bends remains approximately constant during the entire evolution of the tube, and equal to $180 - \zeta_{R}$ degrees (same as in the uniform case).  

The BOTTOM reconnection case simulation (shown in left panel of Figure \ref{fig:BOTTOM_JES_sim}) presents the same characteristics as the CENTER one, but the time it takes for the tube to arrive at the top of the arcade is much shorter ($\sim 16$ s), as the reconnection site is closer to the apex of the arcade.

On the other hand, the TOP reconnection case (right panel, Figure \ref{fig:TOP_JES_sim}) presents some differences. The general concavity sign of the tube changes with time, and this change occurs much later in the evolution of the tube compared to the corresponding JES (left panel). Initially, the bends accelerate in the vertical direction (the local Alfv\'{e}n speed increases toward the center of the CS) and move faster than the center of the tube, leading to a downward concavity (same as in JES). After the bends reach the center of the CS, they start decelerating as they move toward the top of the arcade. The tube's center accelerates continuously, catching up with the bends after approximately $70$ s, where the tube is almost flat. Meanwhile, the perpendicular viscosity force density increases the curvature near the locations of the shocks, curving the center part in the opposite direction with respect to the general curvature of the tube. For this TOP case, the center of the tube is not the first part to arrive at the top of the arcade. The ratio between some tube elements' perpendicular speed and the magnitude of the local Alfv\'{e}n speed reaches values above $1.7$.

%
%
%
\subsection{Parallel Dynamics}
  \label{sec:parall_dyn}

The parallel dynamics along the reconnected tubes presents complex features such as shocks and coupled nonlinear waves. There is no simple equilibrium counterpart to compare it with, as there was for the perpendicular motion. Nevertheless, its complexity is what makes the parallel dynamics very rich in behavior, and this section is devoted to its description. 

The bends accelerate plasma supersonically toward the center of the tube where they collide and are stopped abruptly. As a result, two strong GDSs are launched outward. \citet{Guidoni_2010} describe in detail their inner structure for the uniform and skewed magnetic field configuration. It consists of one thermal front, where most of the heating occurs, and an isothermal sub-shock, where most of the compression occurs. The thickness of the GDSs increases with time and can be of order of tens of megameters, which is comparable to the size of the entire reconnected tube and larger than the particle's mean free path. The large and small plus signs in the right panels of Figures \ref{fig:CENTER_JES_sim}--\ref{fig:TOP_JES_sim} represent the beginning (heat front) and the end (sub-shock) of the GDSs, respectively, for each simulation time. 

For the three reconnection cases presented in this chapter, the GDSs are ahead of the bends at early times after reconnection. Initially, the shocks develop their inner structure quite rapidly until they achieve a near-equilibrium thickness. Meanwhile, the central part of the tube is increasing in length due to the difference in vertical speed between the center and the bends. Therefore, the central portion eventually becomes larger than the distance covered by the shocks, and the bends move ahead. For the TOP reconnection case, this situation is reversed after the bends move to the lower half of the CS, where they slow down and the shocks overrun the bends again. 

\begin{figure}
  \centering
   \resizebox{5.9in}{!}{\includegraphics{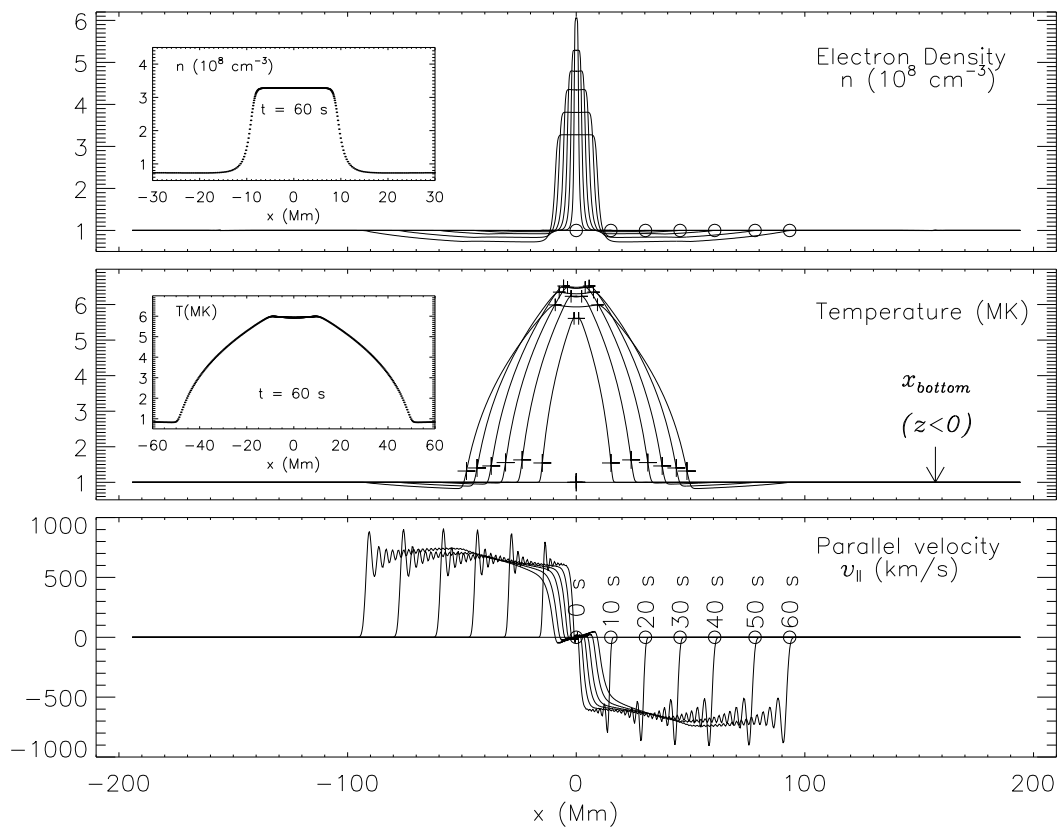}}
   \caption[DOWNWARD moving tube for the CENTER reconnection case]
   {DOWNWARD moving tube for the CENTER reconnection case. All panels share the same horizontal axis. Top panel: electron density along the tube for different times in the simulation. The $x$-positions of the bends (only for the right part of the tube) are shown as circles for each chosen time. Lower densities in the central part of the tube correspond to later times in the simulation. The inset figure repeats the electron density profile for the time $t=60$ s, where each point of the curve corresponds to a grid point of the tube, to show that the sub-shocks are well resolved by the DEFT computer program. Middle panel: temperature along the tube for the same times as in the top panel. Large and small crosses indicate the beginning (heat front) and end (sub-shock) of the gas-dynamic shocks, respectively. The inset figure repeats the temperature profile for the time $t=60$ s, where each point of the curve corresponds to a grid point of the tube, to show that thermal fronts are well resolved by the DEFT computer program. The arrow points to the location where the back side of the tube ($z < 0$) becomes tangent to the edge of the current sheet ($x_{\mathrm{bottom}}$). Bottom panel: parallel (along the tube) speed profile for the same times as in the top panel. Circles are also the same as in the top panel. Times in seconds are shown for each bend position}
   \label{fig:Center_variables_sim} 
\end{figure}

The electron density has a sharp increase at the sub-shocks where plasma is extremely compressed. Here, the second term of the right-hand side of Equation \eqref{eqn:density change} is the most relevant to determine the electron density. The top panel of Figure \ref{fig:Center_variables_sim} shows density profiles for the CENTER reconnection case where each solid line corresponds to a different time in the simulation. At early times, the density at the center of the tube is over six times the background density. Even though the theoretical density jump limit for steady-state GDSs is $4$, we achieve higher values in our simulations because there is an initial transient overshoot of density at the center of the tube. The plasma requires a finite time to adjust to the correct steady-state shock strength. Density decreases as times goes by, although never reaching the equilibrium value. In the figure, we include an inset to the left that reproduces the $t=60$ s electron density profile, where each point of the graph corresponds to a grid point of the tube. This shows that shocks are well resolved by the DEFT program. 

%
%
%

Density enhancements by shocks are a common feature in any standard reconnection model; this compression occurs at slow-mode shocks or switch-off shocks, which are close relatives of the gas-dynamic shocks of our paper \citep{Longcope_2010_I}. Therefore, dense outflows are usually expected from reconnection. In our patchy reconnection case, central long-lasting, high-density post-shock regions, or plugs, expand and move outwardly from the reconnection site. Their increase in size can be seen in the right panels of Figures \ref{fig:CENTER_JES_sim}--\ref{fig:TOP_JES_sim} as tubes descend. The section of the tubes limited by the small crosses corresponds to the plug. This feature should be observable by an X-ray instrument of sensibility suitable to the emission measure. In fact, bright X-ray features apparently shrinking have been observed \citep{McKenzie_2000, McKenzie_2001}. \citet{Sheeley_2002} also describe some coronal inflows as barely visible density enhancements that move sunward, with dark tails forming behind them (sinking columns). 

Between the bends and the sub-shock, the electron density decreases before increasing again at the sub-shocks. This decrease in density is caused by the lateral expansion of the tube as it moves into regions of lower magnetic field (first term in the right-hand side of Equation \eqref{eqn:density change}), as well as by the expansion of the curved tube in its perpendicular direction, although the latter effect is smaller than the former. In the figure, the positive $x$-positions of the bends are shown as circles for each chosen time. After $60$ s, the plasma density in these regions is less than $70$\%\ that of the background density. If observed, this tube would present dark and elongated thin regions (low emission measure) that get darker, descend toward the arcade, and grow in length as the bends move along the legs. This resembles the sinking columns or the SADs mentioned in the introduction. By the time the tube arrives at the top of the arcade, it would be mostly dark, and approximately half of the dark region would be cold, and the other half would have a range of higher temperatures as the heat fronts move outwardly. 

The GDSs are strong and the temperature jump across them may exceed an order of magnitude. The CENTER reconnection case's temperature profiles are shown in the middle panel of Figure \ref{fig:Center_variables_sim} for the same times as in the top panel. The temperature across the shocks initially increases with time, achieving values higher than $7$ MK, and later decreases. The large and small cross signs are the positions of the beginning and end of the thickness of the shocks, respectively. The length of the thermal fronts increases with time as they move outwardly from the center of the tube. The inset to the left reproduces the $t=60$ s temperature profile, where each point of the graph corresponds to a simulation point in the tube.  

The thermal pressure gradient in Equation \eqref{eqn:TFT_mom} is crucial for the generation of the shocks. If it were erroneously neglected in this equation, the plasma would continuously pile up at the center of the tube without limit. However, this term is not the only factor in the determination of parallel dynamics. The parallel, inviscid part of Equation \eqref{eqn:TFT_mom} can be re-written in the following way 
%
\begin{eqnarray}
   \label{eqn:par_vel} 
      \frac{D v_{\parallel}}{Dt} & = & \mathbf{v}_{\perp} \cdot \frac{D \widehat{\mathbf{b}}}{Dt} - \frac{1}{\rho} \frac{\partial P}{\partial l},
\end{eqnarray}
%
%
showing two possible sources for the change in tube element's parallel velocity (except perhaps inside the sub-shocks where the viscosity contribution may also be large). The first term on the right-hand side is a fictitious force caused by the change in direction of the unit parallel vector as tube elements move. The solid dark squares in Figures \ref{fig:CENTER_JES_sim}--\ref{fig:TOP_JES_sim} show the position of a generic tube element as it moves along the tube. 

The parallel velocity profile for the CENTER reconnection case is shown in the bottom panel of Figure \ref{fig:Center_variables_sim}, for the same times as in the other two panels. The positive $x$-positions of the bends are also shown as circles for each chosen time. There, the plasma at rest is abruptly accelerated to supersonic speeds, therefore the slope of the parallel velocity curve is almost vertical at these points. The same occurs at the bends on the other side of the tube where the inflows are directed to the right (positive speed). Since the angle at the bends remains mostly the same as time goes by, the value of the speed achieved at the bends is also approximately constant. The last time shown in this figure is $60$ s, before any part of the tube has arrived to the top of the arcade. 

The parallel velocity after the bends slightly increases due to a small predominance of the first term in Equation \eqref{eqn:par_vel}. This term also nonlinearly couples the parallel dynamics (sound waves) to the perpendicular one (Alfv\'{e}n waves), which introduces dispersion. The oscillations in parallel velocity that follow the bends are not an artifact of the simulation; they are real dispersive waves due to this coupling. This effect was not present in Guidoni10 since the tube segments were straight at all times. These waves are smoothed out by diffusive processes, mostly related to thermal conduction. 

The following slow parallel deceleration coincides with the thermal front positions, where the second term of Equation \eqref{eqn:par_vel} becomes relevant. There, the inflows are decelerated when tube elements encounter the heat front, until they are suddenly stopped at the sub-shocks (sharp decrease in parallel velocity near the center of the tube). In the post-shock region, the flows are reversed in direction with respect to the post-bend velocity.  

The BOTTOM reconnection case presents the same general features as the CENTER reconnection case for the three quantities plotted in Figure \ref{fig:Center_variables_sim}. These two cases are similar because their dynamics corresponds purely to the lower part of the CS where the background magnetic field decreases monotonously toward the edge of the CS. 

\begin{figure}[ht]
  \centering
   \resizebox{5.5in}{!}{\includegraphics{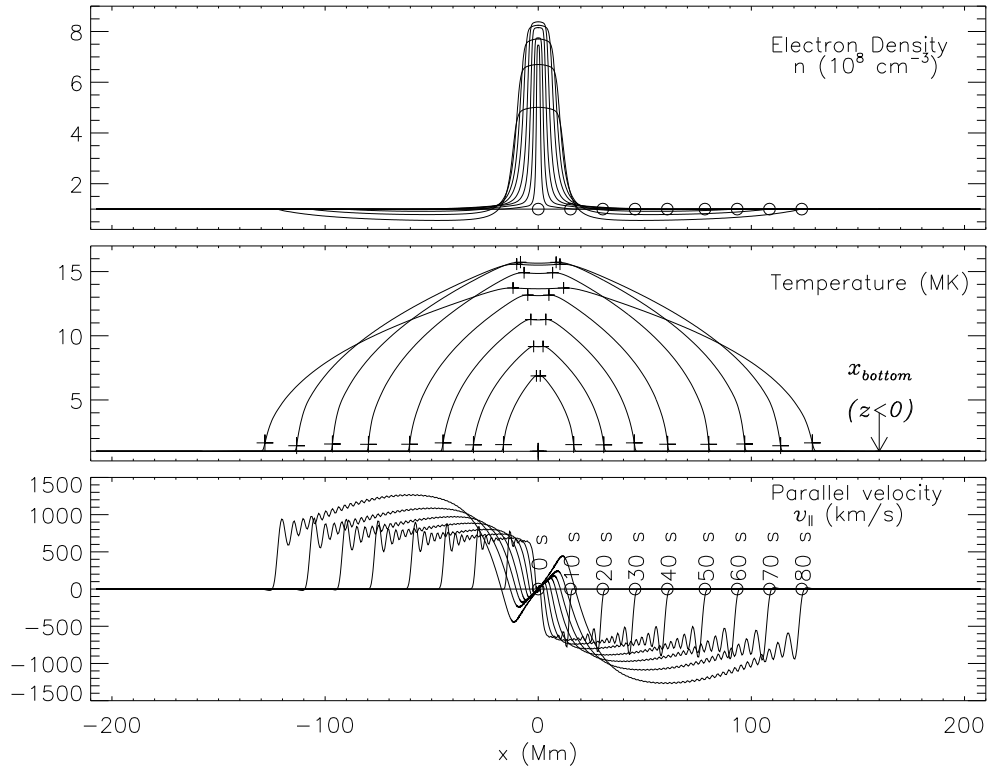}}
   \caption[DOWNWARD moving tube for the TOP reconnection case]
   {DOWNWARD moving tube for the TOP reconnection case, with the same format as Figure \ref{fig:Center_variables_sim}}
   \label{fig:TOP_variables_sim} 
\end{figure}

The TOP reconnection case presents some differences with respect to the other cases. The three panels of Figure \ref{fig:TOP_variables_sim} show the same type of graphs as in Figure \ref{fig:Center_variables_sim}, for the TOP reconnection case. The post-shock electron density (almost one order of magnitude higher than the ambient density) increases while the plug is in the top part of the CS (the tube is being squeezed by the ambient pressure) and decreases after crossing the center of the CS. The squeezing increases the electron density after the bends while those regions are in the top half-plane of the CS, and a pronounced decrease in density occurs after the bends cross the center of the CS. Those regions below the center of the CS expand as the ambient pressure decreases. The decreased density after the bends can be half that of the background density by the time the tube arrives at the arcade. As the tube descends, a very dense and hot central region develops and moves downward toward the top of the arcade, while very dark and extended regions grow in length to its sides. 

%
%
%

The maximum post-shock temperature is high, over $16$ MK. The thermal fronts are almost as long as the entire tube, as shown in the middle panel of the figure. There, the fictitious force is several times larger than the pressure gradient force and has opposite sign. After the bends, plasma is continuously being accelerated toward the center of the tube and stopped at the sub-shock. For a given tube element, it is possible to roughly estimate the value of the fictitious force, for a short period of time, as $\langle \left| \mathbf{v}_{\perp} \right| \rangle \left|\Delta \theta \right|$, where $ \langle \left| \mathbf{v}_{\perp} \right|  \rangle $ represents the average magnitude of the perpendicular velocity in the considered period of time, and $\Delta \theta$ is the corresponding angle change in the parallel unit vector. We have estimated this term for this case, and it is several times larger than the corresponding pressure gradient term in Equation \eqref{eqn:par_vel}. 

The speed achieved at the bends is almost unchanged over the entire simulation as the angle at the bends does not change considerably through the simulation. We have run simulations with different reconnection angles and this angle seems to always remain approximately equal to $180 - \zeta_{R}$ degrees. The post-shock parallel velocities are reversed with respect to the direction of the post-bend velocities, and the resulting speeds are high, almost equal in magnitude with respect to the speeds achieved at the bends.

A zoom near the bends in the right panels of Figures \ref{fig:CENTER_JES_sim}--\ref{fig:TOP_JES_sim} would show small oscillations in the shape of the tubes. There, the fictitious term is much larger than the second term in the right-hand side of Equation \eqref{eqn:par_vel}, as the RDs do not change the pressure of the plasma. 

\subsection{Top of the Arcade}
  \label{sec:Top_arcade}

In the previous sections, the analysis of the evolution of the tubes was done for times when the tubes were above the edge of the CS. The DOWNWARD moving tubes will arrive at the top of the arcade after a finite time. For example, the middle point of the CENTER reconnection tube arrives at the top of the arcade shortly after a minute of travel. This arrival is usually referred in the literature as the place of the fast-mode shock or termination shock \citep{Forbes_1983, Forbes_1986, Forbes_1986_I}. Here, the reconnected flux tubes collide with the underlying arcade. 

\begin{figure}
  \centering
   \resizebox{5.5in}{!}{\includegraphics{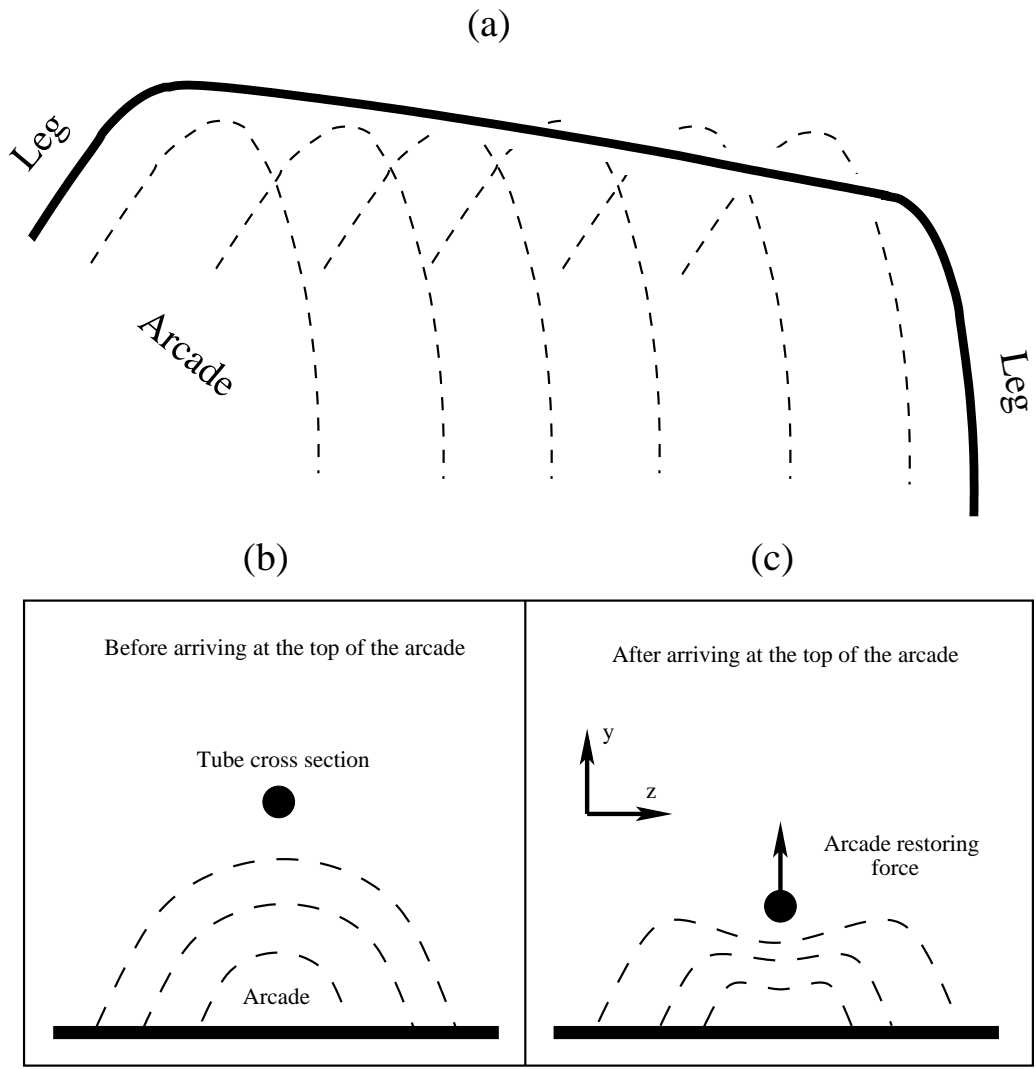}}
   \caption[Tube arrival at the top of the arcade cartoon]
   {Tube arrival at the top of the arcade cartoon. (a) Reconnected tube (dark solid line) lies on top of the arcade after having descended from the reconnection site. Dashed lines represent the underlying arcade. The legs of the tube connect the parts of the tube that were close to the CS with the tube's footpoints at the solar surface. (b) Cross section of the tube (solid black circle) before its arrival at the top of the arcade (dashed parabolic lines). (c) Arcade restoring force. The underlying arcade gets compressed and lightly curved due to the arrival of the reconnected tube, and these stronger field lines exert an upward force on the tube}
   \label{fig:Top_arcade} 
\end{figure}

It is likely that the parallel dynamics will continue evolving along the legs of the tube after the tubes had arrived to the top of the arcade. The legs extend downward to the solar surface and do not lie in the same plane of the CS, therefore the simple two-dimensional analysis we have done so far for the evolution of the tube cannot be applied. An exact analysis of the dynamics of the tube after the tube arrives at the edge of the CS escapes the scope of this paper. Nevertheless, we can make some approximations to study it. The decrease in length of the tubes after they arrive at the top of the arcade is probably much smaller than the decrease they undergo since reconnection, as the underlying arcade prevents them from shortening. By the time the tubes have arrived to the top of the arcade, their total lengths decreased by $7$\%, $15$\%, and $28$ \%\ for the BOTTOM, CENTER, and TOP reconnection cases, respectively. They may also change their length, only by a small fraction, by decreasing the angle between the legs of the tube and its flat central part (panel (a) of Figure \ref{fig:Top_arcade} shows a schematic where a reconnected tube (thick line) is lying on top of the underlying arcade).

%
%
%

To simplify the analysis, we can assume that this length is fixed after the arrival at the arcade apex. The perpendicular motion of the tube is assumed to have stopped, and only the parallel dynamics along the legs would continue, mostly independently of how curved the tube is. To study this parallel dynamics, we will approximate the tube legs as straight line extensions on the sides of the tubes at the edge of the CS and parallel to the CS plane. Therefore, in our simulations an extra straight segment is added to the end of the tubes to simulate the legs; then the tubes are extended in the horizontal axis near the edge of the CS, as shown in the left panel of Figure \ref{fig:CENTER_JES_sim} (dashed lines). 

Most of the observed dark voids slow to a stop when they arrive at the top of the arcade \citep{Sheeley_2004}. \citet{Linton_2009} show through three-dimensional MHD simulations that reconnected flux tubes decelerate rapidly when they hit the Y-lines of a skewed Syrovatski\v{i} CS and the sheared arcade beyond them. Therefore, we chose in this first analysis to ignore the possibility of full ``bounce'' back of the tube toward the reconnection site. We propose a scenario where tubes are slowed down and stopped there by a damping force. We simulate this effect as a restoring force, as shown in panel (c) of Figure \ref{fig:Top_arcade}. The underlying arcade gets compressed and lightly curved due to the arrival of the reconnected tube, and these stronger field lines exert an upward force on the tube. We add a perpendicular damping and spring force densities to Equation \eqref{eqn:TFT_mom}, active only below the edge of the CS
%
\begin{eqnarray}
   \label{eqn:Damp_force} 
      \mathbf{f} & = & \rho \widehat{\mathbf{n}} \left(   \Omega^{2} y - \nu  \widehat{\mathbf{n}} \cdot \mathbf{v} \right) \hbox{   for } y^{\prime}  < -1. 
\end{eqnarray}   
%
Here, $\widehat{\mathbf{n}}$ indicates a unit vector perpendicular to the tube, $\Omega^{2}$ is the spring constant, and $\nu$ is the damping coefficient. This force only acts in the perpendicular direction, slowing down the tube and stopping it. 

The above force density will result in overdamped oscillations when the ratio between the damping coefficient and the spring force, $\xi = \nu / 2  \Omega$, is larger than one. We opted for this in our simulations in Section \ref{sec:simul}. The cases $\xi = 1$ and $\xi < 1$ correspond to critical and underdamped oscillations, respectively. 

Different arcade damping coefficients lead to different damping times, but do not considerably affect the parallel dynamics of the tube. We ran simulations with coefficients differing by orders of magnitude and all results were similar. For the results presented in this paper, the simulations were carried out with coefficients equal to $\xi = 2.0$ and $\Omega \simeq 0.49$ s\textsuperscript{-1}. 

In our simulations, the reconnected tubes retract from the reconnection site to the top of the arcade, and slow down due to the damped spring force from the arcade, until they lie flat on top of it. Then, the final shape of a DOWNWARD moving tube would be a straight line on top of the arcade (see right panels of Figures \ref{fig:CENTER_JES_sim}--\ref{fig:TOP_JES_sim} for late times in the simulation). 

Since the bends decelerate in the lower part of the CS, they are not the first sections of the tube to arrive at the top of the arcade. The first regions to do so become flat and grow in size as more tube elements descend atop the arcade (compare, for example, $t=65$ and $70$ s of the CENTER simulation in Figure \ref{fig:CENTER_JES_sim}). We will call the outer edges of these flat regions ``second bends.'' 

\begin{figure}[t]
  \centering
   \resizebox{5.5in}{!}{\includegraphics{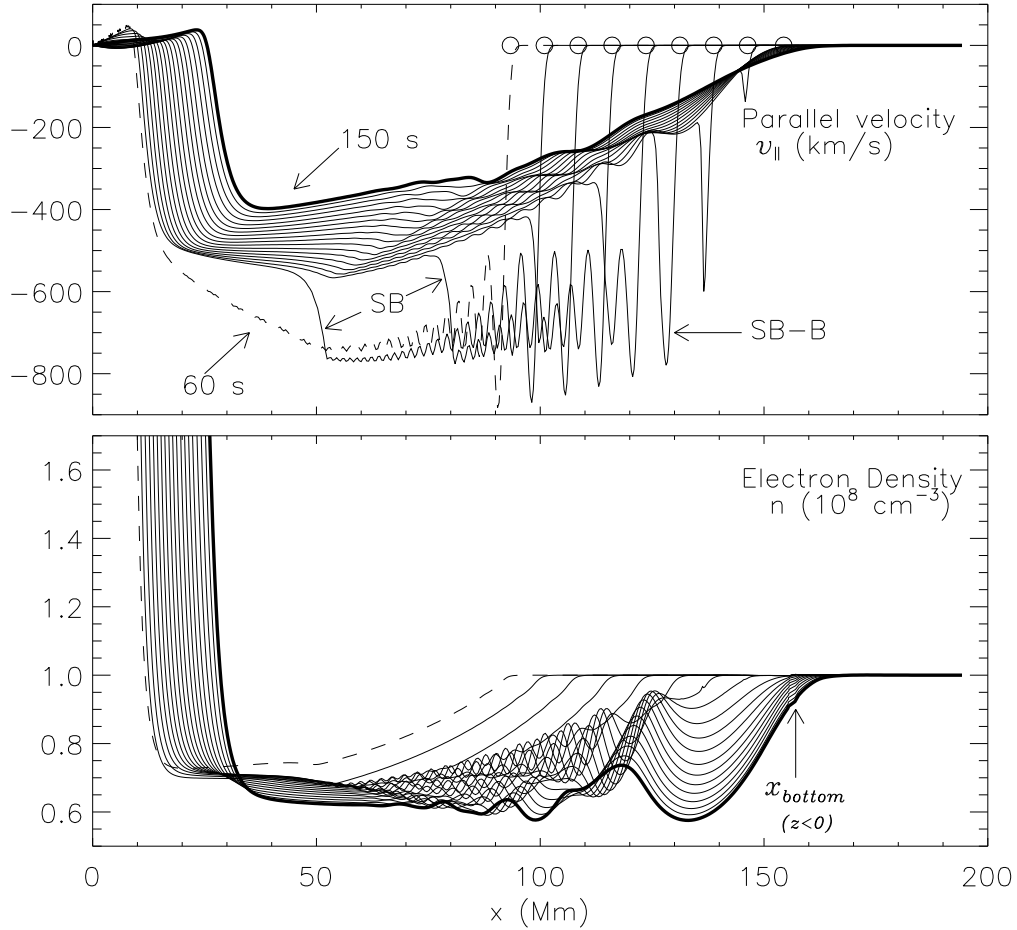}}
   \caption[DOWNWARD moving tube for the CENTER reconnection case]
   {DOWNWARD moving tube for the CENTER reconnection case. The two panels share the same horizontal axis that corresponds to the right side of the reconnected tube. Top panel: the parallel component of the speed for times between $60$ and $150$ s, at 5 s intervals. The first time (dashed line) corresponds to $t = 60$ s, which is the last time shown in Figure \ref{fig:Center_variables_sim}. Time $t = 150$ s is shown with a thicker line. Circles show the $x$-position of the bends. The ``SB'' arrows point toward the second bend locations at $t=65$ s and $t=70$ s. The ``SB-B'' arrow indicates a time when the second bend and the original bend are close. Bottom panel: electron density profile for the same times as in the top panel. The arrow points to the location where the tube becomes tangent to the edge of the current sheet}
   \label{fig:Center_rarefaction} 
\end{figure}

Some of the parallel velocity change is due to the perpendicular dynamics captured in the first term on the right-hand side of Equation \eqref{eqn:par_vel}. After a section of the tube arrives at the top of the arcade, its perpendicular velocity is suddenly decreased by the damping spring force. Therefore, the first term of Equation \eqref{eqn:par_vel} becomes unimportant compared to the other terms in the equation. The top panel of Figure \ref{fig:Center_rarefaction} shows parallel velocity profiles for late times in the CENTER reconnection simulation (only the positive $x$-axis side). The dashed line corresponds to simulation time equal to $60$ s -the earliest time shown in this graph and the last time shown in Figure \ref{fig:Center_variables_sim}, just before any part of the tube arrives at the arcade. The sharp discontinuity on the left shows the deceleration at the sub-shock. The next times shown in the graph present a new discontinuity at the locations of the second bend, where the parallel velocity is sharply decreased by the sudden reduction in perpendicular speed. The ``SB'' arrows point toward the second bend locations at $t=65$ s and $t=70$ s. 

%
%
%

When the second bend and the original bend become close (see arrow labeled ``SB-B'' in Figure \ref{fig:Center_rarefaction}), the angle at the latter one decreases until it finally becomes $180$\degree\ when the bend arrives at the top of the arcade.  Then, the inflow speed is drastically reduced because it depends strongly on this angle. The gas becomes rarefied and the electron density decreases rapidly behind the bends. Here, previously accelerated plasma continues moving toward the center of the tube leaving a density depletion behind. By the last time of our simulation, the electron density decrease is more than $40$\% of the background density. The bottom panel of Figure \ref{fig:Center_rarefaction} shows the electron density evolution for the same times shown in the top panel. The almost vertical slopes to the left indicate the sub-shock positions that continue moving outwardly, even after the bends arrive at the top of the arcade. For the TOP reconnection case, the maximum electron density decrease achieved by our simulations is more than $50$\%\ of the background.

The sub-shocks will continue moving outwardly and interact with the vacated regions. Although our simulations stop before this occurs, we hypothesize that the electron density will continue decreasing for $|x| < |x_{\mathrm{bottom}}|$ and the central hot plug will mix with these rarefaction waves, generating secondary rarefaction waves that would move toward the center of the tube and ultimately will disassemble the hot plug, as described by \citet{Longcope_2010}.  

\clearpage

\subsection{Emission Measure and Mean Temperature}
  \label{sec:rarefaction}

\begin{figure}[ht]
  \centering
   \resizebox{5.5in}{!}{\includegraphics{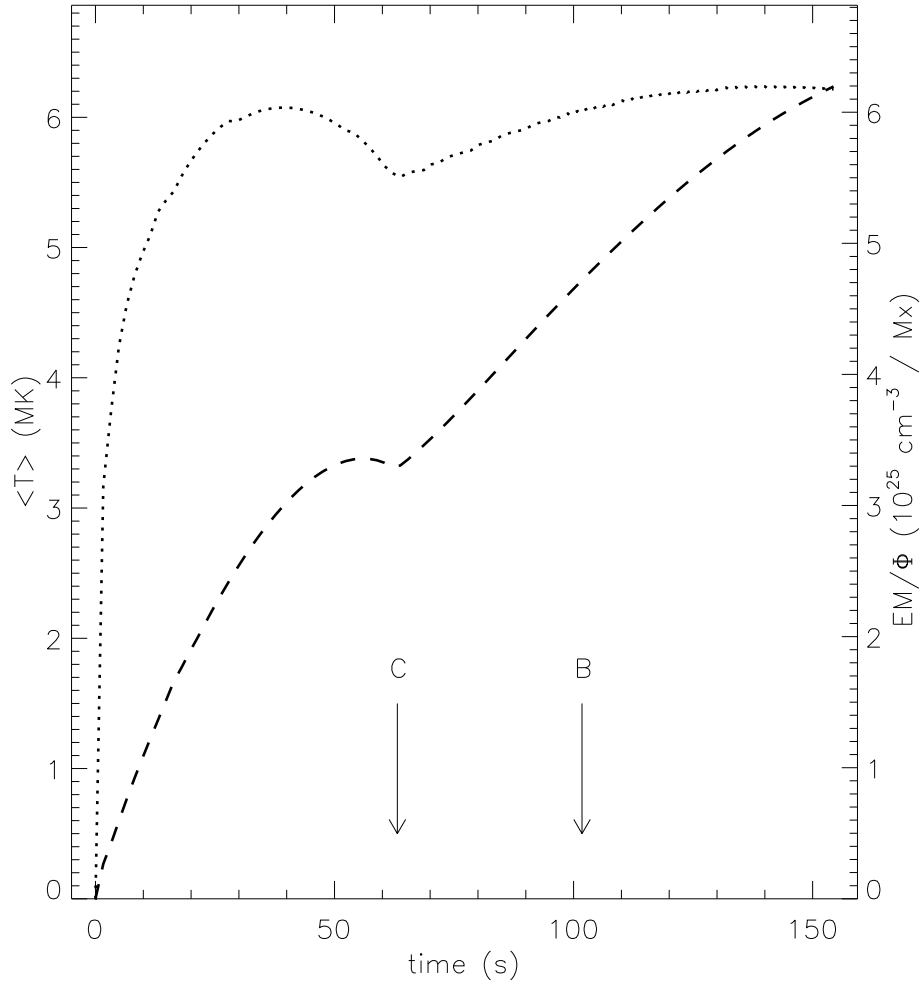}}
   \caption[DOWNWARD moving tube for the CENTER reconnection case]
   {DOWNWARD moving tube for the CENTER reconnection case. The time evolution of the mean temperature of the tube (left vertical axis) is shown with a dotted line. The total emission measure per flux of the tube (right vertical axis) is also shown for the same times with a dashed line. Arrow ``C'' indicates the time when the center of the tube arrives at the top of the arcade. Arrow ``B'' indicates the time when the bends arrive at the top of the arcade}
   \label{fig:Center_EM_T_evol} 
\end{figure}

The differences in dynamics between each reconnection case suggest an observational signature of where in the CS a flux tube had been reconnected. The tube's average temperature, weighted by the differential emission measure per flux, DEM(T), can be calculated as
%
\begin{eqnarray}
   \label{eqn:mean_T} 
      \langle T \rangle & = & \frac{ \sum\limits_{T>T_{e}} DEM(T) \hbox{ } T \hbox{ } \Delta T}{\sum\limits_{T>T_{e}} DEM(T) \hbox{ }  \Delta T}, 
\end{eqnarray}   
%
where the DEM(T) is computed for small temperature bins of size $\Delta T$ along the tube, as
%
\begin{eqnarray}
   \label{eqn:DEM} 
      DEM(T) & = & \frac{\sum\limits_{\left |T(l) - T\right| <\frac{\Delta T}{2}} EM(l)}{\Delta T}. 
\end{eqnarray}   
%
Only tube elements whose temperature is inside the range of the bin are included in the sum. $\langle T \rangle$ refers to the heated material of the tube; it excludes segments at or below ambient temperature. Figure \ref{fig:Center_EM_T_evol} shows the time evolution of the tube's average temperature (dotted line) for the CENTER reconnection case. 

%
%
%

The emission measure per flux of a given tube element, EM(l), is computed in the following way
%
\begin{eqnarray}
   \label{eqn:EM_l} 
      EM(l) & = & \frac{n^{2}(l) \hbox{ }  \delta l} {B_{e}(l)}. 
\end{eqnarray}   
%
With the above definition, the tube's total emission measure per flux becomes
%
\begin{eqnarray}
   \label{eqn:EM_flux} 
      \frac{EM}{\phi} & = & \sum\limits_{T(l)>T_{e}} EM(l). 
\end{eqnarray}   
%
The above quantity considers only the emission measure of the heated parts of the tube ($T > T_{e}$), excluding cool sections like the ones following the bends and preceding the heat fronts, shown in the middle panel of Figure \ref{fig:Center_variables_sim}. Ambient plasma emission measure is not included either.  Figure \ref{fig:Center_EM_T_evol} shows the time evolution of the tube's total emission measure per flux (dashed line) for the CENTER reconnection case.  

In the figure, the arrow labeled ``C'' indicates the time at which the center of the tube arrives at the top of the arcade, and the one labeled ``B'' represents the arrival of the bends at the top of the arcade. Initially, the mean temperature of the tube increases at a faster rate than the emission measure. They both achieve a local maximum before time ``C''. The following negative slope for both curves indicates that these quantities would rapidly decrease if the tube were left to retract without the arcade interrupting its motion. Remarkably, they increase after this point. The length of the tube does not change considerably while lying on top of the arcade, but the heat fronts continue moving along the tube, increasing its average temperature. Our simulations end before the heat fronts arrive at the footpoints, but we hypothesize that the average temperature of the tube will continue increasing until they do so.

\begin{figure}[ht]
  \centering
   \resizebox{5.5in}{!}{\includegraphics{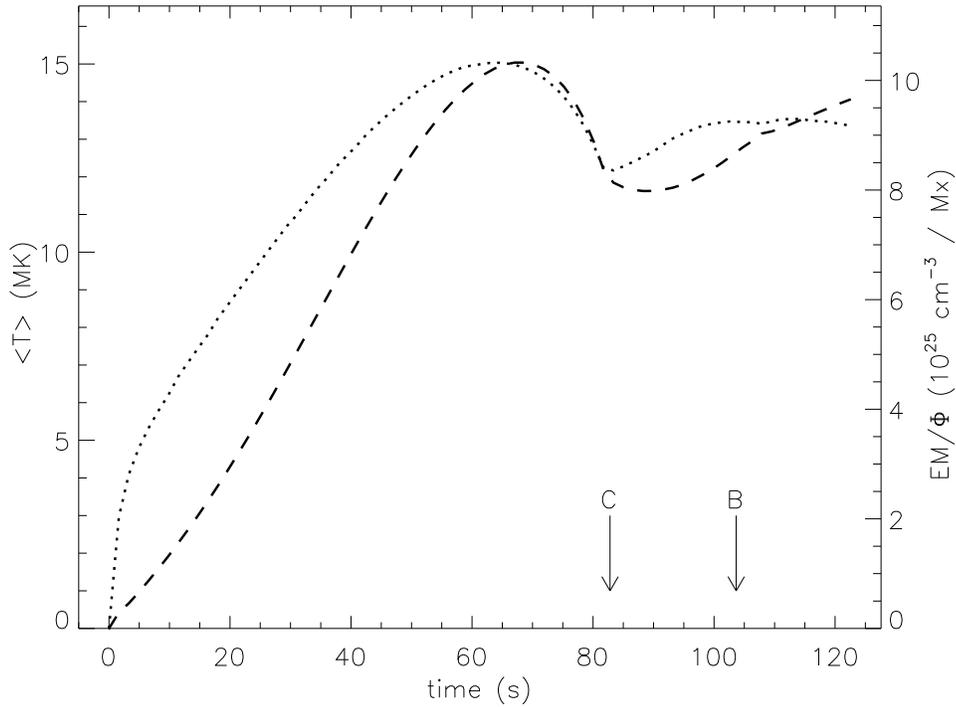}}
   \caption[DOWNWARD moving tube for the TOP reconnection case]
   {DOWNWARD moving tube for the TOP reconnection case, with the same format as Figure \ref{fig:Center_EM_T_evol}}
   \label{fig:TOP_EM_T_evol} 
\end{figure}

The central hot plug lasts longer than a conductive cooling time of a region with similar temperature and size \citep{Longcope_2010} as it is maintained by the inflows. These quantities are therefore long lasting. The TOP reconnection case presents some differences. The mean temperature and emission temperature grow together, as shown in Figure \ref{fig:TOP_EM_T_evol}. Here, the arcade also prevents a decrease in both quantities.

%
%
%

In order to provide a general method to determine where in the CS reconnection happened for a given flux tube, we normalized the mean temperature and emission measure by their maximum achieved value during the entire simulation. For example, the maximum achieved mean temperature during the simulation for the CENTER case is $\langle T \rangle_{\mathrm{MAX}} \simeq 6.2$ MK (see Figure \ref{fig:Center_EM_T_evol}). Then, for each time in the simulation, the normalized mean temperature is defined as
\begin{eqnarray}
   \label{eqn:EM_norm} 
      \langle T \rangle \hbox{(normalized)}& = & \frac{\langle T \rangle}{\langle T \rangle_{\mathrm{MAX}}}. 
\end{eqnarray}   
The normalized emission measure is defined in an analogous way.

\begin{figure}[ht]
  \centering
   \resizebox{5.5in}{!}{\includegraphics{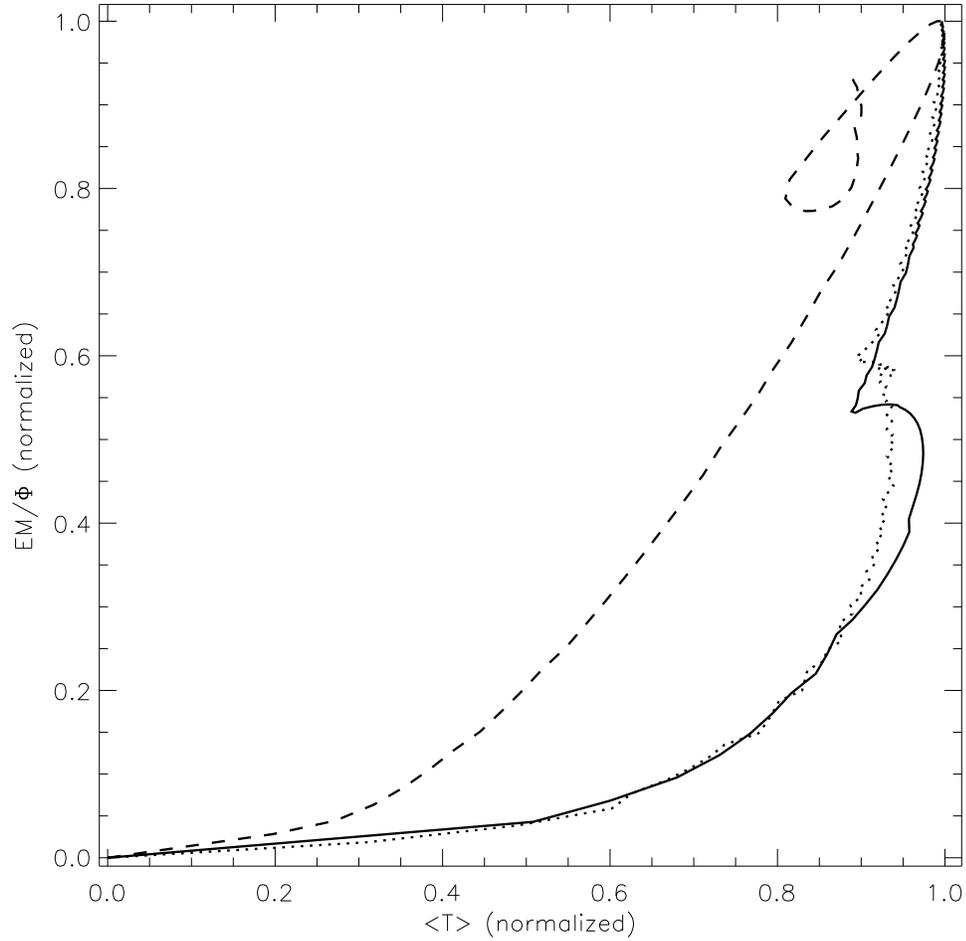}}
   \caption[Normalized temperature and emission measure]
   {Normalized emission measure as function of the normalized mean temperature of the tube for DOWNWARD moving tubes. The solid line corresponds to the CENTER reconnection case, the dotted line to the BOTTOM reconnection case, and the dashed line to the TOP reconnection case}
   \label{fig:T_EM_relation} 
\end{figure}

Figure \ref{fig:T_EM_relation} shows the dependence of normalized emission measure with normalized mean temperature for each of the reconnection cases. In this figure, for each time in the simulation, the value of the normalized emission measure is plotted as a function of the value of the mean temperature at that same time. BOTTOM and CENTER cases (dotted and solid lines, respectively) are very similar to each other. For these cases, the mean temperature achieves its maximum before the emission measure does. On the other hand, the TOP case presents a different dependence. Both quantities initially increase simultaneously and curl at the top end of the curve. 

%
%
%

\section{Discussion}
  \label{sec:conclusions_ch2}

We have presented a model of tube dynamics, accompanied by simulations, following transient and localized magnetic reconnection in a realistic coronal background configuration (Syrovatski{\v i} CS). We have shown that retracting reconnected flux tubes may present elongated regions devoid of plasma, as well as long lasting, dense central hot regions. The latter are created by GDSs at the center of the tube, consisting in long thermal fronts followed by an isothermal sub-shock. In general, the jump in density across the shock exceeds the maximum value predicted by Rankine--Hugoniot \citep{Rankine_1870,Hugoniot_1887} conditions. These jump conditions are calculated assuming the shock is in steady state, which is not the case here. For the TOP reconnection case, the jump in density can be almost an order of magnitude. This descending plasma plug, although very thin, would be extremely bright. It is long lived compared to a free expansion of similar temperature and size region since it is maintained by the inflows generated at the bends.     

GDSs are also present in reconnected tubes sliding through a CS with uniform skewed fields \citep{Guidoni_2010}. However, in our present model, these hot plugs respond to the change in background magnetic pressure that compresses or expands them laterally, depending on which region of the CS the tube is sliding through. For Syrovatski{\v i} CSs, the magnetic field decreases toward the edges of the CS (Y-points). There, its magnitude is minimum (it is zero if there is no guide field). Therefore, it is possible for the tubes to move through regions of a decreasing magnetic field as they descend toward the Sun. This seems counterintuitive because usually magnetic fields in the corona are assumed to increase at lower heights. 

We have presented only three reconnection locations with a reconnection half angle of 45\degree\ for illustrative effects. We used generic dimensional values, but results scale with $B_{e}$, $\rho_{e}$, and $L_{e}$. For all the cases presented, the maximum plasma-$\beta$ achieved never exceeded unity.

RDs (bends) move at the local Alfv\'{e}n speed along the legs of the tubes. Elongated low density regions are generated by lateral expansions behind the bends, when tubes move through the lower half plane of the CS. There, background magnetic pressure decreases toward the edge of the CS. The amount of density depletion depends on how curved the tube is, which is directly related to the perpendicular gradient in the background magnetic field (reconnection angle). On the contrary, in uniform background fields, as shown in Guidoni10, tubes present only straight sections, regardless of the reconnection angle. No density depletion is expected in this case. When there is no significant density depletion, and the GDSs are allowed to extend for a detectable length, an observed signature of these tubes would be a descending hot plug with high density. A signature in spectra would also be expected from the parallel flows that might be detectable as Doppler shifts.  

The achieved decrease in density could be as much as $30$\%--$50$\%\ of background (pre-flare) values. This level of depletion agrees with observations \citep{McKenzie_1999, Sheeley_2002}. These percentages are lower bounds since we have not considered any increase in the surrounding plasma density by, for example, chromospheric evaporation from previous reconnection episodes. 

The appearance of the retracting tubes changes considerably with the direction of the line of sight and with the location of the reconnection episode. If reconnected tubes (their observable parts) are seen from the same view as in panel (a) of Figure \ref{fig:flare_CS}, they would have loop shapes, although their orientation would be perpendicular to the arcade since they move in a plane that is parallel to the arcade axis. On the other hand, if the reconnected tubes are seen from the view shown in panel (b) of Figure \ref{fig:flare_CS}, they would be seen as ``hairpins'' because they are tangent to the plane of the CS. 

For the TOP reconnection case, the concavity of descending loops when they move in the top part of the CS plane is the one that would be expected from a cusp-shaped loop. In this region, these loops would be bright at their sides (they are being squeezed by ambient plasma) and even brighter at their center. If the line of sight is parallel to the $x$-direction, this loop would appear like a bright small region, descending at Alfv\'{e}nic speeds. After the bends cross the center of the CS, dark regions will develop behind them, and this view will change to vertical dark regions preceded by a bright region, until the center catches up with the bends and the situation is reversed. The dark and bright regions are hot since the heat fronts move at speeds similar to the bend speeds. Their sizes are comparable to the entire tube. 

For BOTTOM and CENTER reconnection cases, their concavity is always U-shaped. If seen from the $x$-direction, a bright region is followed by an elongated vertical dark one that might be hot or not, depending on the location of the heat front.  

Observed coronal inflows can be bright or dark, or both \citep{Savage_2011}; they may also have loop- or tadpole-shapes \citep{McKenzie_2000, McKenzie_2001,Sheeley_2002}. It is possible that many of them are the manifestation of the same three-dimensional phenomena observed from different lines of sight. The location of the tube with respect to the CS also matters. If an observation occurs, for example, while a tube is in the top half-plane of the CS, only bright regions would be observed. On the other hand, if a tube is moving through the lower portion of the CS, it would present dark regions, sometimes preceded by bright regions, or sometimes followed by bright regions, as described above.

A dense, hot region may not always be observed due to its small size, or its high temperature. For reconnection angles larger than $45$\degree, it is possible to achieve post-shock temperatures that are much higher than the usual instrument passbands. On the other hand, after the tube's arrival at the top of the arcade, this region continues growing in time, and may become visible. The pile up of several newly reconnected tubes may also increase the size of this region to observable size and would manifest as a stationary bright region on top of the arcade.

Remarkably, signatures of reconnection persist in a single tube, longer than the Alfv\'{e}nic transit-time required for the tube to relax. As the tubes move downward, they encounter the top of the arcade that lies beneath the bottom edge of the CS. Here, the downward motion is halted by the arcade. We simulated this effect with a perpendicular damped spring force exerted by the compressed arcade. With this force, the tube comes to rest on top of the arcade as a straight line. The parallel dynamics continues along the legs. After they stop, the tubes cannot decrease their length any longer, and the RDs are shut down. Then, the gas gets rarefied even more because the already accelerated plasma continues moving toward the center of the tube, increasing the density depletion behind it. 

Although our simulations stop before this, we hypothesize that the hot post-shock regions will continue moving along the tube and interact with the rarefaction waves, subsequently disassembling the hot plug. The legs of the tube would also brighten up as the thermal fronts descend toward the footpoints.  

The tube's parallel and perpendicular dynamics are nonlinearly coupled to each other, which can be seen, for example, in the oscillations of the parallel velocity profiles. In addition, when the tube arrives at the arcade, the perpendicular dynamics is halted, and changes in the parallel speed are evident at the ``second bend''. The parallel velocity profiles are determined by pressure and fictitious forces due to the motion of tube elements along curved paths. This can have important consequences for Doppler-shift observations. For example, the post-shock region in the TOP reconnection case has very strong flows in a direction almost parallel to the surface of the Sun. This would manifest as coincident blue- and red-shifts along the line of sight, if seen from the $x$-direction, and perpendicular to the line of sight if seen from the $z$-direction. Perpendicular velocities can be higher or lower than the local background Alfv\'{e}n speed, which can be related to the generation of secondary shocks as the tubes move (the surrounding plasma may be shocked by the passing tubes).

 We described the temporal behavior of the total emission measure and mean temperature of the heated parts of the tube and provide an observational method that may indicate where in the CS the tube has been reconnected. Tubes that were reconnected in the TOP half of the CS present total emission measure that grows simultaneously with the mean temperature of the tube. On the other hand, tubes that have been reconnected in the BOTTOM part of the CS achieve a maximum in their mean temperature much earlier than the emission measure does. For all the reconnection cases, the emission measure and mean temperature duration is extended by the arrival of the tubes at the top of the arcade. After the arrival, the length of the tube remains the same but the heat fronts continue heating the plasma as they move along the legs. 

If observed downflows (SADs and SADLs) are reconnection outflows, the question of why their speeds are lower than the assumed Alfv\'{e}n speed still remains. We restricted ourselves to standard reconnection scenarios where the outflows are Alfv\'{e}nic. Nevertheless, if dark voids are related to the dark regions generated behind the bends, their speeds near the edge of the CS are decreased considerably from the one they had at the center of the CS. For instance, in the CENTER reconnection case, the bend speeds are reduced by a fourth of what it was at the reconnection site. This decrease in speed of the bends is not enough to explain the observed descending voids that move at half of the presumed reconnection Alfv\'{e}n speed. However, if the half reconnection angle is greater than $60$\degree, the ratio of the Alfv\'{e}n speed at the center of the CS to the edge of the CS is larger than $1/2$. 

HXR sources have been observed near top of SXT arcades, at the same time as the SADs descend to the top of the arcade \citep{Krucker_2003,Asai_2004}. This may be the result of the interaction between downflows and the fast shock, or could be related to reconnection at the arcade apex between the retracting tubes and the arcade. Apex reconnection is a possibility in our three-dimensional model, but one we have not yet explored. Field lines that arrive at the top of the arcade have a different angle than the underlying arcade (panel (a) of Figure \ref{fig:Top_arcade}) and arrive there at Alfv\'{e}nic or super-Alfv\'{e}nic speeds. This is not an option in purely two-dimensional models where the new reconnected field lines arrive at the arcade with magnetic field direction parallel to the arcade.

Maintaining a fast shock on top of the arcade may not be feasible at sub-Alfv\'{e}nic speeds. The standing fast shocks exist as long as the reconnection jets exists and is super-magnetosonic, therefore it is a feature of a flare gradual phase \citep{Forbes_1986_I}. Most of the SADs have been reported during the decay phase (there are some exceptions to this: see \citet{Asai_2004}), but their speeds are not Alfv\'{e}nic. If SADs are the only reconnection outflows, it is not clear if the termination shocks would still be located at the top of the SXT loops.

Another possibility could be that the termination shock is located further up from the top of the SXT loops. If loops hotter than the SXT loops (that cannot be seen in pass bands at lower temperatures) are located above them, the termination shock could be located higher in the corona. In this case, the downflows could be localized downstream flows from the fast shock, which would explain their low speeds. The models for fast shocks are two dimensional and it is not clear how this would apply to localized and time-dependent flux tubes like the ones described here. The down side to this explanation is that a fast shock certainly would compress the plasma downstream instead of vacating it. 

We have not considered other interactions between the reconnected tubes and their ambient plasma, besides pressure balance between them. The addition of a drag force to our equations could be an important improvement to our model that would contribute to the decrease in speed, as well as to make the model more realistic. We expect to include this effect in future work. The assumption of initial uniform density could also be improved by assuming a stratified atmosphere. Here, tubes that have been reconnected in the corona will descend in regions of higher density, and seen darker than the background, as suggested by \citet{Savage_2010}, although this effect has not been corroborated by theoretical analysis. Another effect that we have not considered is the possible tangling of field lines \citep{Klimchuk_1997}. For example, an UPWARD moving tube reconnected at a given location might run into a DOWNWARD moving tube retracting from a higher reconnection site. 

We have neglected any non-fluid effects and therefore neglected non-thermal particles and their acceleration. 

We do not claim that the depleted regions from our model can completely explain the observed dark voids, but it is suggestive that several observed phenomena have some similarities with our model. \citet{Hudson_2001} stated that ``Dark outflows seem inconsistent with the idea of heating by reconnection''. We believe our work demonstrates the contrary, and thus makes the role of reconnection in flares still more plausible.

%

\chapter{TESTING THE THIN FLUX TUBE MODEL}
   \label{chap:chap_3}

Our model for reconnected thin flux tubes appears to explain some observed phenomena in the Solar Corona, as described in the previous chapter. However, it is still not clear how realistic or restrictive its assumptions are. The TFT model assumes that, if the reconnection region is small and short-lived, a reconnected tube will retract from this region and bends will move along its legs at the local Alfv\'{e}n speed. The model also predicts that shocks will travel outward from the center of the tube, heating and compressing its plasma. The tube is assumed to be isolated from its surroundings and to slide freely between the current sheet layers. The only interaction with its surroundings is through pressure balance. 

One way to test the validity of the foregoing assumptions is to compare results of the model with those from fully three-dimensional solutions of the general MHD equations. In this chapter, we present three-dimensional MHD simulations carried out in a volume filled with plasma with a Green-Syrovatski{\v i} \citep{Green_1965,Syrovatskii_1971} CS, like the one depicted in Figure \ref{fig:flare_CS}. The reconnection episode is triggered by imposing a short-lived and localized magnetic resistivity patch on the top half of the current sheet. The results of these simulations are compared to the corresponding DEFT simulation. 

For such comparison, the possible options from the parameter space are limitless: reconnection location, field line angle at each side of the current sheet, plasma unitless numbers, etc. However, given the complexity of the MHD simulations, we only chose one particular case for our comparison. Typical  simulations keep track of millions of cells that describe a plasma in a given volume. They are computationally demanding, and can only be executed in reasonable amounts of time on super-computers. In the near future, more of such comparisons will be performed.

\section{3D MHD Simulations}
  \label{sec:Comparison}

The fully three-dimensional MHD simulations were carried out by Dr. Mark Linton at the Naval Research Laboratory. The computer program used was the Adaptively Refined Magnetohydrodynamics Solver (ARMS) developed by Dr. C. Richard Devore. The ARMS computer program is a time-explicit, conservative, monotone MHD model. The MHD equations are advanced using flux-corrected transport techniques \citep{Devore_1991}, and the adaptive mesh refinement is managed by the PARAMESH toolkit \citep{MacNeice_2000}. The grid adapts to the spatial variations of the magnetic field magnitude and electric current density, therefore the region of the current sheet is refined at the highest resolution, and it is effectively one cell wide. Newly created grid points are linearly interpolated with a monotone limiter. 

The three-dimensional Syrovatski\v{i} current sheet is described by the following equation \citep{Priest_2000}
%
%
\begin{eqnarray}
   \label{eqn:3D_Sirovatskii}
      B_{z} + i B_{y} & = &  B_{s} \sqrt{\frac{\omega^{2}}{L_{e}^{2}} - 1},
\end{eqnarray}
%
%
where $i$ is the positive value of $\sqrt{-1}$, and $\omega \equiv y + i z$. In this configuration, the current sheet plane lies on the $z=0$ plane. To allow for a finite angle between field lines on each side of the current sheet, a uniform guide field $B_{ex}$ is added.

Extrapolative, zero gradient, open boundary conditions are imposed in the $y$- and $z$-directions, while periodic boundary conditions are imposed in the guide field direction. The maximum resolution in the $x$, $y$, and $z$ directions is $256 \times [3,1,1]$ cells, respectively.

The program solves ideal MHD Equations \eqref{eqn:MHD_mass}, \eqref{eqn:MHD_momentum}, and \eqref{eqn:MHD_energy}, with the addition of an explicit magnetic resistivity ($\lambda$) term in the right hand side of induction equation \eqref{eqn:MHD_induction}
%
%
\begin{eqnarray}
   \label{eqn:MHD_mag_res_3D}
      - \nabla \times \left( \lambda \nabla \times \mathbf{B} \right),
\end{eqnarray}
%
%
and the corresponding ohmic heating term added to the energy equation \eqref{eqn:MHD_energy}
%
%
\begin{eqnarray}
   \label{eqn:MHD_ohmic_3D}
      \frac{\lambda }{4 \pi} \left| \nabla \times \mathbf{B} \right|^{2}.
\end{eqnarray}
%
%

The simulations are in non-dimensional form. The density and gas pressure are initially set to be uniform at $\rho_{i} = 1/2$ and $P_{i}=20/3$. The chosen magnitude of the magnetic field at $(x,y,z) = (0,0,\sim 0)$ is $B_{i}=20$. With these initial conditions, the Alfv\'{e}n speed and the plasma-$\beta$ at the center of the current sheet are $v_{ai} = 20/\sqrt{2 \pi}$, and $\beta_{i} = 0.42$. The resistivity patch (reconnection region) is centered at $y^{\prime}_{R} = 2/3$, close to the upper Y-line, so the upflowing reconnected lines quickly hit that line. The chosen angle at the center of the CS is $\zeta_{0} \sim 45$\degree. This implies a reconnection angle equal to $\zeta_{R} \sim 37$\degree, and a plasma-$\beta$ there equal to $0.54$. The unitless half-length of the current sheet is $L_{i} = 1$. 

Simulations are run with uniform background resistivity, $\lambda_{e}$, locally enhanced at the reconnection location within a sphere or radius $\delta = 0.175$. This enhancement is short lived, for the first $t^{*} = 0.1$ of the simulation, and has the form  
%
\begin{eqnarray}
   \label{eqn:MHD_res_enhan_3D}
      \lambda = \lambda_{e} \left( 1 + C e^{-r^{2}/\delta^{2}} \right), 
\end{eqnarray}
%
%
for $r = \sqrt{x^{2}+ \left(y - y_{R}\right)^{2} +z^{2}} < 2 \delta$, and $1+C = 10^{5}$. The background Lundquist number, $S_{\lambda} = \delta v_{ai} / \lambda_{e}$, is approximately equal to $1.4 \times 10^{6}$, and the Lundquist number for the peak resistivity at the center of the reconnection region is approximately $14$. 

Including realistic transport coefficients, such as viscosity and thermal conduction, in a fully three-dimensional simulation is extremely difficult. The viscous stress tensor components in the direction parallel to the magnetic field are many orders of magnitude larger than the ones in the perpendicular direction. The same is true for the heat conduction. The ARMS computer program does not include these coefficients explicitly, but there is numerical resistivity due to the finite grid-scale of the simulation. This is analogous to having transport coefficients that are temperature-independent with a Prandtl number equal to $1$. These limitations are quite drastic since this transport coefficient temperature dependence is non trivial in establishing the thickness of the shocks, as shown in chapter \ref{chap:chap_1}.

\begin{figure}[htbp]
   \centering
   \begin{center}$
      \begin{array}{c}
         \resizebox{5.5in}{!}{\includegraphics{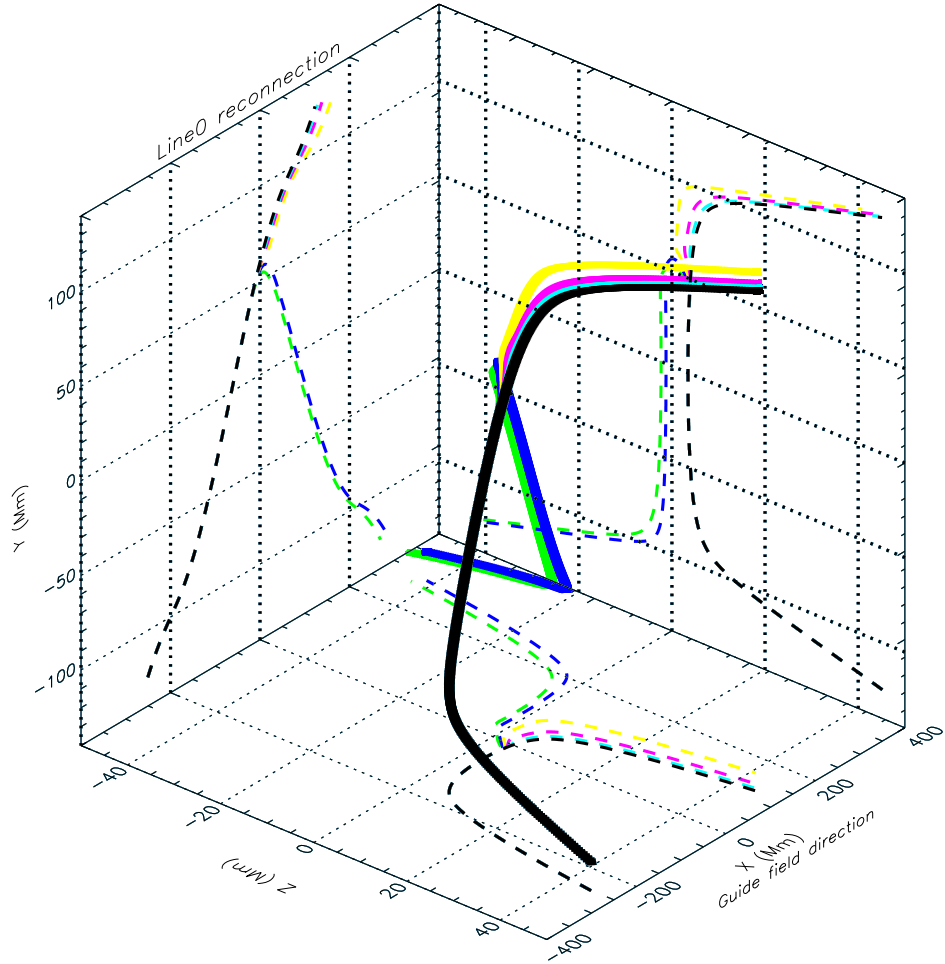}} \\
      \end{array}$
   \end{center}
   \caption[Line0 evolution at early times in the three-dimensional simulation]
   {Line0 evolution for early times in the three-dimensional simulation. The colored solid lines are the field line curves for times at $30$ s intervals: $0$ s (black), $30$ s (cyan), $60$ s (magenta), $90$ s (yellow), $120$ s (blue), $150$ s (green). The plane of the current sheet is located at $z=0$. Line0 bottom footpoint at $t=0$ s is point $(x,y,z) = (-251,-125,-45)$ Mm. Its top endpoint at $t=0$ is located at $(x,y,z) = (123,125,-45)$ Mm. Colored dashed lines are the field line projections on the box walls of the same color three-dimensional curve}
   \label{fig:rec_line0} 
\end{figure}

A set of Lagrangian test particles can be followed dynamically during the simulation, and field lines are traced from these particles \citep{Linton_2009}. In particular, field lines that originally belonged to one side of the current sheet, are anchored at the bottom of the arcade, and intersect the sphere of high resistivity at the start of the simulation, can be traced in time. The particles that initially lie in the high resistivity region are not frozen onto field lines, but they follow the field line evolution after they leave the reconnection region, or after the resistive sphere is turned off. Figure \ref{fig:rec_line0} shows the evolution of one such field line, where each solid color corresponds to a different time. The $z$-coordinate has been enhanced for clarity, so the figure is not to scale.  For concreteness, we will present our results assuming a current sheet half-length of $L_{e}=100$ Mm, initial uniform electron density $n_{e}=$ \realnumdens, temperature $T_{e}=1$ MK, and magnetic field magnitude at the reconnection site, $B_{e}\sim 1.1$ G. In this units, the reconnection sphere radius is $\delta \cong 17.5$ Mm.

The dashed lines on the coordinate box walls are the projections of the field lines on those planes for the corresponding colors. The thick black field line is the field line at $t=0$. At this time, the field line is completely on the front side ($z>0$) of the current sheet plane. This is easy to see on its projection on the bottom box wall. This field line intersects the reconnection sphere with center at $(x_{R},y_{R},z_{R}) = (0,66,0)$ Mm, and its bottom footpoint is anchored at $(x,y,z) = (-251,-125,45)$ Mm. From this point, the field line is traced for different times. We will call this field line, Line0. Its top endpoint is initially located at $(x,y,z) = (123,125,45)$ Mm. 

%
%

\begin{figure}[htbp]
   \centering
   \begin{center}$
      \begin{array}{c}
         \resizebox{5.5in}{!}{\includegraphics{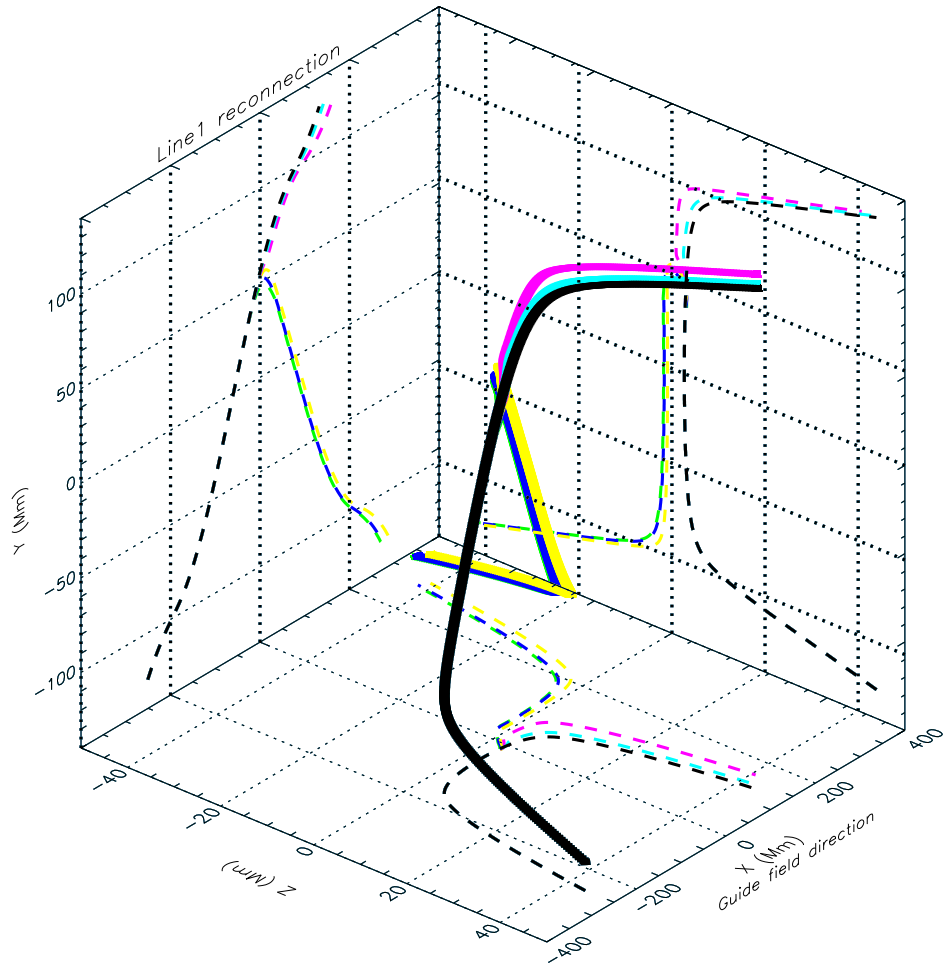}} \\
      \end{array}$
   \end{center}
   \caption[Line1 evolution]
   {Line1 evolution for the same times as in Figure \ref{fig:rec_line0}. Line1 bottom footpoint at $t=0$ is point $(x,y,z) = (-251,-125,44)$ Mm. Its top endpoint at $t=0$ is located at $(x,y,z) = (132,125,44)$ Mm. Colors represent the same times as in Figure \ref{fig:rec_line0}. Dash lines are also the field line projections on the walls} 
   \label{fig:rec_line1} 
\end{figure}

\begin{figure}[htbp]
   \centering
   \begin{center}$
      \begin{array}{c}
         \resizebox{5.5in}{!}{\includegraphics{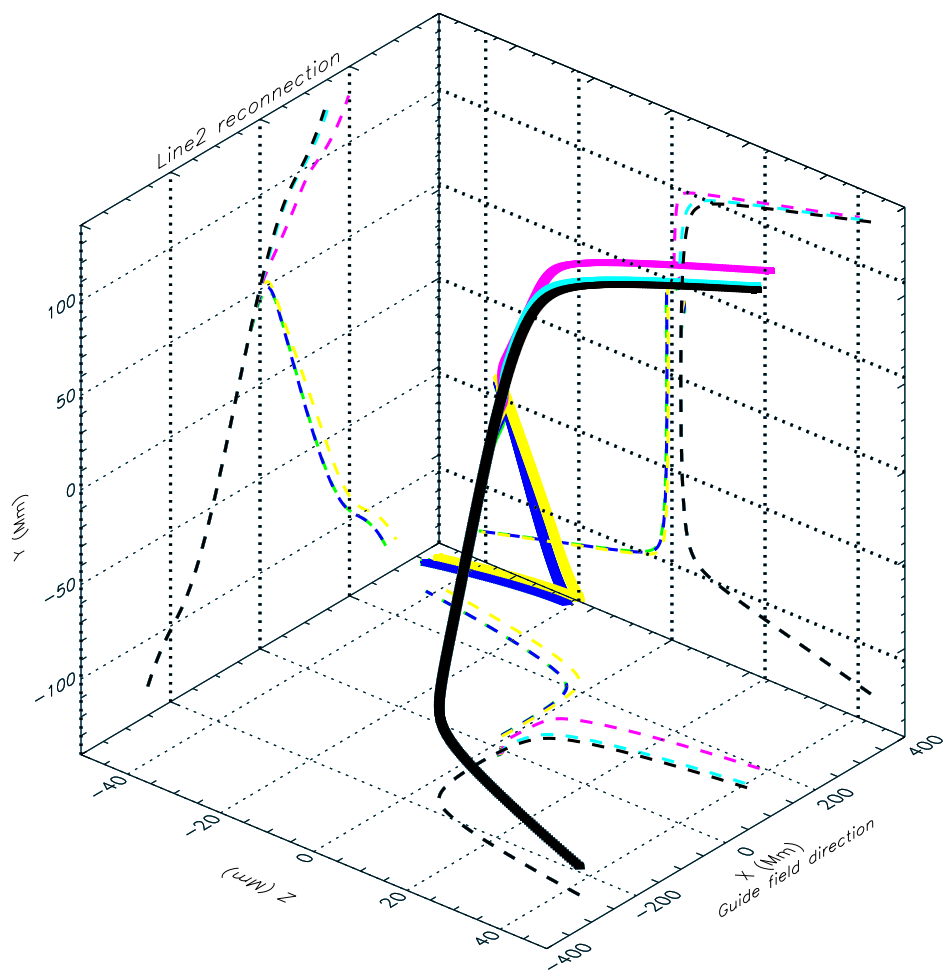}} \\
      \end{array}$
   \end{center}
   \caption[Line2 evolution]
   {Line2 evolution for the same times as in Figure \ref{fig:rec_line0}. Line2 bottom footpoint at $t=0$ is point $(x,y,z) = (-251,-125,43)$ Mm. Its top endpoint at $t=0$ is located at $(x,y,z) = (144,125,43)$ Mm. Colors represent the same times as in Figure \ref{fig:rec_line0}. Dash lines are also the field line projections on the walls}
   \label{fig:rec_line2} 
\end{figure}

\begin{figure}[htbp]
   \centering
   \begin{center}$
      \begin{array}{c}
         \resizebox{5.5in}{!}{\includegraphics{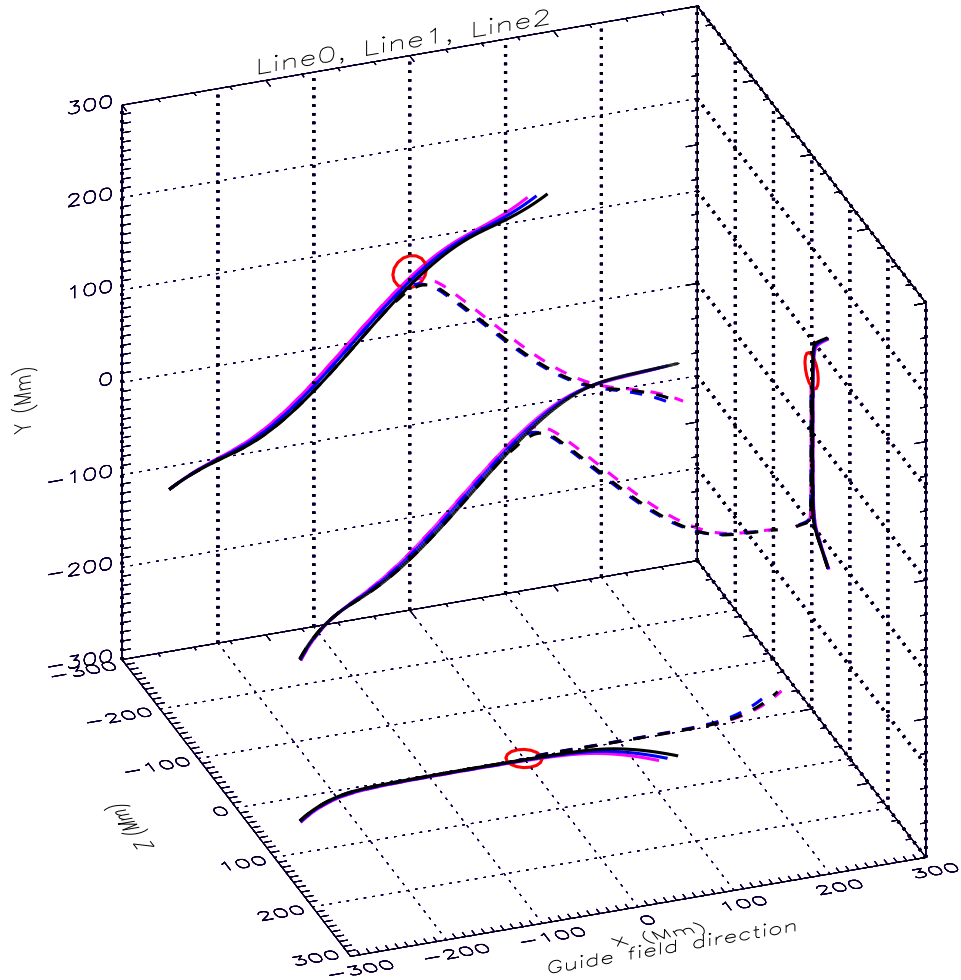}} \\
      \end{array}$
   \end{center}
   \caption[Line0, Line1, Line2 to scale]
   {Line0, Line1, Line2 to scale. Solid lines represent field lines at $t=0$, as well as their projections on the box walls, and dashed lines are the same color field line at $t=120$ s, as well as their corresponding projections. Magenta color is used for Line0, blue for Line1, and black for Line2. Red circles are the projection on the box walls of a sphere with radius $\delta = 17.5$ Mm}
   \label{fig:rec_line0_scale} 
\end{figure}

While the resistive sphere is active, the top footpoint moves as the field line ``diffuses''. This artificial motion of the footpoint continues, and later on the field line is connected to the other side of the current sheet (the blue curve has one footpoint in the $z>0$ region). Figures \ref{fig:rec_line1} and Figures \ref{fig:rec_line2} show the same process for field lines with bottom footpoints at $(x,y,z) = (-251,-125,44)$ Mm (we will call this line, Line1) and $(x,y,z) = (-251,-125,-43)$ Mm (Line2), respectively. These field lines also cross the reconnection sphere. While Figures \ref{fig:rec_line0}, \ref{fig:rec_line1}, and \ref{fig:rec_line2} have an enhanced $z$-axis, Figure \ref{fig:rec_line0_scale} shows field lines Line0, Line1, and Line2 to scale at $t=0$ s (before reconnection) and at $t=120$ s (after reconnection); they change their connectivity as they reconnect across the current sheet. The red circles are the projections on the box walls of a sphere of radius $\delta = 17.5$ Mm, representing approximately the region of high resistivity. After reconnection, the field lines are symmetric around the $x=0$ plane. The sections of these field lines whose $y$-coordinates lie between $y=-100$ Mm and $y=+100$ Mm (the edges of the current sheet) are almost flat and parallel to the plane of the current sheet. The sections outside these edges are the tube legs and are not parallel to the current sheet.

%
%

\begin{figure}[htbp]
   \centering
   \begin{center}$
      \begin{array}{c}
         \resizebox{5.5in}{!}{\includegraphics{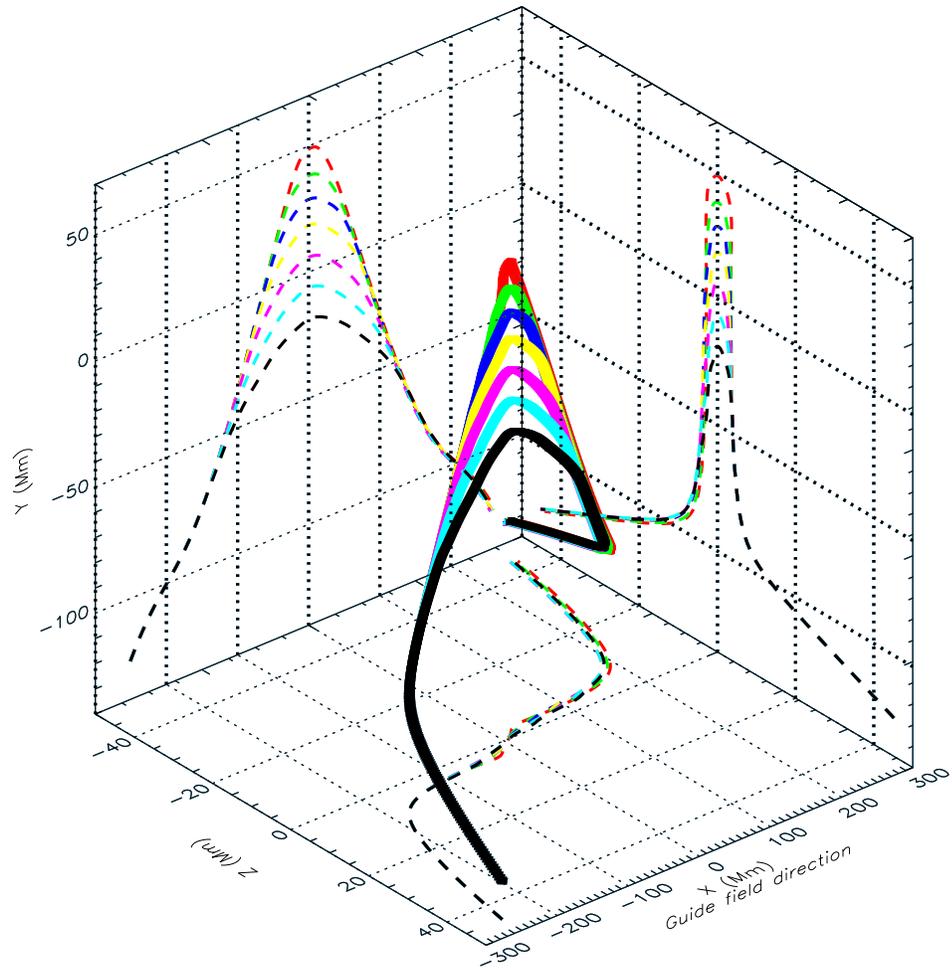}} \\
      \end{array}$
   \end{center}
   \caption[Line0 retraction]
   {Line0 retraction from the reconnection site. The colored solid lines are the field line curves for times at $90$ s intervals: $180$ s (red), $270$ s (green), $360$ s (blue), $540$ s (yellow), $630$ s (magenta), $720$ s (cyan), and $810$ s (black). Colored dashed lines are the field line projections on the box walls of the same color three-dimensional curve}
   \label{fig:rec_line0_evol} 
\end{figure}

Once a field line is completely reconnected, its newest footpoint remains almost stationary, and the field line retracts from the reconnection region, decreasing its length. The evolution of Line0 after the resistive sphere has been turn off is shown in Figure \ref{fig:rec_line0_evol}. Most of the retraction occurs in the plane of the current sheet, as can be seen from the field line projection on the bottom box wall. It was a principal assumption of the TFT model underlying previous chapters that  post-reconnection motion was confined to the current sheet. 

%
%

If Line0 is observed from the $x$-direction as it moves, it would look like a retracting hairpin (this can be seen in its projections on the background walls in Figures \ref{fig:rec_line0} and \ref{fig:rec_line0_scale}). On the other hand, if the field line is observed from the $z$-direction, it would look like a retracting loop (as can be seen from the projections on the other background wall). This is very similar to what we described in our conclusions in the previous chapter. 

%
%

\begin{figure}[htbp]
   \centering
   \begin{center}$
      \begin{array}{c}
         \resizebox{5.5in}{!}{\includegraphics{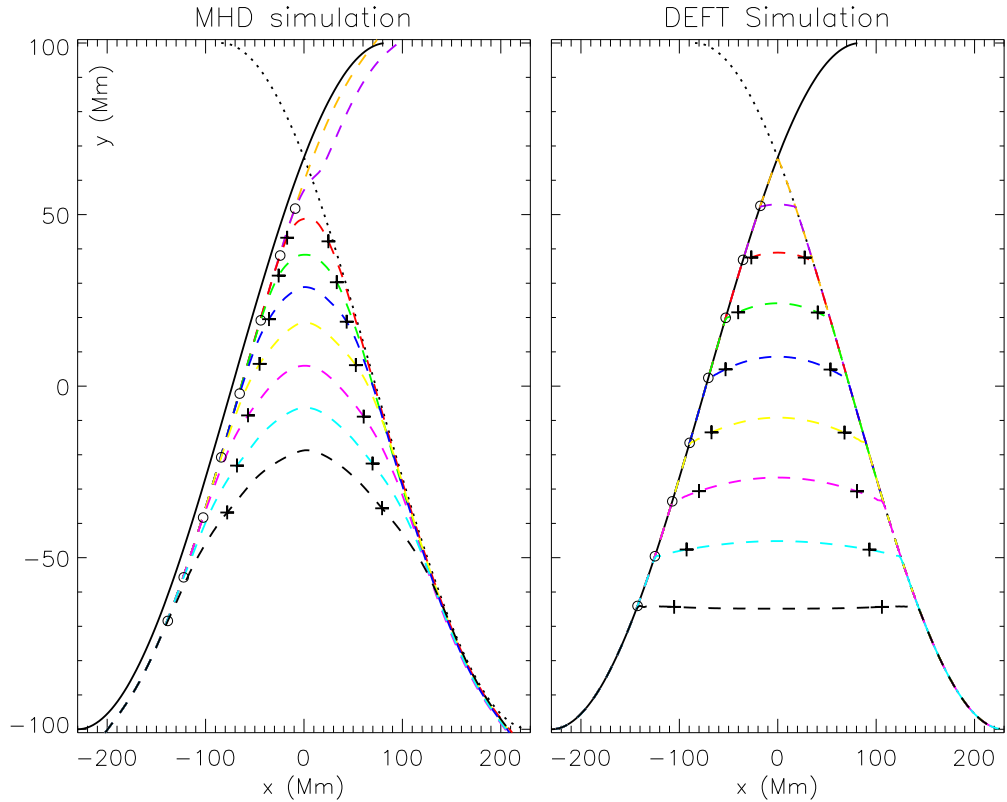}} \\
      \end{array}$
   \end{center}
   \caption[Current sheet plane. Line0 MHD evolution and DEFT simulation]
   {Current sheet plane. Both panels share the same $y$-axis. Left panel: MHD Line0 evolution projection on the current sheet plane. Solid (dotted) black line is the theoretical initial field line on the front (back) side of the current sheet. Orange dashed field line corresponds to $t=0$ s and violet dashed field line corresponds to $t=90$ s. The rest of the colored dashed lines correspond to the same times and colors as in Figure \ref{fig:rec_line0_evol}. Circles indicate the positions of the bends for each field line on the front side of the current sheet. No circles have been plotted for the initial field line. Crosses indicate the locations of the gas-dynamic shocks. Right panel: DEFT flux tube simulation. Each line represents the shape of the tube for the same times as in the left panel. Circles are the positions of the bends and crosses indicate the locations of the gas-dynamic shocks} 
   \label{fig:line0_shape_comp} 
\end{figure}

The limit of Equation \eqref{eqn:3D_Sirovatskii} when $z \rightarrow 0$ corresponds to the magnetic field components in the plane of the current sheet, and it is equivalent to the one described in the previous chapter by Equation \eqref{eqn:B_syrovatskii} with $B_{s} = B_{ey}$, and uniform guide field $B_{ex}$. Therefore, the axes of the initial three-dimensional tubes that cross the reconnection region on each side of the current sheet are well described by the two-dimensional field lines \eqref{eqn:initial_tube}, at least while the tubes are close to the CS plane. The tubes have legs that do not lie on this plane and have footpoints at the base of the arcade, but their retractions occur mostly while the field lines lie within the plane of the current, until they arrive to the top of the arcade, as described in the previous chapter. The left panel of Figure \ref{fig:line0_shape_comp} shows the evolution of Line0 projected on the plane of the current sheet. The solid (dotted) black line is the theoretical initial field line on the front (back) side of the current sheet (Equation \eqref{eqn:initial_tube} for a reconnection location $y_{R}^{\prime}$ and $\zeta_{0}=45$\degree). Each color represents a different time as described in the caption. Circles indicate the positions of the bends on the front side of the current sheet.

%
%

To be compared with the above three-dimensional simulations, we run DEFT simulations with the following parameters: reconnection location $y_{R}^{\prime} = 2/3$, reconnection angle $\zeta_{R}=37$\degree, and plasma-$\beta$ there equal to $0.54$. We implement temperature-independent viscosity and thermal conduction. The Reynolds number of this simulation is 300, and the Prandtl number is $1$. We divided the tube in $1000$ segments. We will present our results in the same units used for the three-dimensional simulation. The right panel of Figure \ref{fig:line0_shape_comp} shows the reconnected flux tube retraction for the DEFT simulation. The initial shape of the tube is similar to the reconnected shape of Line0. 

%

\begin{figure}[htbp]
   \centering
   \begin{center}$
      \begin{array}{c}
         \resizebox{!}{4.0in}{\includegraphics{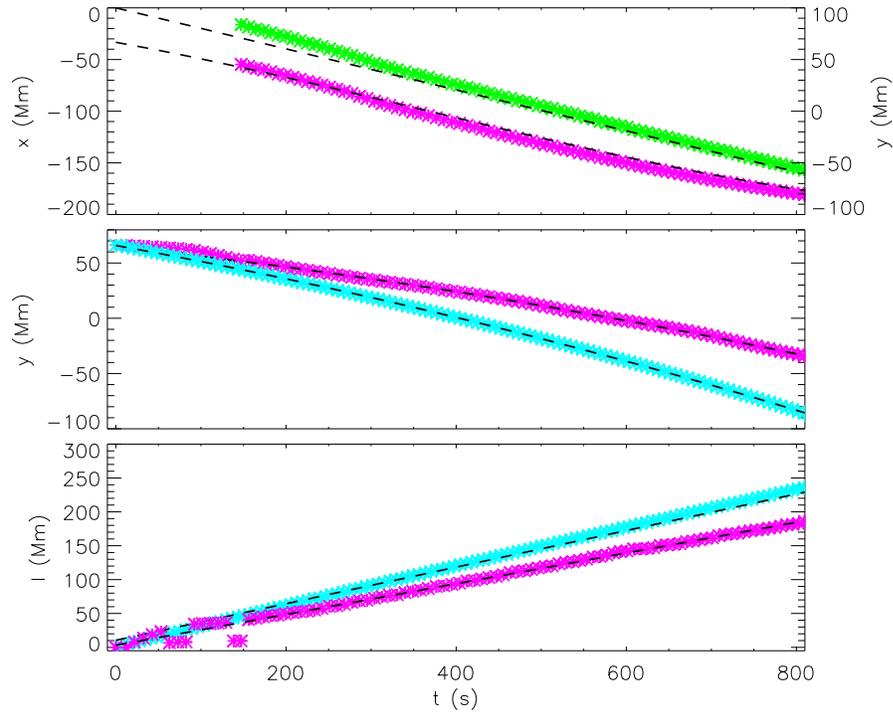}} \\
      \end{array}$
   \end{center}
   \caption[Line0 speed comparison]
   {Line0 speed comparison. All panels share the same horizontal axis (time in seconds). Top panel: the green symbols (left coordinate axis) show the $x$-position of the bend on the front side of the current sheet as time goes by for the MHD simulation of Line0. The dashed black line close to the green symbols represents the theoretical $x$-positions of the bend if it were moving along the legs at the local Alfv\'{e}n speed. The magenta symbols (right coordinate axis) show the $y$-position of the bend on the front side of the current sheet as time goes by for the MHD simulation of Line0. The dashed black line close to the magenta symbols represents the theoretical $y$-positions of the bend if it were moving along the legs at the local Alfv\'{e}n speed. Times before approximately $t=150$ s are not shown since the position of the bends are not well-defined while the tube reconnects. Middle panel: the magenta symbols show the $y$-position of the center of the tube as time goes by for the MHD simulation of Line0. The dashed black line close to the magenta symbols represents a quadratic fitting of the magenta curve, $y(t) = 65$ Mm $-0.08$ Mm s\textsuperscript{-1} $t-4.8 \times 10^{-5}$ Mm s$^{-2}$ $t^{2}$. The magenta symbols show the $y$-position of the center of the tube as time goes by for the DEFT simulation. The dashed black line close to the cyan symbols represents a quadratic fitting of the cyan curve, $y(t) = 66$ Mm $-0.13$ Mm s\textsuperscript{-1} $t-5.9 \times 10^{-5}$ Mm s$^{-2}$ $t^{2}$. Bottom panel: the magenta symbols show the length of the post-shock region, $l$, as time goes by for the MHD simulation of Line0. The dashed black line close to the magenta symbols represents a linear fitting of the magenta curve, $y(t) = 3.14$ Mm $+0.23$ Mm s\textsuperscript{-1} $t$. The cyan symbols show the length of the post-shock region, $l$, as time goes by for the DEFT simulation. The dashed black line close to the cyan symbols represents a linear fitting of the cyan curve, $y(t) = 10.0$ Mm $+0.27$ Mm s\textsuperscript{-1} $t$} 
   \label{fig:line0_speed_comp} 
\end{figure}

As times goes by, the concavity of the DEFT tube is less pronounced than in the run for Line0. This appears to be due to a smaller initial velocity of the center of the tube in the three-dimensional case and could be related to the finite time it takes to create the reconnected tube by the diffusion process. In the DEFT simulations, reconnection is assumed to happen instantaneously. The bends in both cases move at the local Alfv\'{e}n speed. The circles in the right panel of Figure \ref{fig:line0_shape_comp} were drawn assuming they do so, and they coincide with the actual bends. The top panel of Figure \ref{fig:line0_speed_comp} shows the $x$- and $y$-positions of the bend on the front side of the current sheet, as well as the theoretical motion they should follow if the move at the local Alfv\'{e}n speed along the legs of the field line (black dashed lines). 

The delay in the motion of the central part of the tube can be seen in the middle panel of Figure \ref{fig:line0_speed_comp}. Magenta color represents the downward motion of the center of the tube for Line0, and the cyan color shows  the motion of the DEFT tube. The dashed lines represent quadratic fits. Their accelerations are similar, but the initial speed for Line0 is significantly smaller than the one for the DEFT tube. In our thin flux tube model, the shape of the tube at the reconnection position has a sharp angle and it immediately retracts after being reconnected at Alfv\'{e}nic speed. The center part of Line0 stays in the top part of the current sheet longer than the DEFT tube does. This has some consequences on the parallel dynamics since the plasma responds to the changes in background magnetic pressure, as described in the previous chapter. Despite the mentioned differences, the retraction is surprisingly similar for both cases. 

%

\subsection{Parallel Dynamics}
  \label{sec:3D_MHD_parallel}

\begin{figure}[htbp]
   \centering
   \begin{center}$
      \begin{array}{c}
         \resizebox{5.5in}{!}{\includegraphics{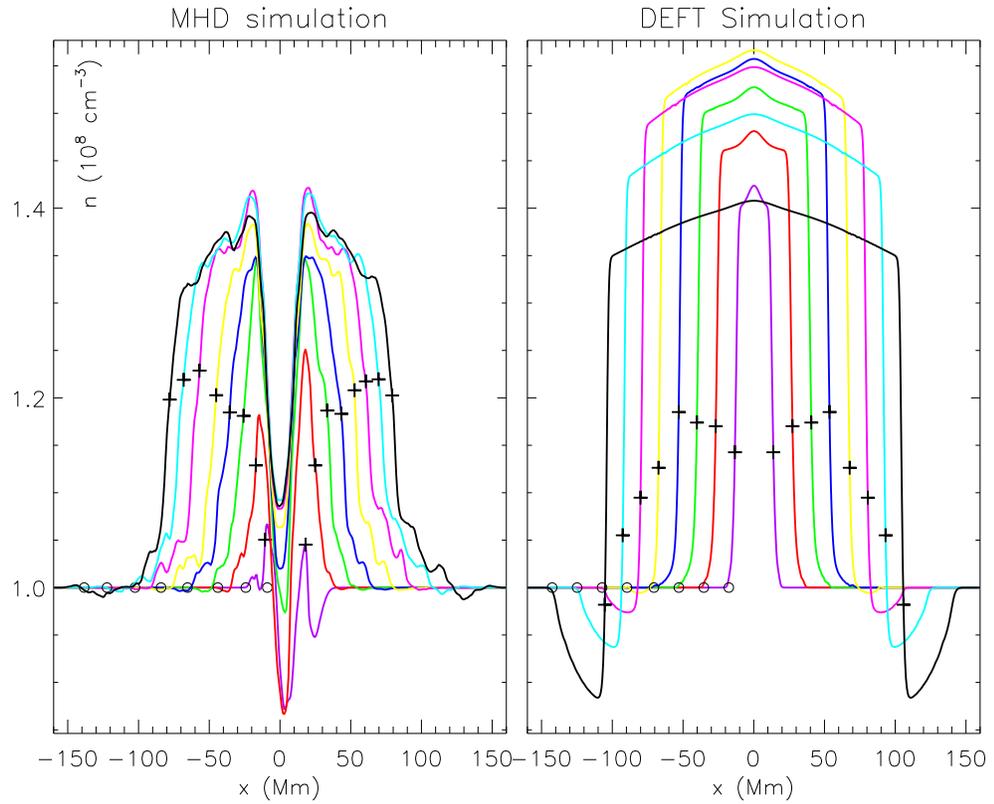}} \\
      \end{array}$
   \end{center}
   \caption[Line0 electron density comparison] 
   {Line0 electron density comparison. Both panels share the same $y$-axis. Left panel: each solid curve represents the electron density at a different time for Line0. Colors represent the same times as in Figure \ref{fig:line0_shape_comp}. Circles and crosses are also the same as for the left panel of Figure \ref{fig:line0_shape_comp}. Right panel: each solid curve represents the electron density at a different time for the DEFT simulation. Circles and crosses correspond to the same ones as in the right panel of Figure \ref{fig:line0_shape_comp}}
   \label{fig:line0_density_comp} 
\end{figure}

\begin{figure}[htbp]
   \centering
   \begin{center}$
      \begin{array}{c}
         \resizebox{5.5in}{!}{\includegraphics{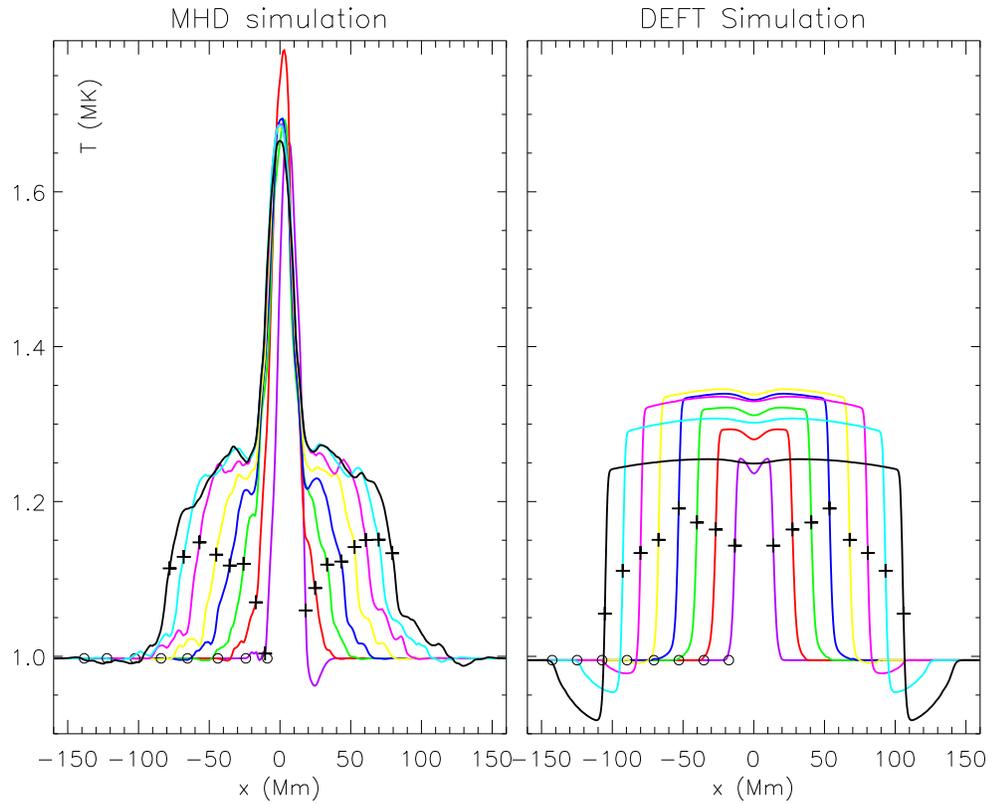}} \\
      \end{array}$
   \end{center}
   \caption[Line0 temperature comparison]
   {Line0 temperature comparison. Both panels share the same $y$-axis. Left panel: each solid curve represents the temperature at a different time for Line0. Colors represent the same times as in Figure \ref{fig:line0_shape_comp}. Circles and crosses are also the same as for the left panel of Figure \ref{fig:line0_shape_comp}. Right panel: each solid curve represents the temperature at a different time for the DEFT simulation. Circles and crosses correspond to the same ones as in the right panel of Figure \ref{fig:line0_shape_comp}}
   \label{fig:line0_temperature_comp} 
\end{figure}

The comparison between the parallel dynamics for both cases can be seen in Figures \ref{fig:line0_density_comp} and \ref{fig:line0_temperature_comp}. The first figure shows the evolution of the electron density for both cases. The left panel corresponds to Line0. The central density depression is caused by the initial reconnection episode. Ohmic heating from the enhanced resistivity creates a hot spot at the center of the tube that cannot be efficiently smoothed out due to the lack of thermal conduction in the ARMS computer program. Initial pressure equilibrium is quickly established through sound waves, leaving a density void in the center of the tube that persists for a long time. This is an artificial effect of the three-dimensional simulations. We hypothesize that this hot spot will quickly disappear in a real plasma due to thermal conduction. The thin flux tube model does not present this behavior since the effects of the ohmic heating are assumed to have already ceased by the time the simulation starts.

Sharp discontinuities in the electron density can be seen moving outward from the center of the tube. These are the gas-dynamic shocks. Crosses show their locations for each time, also shown in Figure \ref{fig:line0_shape_comp}. The density jumps by approximately $40$\%. The post-shock density increases while the center part of the tube is in the top part of the current sheet and decreases while this region moves through the lower half of the plane (cyan and black colors). This is as expected from our conclusions from the previous chapter. A decrease in density can also be seen following the bends (circles in the figure) for late times as the bends move through the lower part of the current sheet. Due to the increased concavity of the tube, this decrease in density is not as pronounced as in its DEFT counterpart. 

The electron density evolution for the DEFT simulation is shown in the right panels of the figure. The jump in density is very similar to the one for the MHD simulation. Its value increases as the central part of the tube moves through the top half of the current sheet and decreases during its motion through the lower half of the plane. The density voids after the bends are more pronounced in this case than in its three dimensional counterpart since the sections of the tube following the bends move through regions of the current sheet with lower magnetic pressure. Although the MHD simulations were stopped before the field lines arrived to the top of the arcade, we expect them to come to a rest at the top of the arcade. Therefore, the sections after the bends will eventually decrease their density as they move through the bottom half of the current sheet. The slowing of the tubes on top of the arcade for three-dimensional simulations has been shown by \citet{Linton_2009}, although no analysis of the parallel dynamics was attempted in that paper.

%

The hot spot at the center of the tube for the three-dimensional simulations can be seen in the left panel of Figure \ref{fig:line0_temperature_comp}. This panel shows temperature profiles for different times. The right panel shows the temperature profiles for the DEFT simulation. The jump in temperature across the gas-dynamic shocks is comparable for both simulations. 

%

The speed of the shocks along the tubes is shown for both cases in the bottom panel of Figure \ref{fig:line0_speed_comp}. The magenta symbols show the Line0 post-shock region length, $l$, as function of time. For each time shown in Figure \ref{fig:line0_temperature_comp}, this corresponds to the distance between crosses. The cyan color corresponds to the post-shock region length for the DEFT simulation. The speeds are similar. The slight difference could be attributed to the fact that the shock speed is related to the pre-shock temperature. These temperatures are different for each case since the regions after the bends have different temperatures, depending on where in the current sheet these sections are moving through. 

\begin{figure}[htbp]
   \centering
   \begin{center}$
      \begin{array}{c}
         \resizebox{5.5in}{!}{\includegraphics{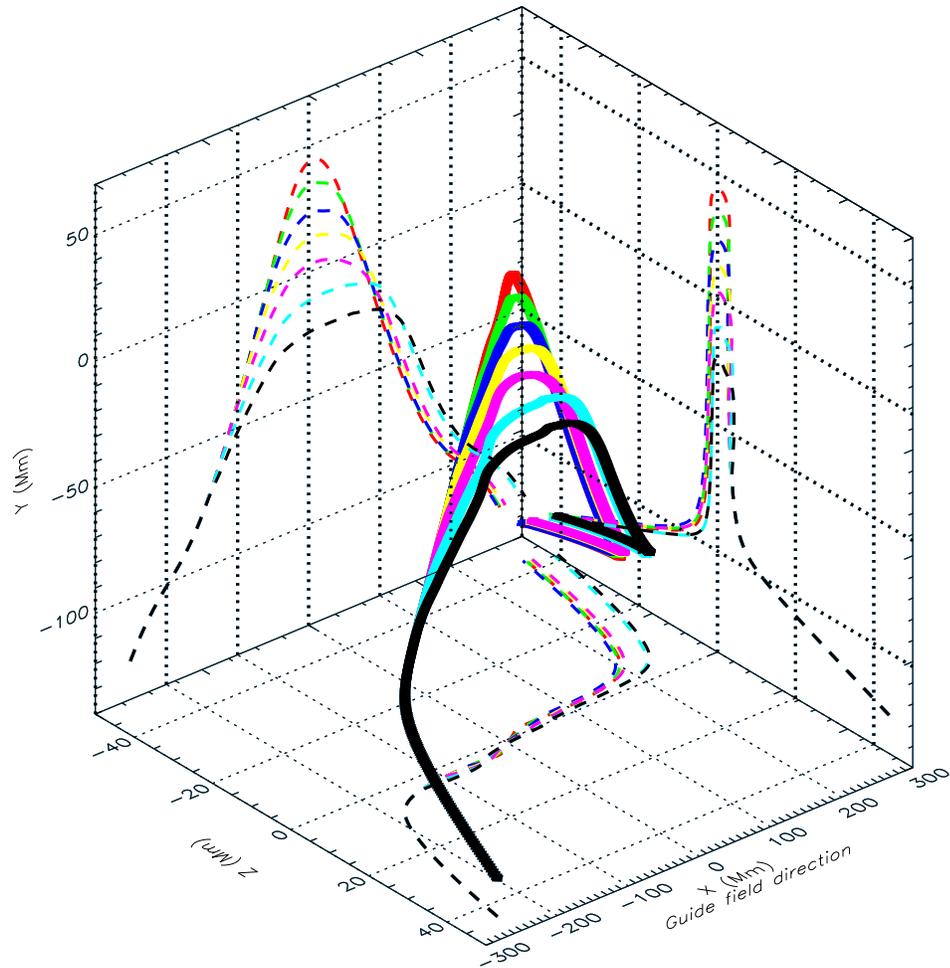}} \\
      \end{array}$
   \end{center}
   \caption[Line1 retraction]
   {Line1 retraction from the reconnection site. The colored solid lines are the field line curves for times at $90$ s intervals: $180$ s (red), $270$ s (green), $360$ s (blue), $540$ s (yellow), $630$ s (magenta), $720$ s (cyan), and $810$ s (black). Colored dashed lines are the field line projections on the box walls of the same color three-dimensional curve}
   \label{fig:rec_line1_evol} 
\end{figure}

\begin{figure}[htbp]
   \centering
   \begin{center}$
      \begin{array}{c}
         \resizebox{5.5in}{!}{\includegraphics{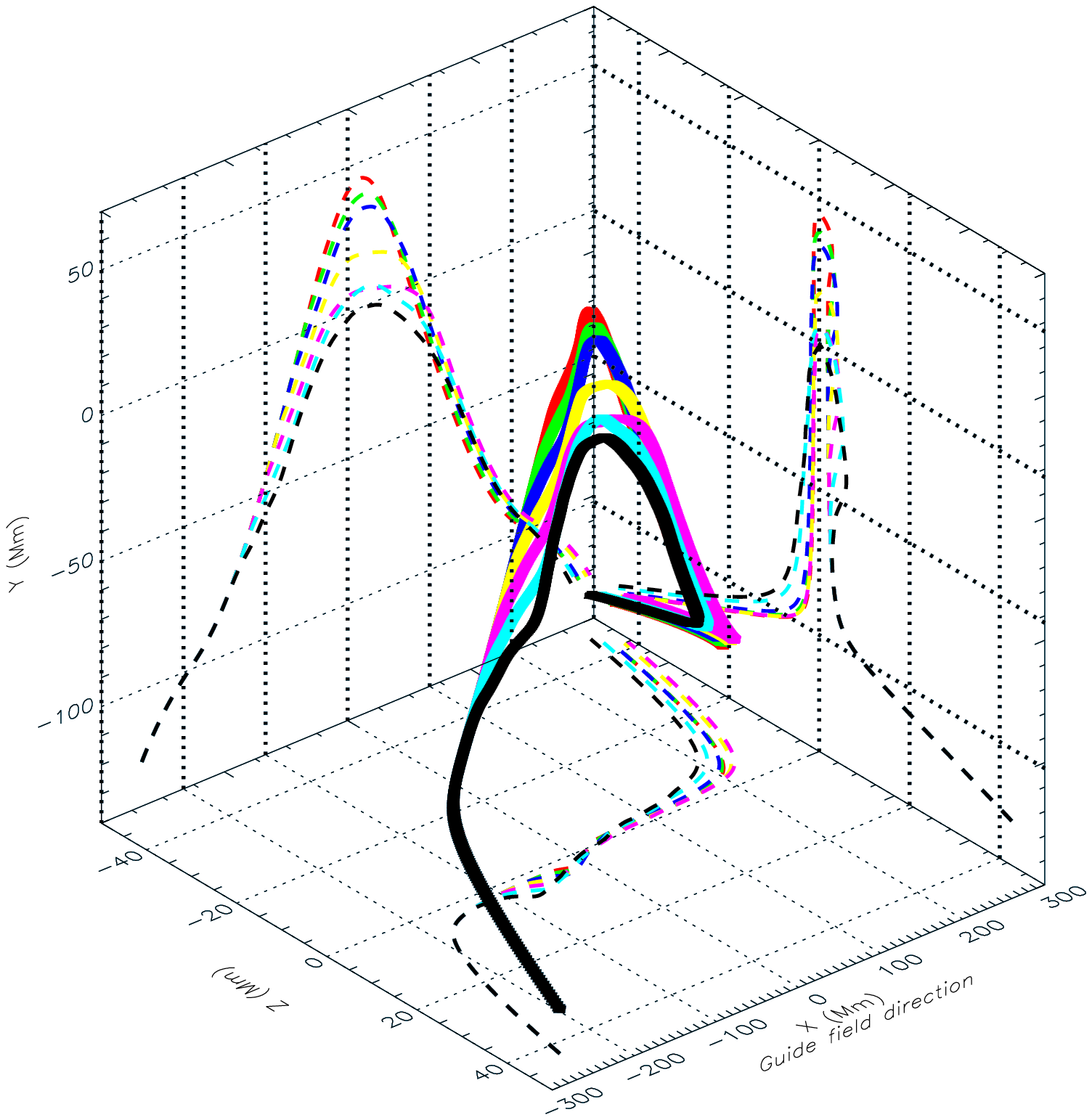}} \\
      \end{array}$
   \end{center}
   \caption[Line2 retraction]
   {Line2 retraction from the reconnection site. The colored solid lines are the field line curves for times at $90$ s intervals: $180$ s (red), $270$ s (green), $360$ s (blue), $540$ s (yellow), $630$ s (magenta), $720$ s (cyan), and $810$ s (black). Colored dashed lines are the field line projections on the box walls of the same color three-dimensional curve}
   \label{fig:rec_line2_evol} 
\end{figure}

Line0 behaves like a typical field line from a flux tube that reconnected at the high resistivity sphere. If all the field lines that cross the high resistivity sphere followed this behavior, the DEFT simulation would approximately reproduce their evolution. The evolutions of Line1 and Line2 are shown in Figures \ref{fig:rec_line1_evol} and \ref{fig:rec_line2_evol}, respectively. Apparently, for these field lines the disturbance caused by the initial reconnection event spontaneously excites patchy reconnection events later in the simulation. This can be seen in the artificial motion of the footpoint on the $z<0$ region. The source of this secondary reconnection may be numerical resistivity \citep{Linton_2009}. We do not know if this would occur in a real plasma. 

%
%

%
%

\section{Conclusions}

We have presented a comparison between three-dimensional MHD simulations and DEFT simulations. It seems that the assumptions and approximations of our thin flux tube model are not too restrictive and describe in reasonable terms the behavior of a flux tube that has been reconnected in a patch, as long as there are not secondary reconnections along the tube. The non-trivial assumption that a small enhanced diffusion region will disconnect field lines from one side of the current sheet and connect them to field lines on the other side is satisfied for the MHD simulations.

As predicted, reconnected field lines retract from the reconnection sphere in an almost two-dimensional motion parallel to the current sheet, and gas-dynamic shocks move along the field lines, heating and compressing the plasma. The jump conditions across the shock predicted by our model seem to be accurate. 

There are some expected differences in the field line shapes and parallel dynamics between the simulations, but they are surprisingly small, given the simplifications of our model. The main differences are a consequence of the initial velocity of the center of the tube. In three-dimensional simulations, the formation of the tube is delayed during the finite amount of time that reconnection takes to occur, and therefore the center region of the field lines spends a longer time on the top half of the current sheet. Since the bends move at the local Alfv\'{e}n speed along the field lines, the concavity of the field lines is much larger than for the DEFT simulations. 

We have only presented one case for the comparison between our model and MHD simulations. The chosen conditions are quite generic, and we expect agreement also for other cases. We hope that in the near future we can extend our analysis to more points in the parameter space. 

The thin flux tube equations are simpler than the fully three-dimensional MHD equations. This has allowed us to predict approximately the strength of the shocks generated by reconnection. The thickness of the shocks, determined by the transport coefficients, is non-trivial for solar coronal conditions. This thickness can be calculated using our model more accurately than through three-dimensional simulations since anisotropic, temperature-dependent viscosity and thermal conduction cannot be easily implemented in three-dimensional MHD simulations. In addition, due to the simplicity of our DEFT code, the exploration of the parameter space is less computationally-expensive and time-consuming.

%

\chapter{CONCLUSIONS}
   \label{chap:chap_4}

The foregoing chapters have given an account of several investigations and developments into the dynamics of magnetic loops following their formation by magnetic reconnection.

We have not only refined and improved a novel, time-dependent patchy reconnection model, but also presented general equations that govern the evolution of thin flux tubes. The ideal part of these equations could be applied to model thin flux tubes in several non-relativistic astrophysical situations. For example, there is evidence for localized and transient nature of magnetic reconnection in the magnetosphere. The first such evidence was found in so-called {\em flux transfer events} identified in situ magnetospheric observations \citep{Russell_1978}. There, reconnection starts and stops in a few minutes resulting on the ripping off of flux tubes. It is also believed that magnetic fields generated by the solar dynamo are convectively collapsed into coherent flux tubes (through processes which are still not well understood), become buoyant in the solar convection zone, and consequently erupt as active regions \citep{Schussler_1994, Nandy_2006}. If non-ideal processes are unimportant, our equations could be used to study the mentioned phenomena.

Through our investigations, we have learned that even for plasmas that are strongly magnetized (like the ones in the lower solar corona), where thermal pressure is much smaller than magnetic pressure (low plasma-$\beta$), the pressure-driven parallel term in the thin flux tube momentum equation cannot be neglected since magnetic forces are only perpendicular to the magnetic field. This term modulates tube's acoustics and prevents unlimited mass pile up inside the tubes.

Standard models of reconnection assume perfectly anti-parallel magnetic fields. We believe this to be too restrictive for the solar corona, where photospheric motions produce coronal fields with varied directions. In addition, the observation of transient and localized features, like SADs and SADLs, and possibly individual loops (like the ones seen with TRACE) seem to indicate that, at least for these cases, reconnection occurs in short-lived patches. The traditional steady-state, two- or $2.5$-dimensional reconnection scenarios are not adequate to describe them.  

We believe our work has shown that the simple addition of a guide field extends considerably the range of possible retracting tube geometries. If only anti-parallel fields are assumed, the shape for post-reconnection outflows corresponds to collimated jets. On the other hand, patchy reconnection with skewed fields not only accommodate jet geometries (if the current sheet is observed edge-on), but also loop-shape geometries (if the plane of the sky is almost parallel to the current sheet plane). SADs and SADLs seem to be examples of each case, respectively.

The parallel dynamics inside the tubes are as rich as the perpendicular dynamics. Super-sonic parallel flows are generated at the RDs that move along the legs of the tubes, ``transmitting'' at the local Alfv\'{e}n speed the information of what has happened at the reconnection location (change in topology). The super-sonic flows collide at the center and result in gas-dynamic shocks that also move along the legs of the tubes, heating and compressing the plasma. This generates two long-lived (confined by the inflows) and hot plasma plugs that move outwardly in opposite directions from the reconnection region. 

The reconnection patch creates two V-shaped reconnected tubes that shorten as they retract in opposite directions, due to magnetic tension. One of them moves upward toward the top edge of the CS, and the other one moves downward toward the top of the underlying arcade. The plasma inside each tube is far from equilibrium and the densities achieved in the plasma plugs can be much higher than the steady-state Rankine-Hugoniot post-shock values. This is only possible for very strong shocks and time-dependent models like the one we have presented in this dissertation. The strength of the shocks is determined by the Mach number of the parallel inflows inside the tubes (Equation \eqref{eqn:Mach_num}). For strongly magnetized plasmas (small plasma-$\beta$), the jump in density across the shocks may be more than one order of magnitude, depending on the half reconnection angle value.

We have also included realistic non-ideal terms in our equations, namely thermal conduction and viscous momentum exchange. We restricted ourselves to strongly magnetized plasmas, for which the magnetic field confines the transport processes to be mostly aligned with its direction. For fully ionized plasmas, like the ones found in the solar corona, the transport coefficients have a strong temperature dependence ($\sim T^{5/2}$). The addition of these diffusive processes has proven to be essential for a self-consisting generation of the gas-dynamic shocks. 

It is generally believed that shocks require a kinetic treatment. We have shown that a simple fluid treatment is adequate whenever the diffusive coefficient's growth with temperature is faster than the particle's mean free path growth with temperature. This condition is satisfied for fully ionized plasmas where Coulomb potentials determine the particle mean free paths ($\sim T^{2}$) and Spitzer's transport coefficients ($\sim T^{5/2}$) are used.  In this case, the thickness of the shocks exceeds the particle mean free paths by at least an order of magnitude. When temperature independent transport coefficients are used, the results are several orders of magnitude smaller than the corresponding mean free paths, which in the past created some concern about the use of fluid equations for shock studies. 

The validity of assuming fluid equations, as opposed to kinetic theory, may also be a concern if the ratio between the diffusion heat flux and the value for streaming particles is larger than a certain value \citep{Campbell_1984}. We tested the collisionality of the plasma in our simulations, and obtained ratios that support the use of continuum fluid equations. 

We theoretically predicted the thickness of the shocks, as well as their internal structure. Our analysis is novel in its focus on solar flare conditions, including very low Prandtl numbers (ratio of viscosity to thermal conductivity). In this case, the shocks present a relatively thin isothermal sub-shock where most of the plasma compression occurs, and thermal fronts that can be as long as the entire tubes and where temperature increases considerably. These fronts move along the legs of the tube and may be relevant to chromospheric evaporation, although we have not yet analyzed this effect. Our model can be used to couple the reconnection regime high in the solar corona with evaporation dynamics at the footpoints. 

Traditional models have not been able to explain why some of the observed SADs appear to be hot and relatively devoid of plasma. We have shown that plasma depletion naturally occurs in flux tubes that retract along regions of decreasing external magnetic pressure. We have demonstrated this effect on non-uniform Syrovatski{\v i}-type CSs with skewed magnetic fields, that seem to be ubiquitous in the solar corona. In this kind of CSs, the background magnetic pressure has its maximum at the center of the CS plane and decreases toward its edges. The hot temperature of the SADs can be explained by the long thermal fronts that heat the plasma in regions of density depletion. 

We have modeled the interaction of the DOWNWARD moving tubes with the underlying arcade by means of a simple damping force. A DOWNWARD moving tube will arrive at the top of the arcade that will slow it to a stop. Here, the perpendicular dynamics is halted, but the parallel dynamics continues along its legs; the RDs are shut down, and the gas is rarefied to even lower densities. After the tubes lie on top of the arcade, rarefaction waves continue decreasing plasma density. The hot post-shock regions continue evolving, determining a long lasting hot region on top of the arcade. We have also provided an observational method based on total emission measure and mean temperature, that indicates where in the CS the tube has been reconnected. 

The DEFT computer program we developed solves the thin flux tube equations and simulates the retraction of post-reconnection tubes for a wide range of background conditions. The program seems to be very robust and resolves the gas-dynamic shocks efficiently. Its output can be used as input for simulations that deal with chromospheric evaporation (for DOWNWARD moving tubes) and with solar ejecta (UPWARD moving tubes). A comparison of DEFT simulations with fully three-dimensional magnetohydrodynamic simulations has been presented to assess the validity of the thin flux tube model. The seemingly restrictive thin flux tube assumptions reproduce the general behavior of the three-dimensional plasma. 

The correct answer to the disparity between observations and models regarding the speed of localized reconnection outflows is still not known, but most likely the slowing down is due to the interaction between tubes and their surroundings, which we have not considered in the present work. Here, the reconnected tubes are assumed to be completely isolated from their background, therefore we have restricted ourselves to standard reconnection scenarios where the outflows are Alfv\'{e}nic. We do not claim that we have explained all SADs' observables, but our model seems to support the idea that they are a by-product of reconnection. 

There are many improvements to the model that can be incorporated and we hope that this work has proven its usefulness and opens the door to such additions.

\backmatter

 \newpage
  \vspace*{9cm}
     {\begingroup
    \normalsize
\begin{center}
\underline{REFERENCES CITED}
\end{center}
  \endgroup
  \addcontentsline{toc}{chapter}{REFERENCES CITED}
  \newpage}

\singlespacing


\appendix{LIST OF ABBREVIATIONS}

\begin{longtable}{p{0.75in} l p{4.25in} @{}}

Al          & -- &     Aluminum \\
AlMg        & -- &     Aluminum Magnesium \\
ARMS        & -- &     Adaptively Refined Magnetohydrodynamics Solver (numerical MHD solver) \\
CME         & -- &     Coronal Mass Ejection             \\
CS          & -- &     Current Sheet             \\
CSHKP       & -- &     Model of flares/CMEs proposed by \citealt{Carmichael_1964, Sturrock_1968, Hirayama_1974, Kopp_1976} \\
DEFT        & -- &     Dynamical Evolution of Flux Tubes \\
DEM         & -- &     Differential Emission Measure (per flux)  \\
DOE         & -- &     Department of Energy \\
EM          & -- &     Emission Measure (per flux)  \\
ESA         & -- &     European Space Agency \\
EUV         & -- &     Extreme Ultraviolet       \\
Fe          & -- &     Iron \\
GDS         & -- &     Gas-dynamic Shock         \\
GEM         & -- &     Geospace Environmental Modeling\\
HXR         & -- &     Hard X-ray \\
JES         & -- &     Joined Equilibrium Solution\\
JPL         & -- &     Jet Propulsion Laboratory \\
LASCO       & -- &     Large Area Solar Coronagraph \\
LHS         & -- &     Left Hand Side \\
MHD         & -- &     Magnetohydrodynamics \\
MK          & -- &     Mega-Kelvin \\
NASA        & -- &     National Aeronautics and Space Administration \\
NSF         & -- &     National Science Foundation \\
RD          & -- &     Rotational Discontinuity  \\
RHS         & -- &     Right Hand Side \\
SAD         & -- &     Supra-arcade Downflow \\
SADL        & -- &     Supra-arcade Downflowing Loop  \\
SMS         & -- &     Slow-mode Shock         \\
SOHO        & -- &     Solar \&\ Heliospheric Observatory \\
SXR         & -- &     Soft X-ray \\
SXT         & -- &     Soft X-ray Telescope on Yohkoh \\
TFT         & -- &     Thin Flux Tube         \\
TRACE       & -- &     Transition Region and Coronal Explorer\\

\end{longtable}

\end{document}